\documentclass[reprint,twocolumn,secnumarabic,amssymb, aps, prx]{revtex4-2}

\usepackage{graphicx}
\usepackage{gensymb}
\usepackage{amsmath}
\usepackage{braket}
\usepackage{lipsum}
\usepackage[colorlinks,linkcolor=black,citecolor=black,urlcolor=black,filecolor=black]{hyperref}
\usepackage{wrapfig}

\newcommand{\phii}{\varphi}

\begin{document}
\title{
Engineering quantum criticality and dynamics on an analog-digital simulator
}

\author{
Alexandra~A.~Geim$^{1}$,
Nazl\i~U\u{g}ur~K\"oyl\"uo\u{g}lu$^{1,2}$,
Simon~J.~Evered$^{1}$,
Rahul~Sahay$^{1}$,
Sophie~H.~Li$^{1}$, 
Muqing~Xu$^{1}$,
Dolev~Bluvstein$^{1,3}$,
Nik~O.~Gjonbalaj$^{1}$,
Nishad~Maskara$^{1}$,
Marcin~Kalinowski$^{1}$,
Tom~Manovitz$^{1,\dagger}$,
Ruben~Verresen$^{4}$,
Susanne~F.~Yelin$^{1}$,
Johannes~Feldmeier$^{1}$,
Markus~Greiner$^{1}$,
Vladan~Vuleti\'c$^{5}$,
Mikhail~D.~Lukin$^{1}$
}
\affiliation{
$^1$Department~of~Physics,~Harvard~University,~Cambridge,~MA~02138,~USA\\
$^2$Harvard~Quantum~Initiative,~Harvard~University,~Cambridge,~MA~02138,~USA\\
$^3$California~Institute~of~Technology,~Pasadena,~CA~91125,~USA\\
$^4$Pritzker~School~of~Molecular~Engineering,~University~of~Chicago,~Chicago,~IL~60637,~USA\\
$^5$Department~of~Physics~and~Research~Laboratory~of~Electronics,~Massachusetts~Institute~of~Technology,~Cambridge,~MA~02139,~USA\\
$^\dagger$Present~address:~Department~of~Physics~of~Complex~Systems,~Weizmann~Institute~of~Science,~Rehovot~7610001,~Israel
}

\begin{abstract}
Understanding emergent phenomena in out-of-equilibrium interacting many-body systems is an exciting frontier in physical science~\cite{anderson_more_1972,girvin_modern_2019,eisert_quantum_2015}. While quantum simulators represent a promising approach to this long-standing problem~\cite{preskill_quantum_2018}, in practice it
can be
challenging to 
directly realize
the required interactions, measure arbitrary observables, and mitigate errors.
Here we use coherent mapping between the Rydberg and hyperfine qubits in a neutral atom array simulator~\cite{bluvstein_fault-tolerant_2025} to engineer and probe complex quantum dynamics. We combine efficient analog dynamics with fully programmable state preparation and measurement, 
leverage non-destructive readout for
loss information
and atomic qubit reuse, and use an atom reservoir for replacing lost atoms.
With this analog-digital approach, 
we first demonstrate dynamical engineering of ring-exchange and particle-hopping dynamics via Floquet driving~\cite{goldman_periodically_2014} and measure the spectral function of single excitations by evolving initial superposition states~\cite{knap_probing_2013}.
Extending these techniques to a 271-site kagome lattice, 
we employ
closed-loop optimization~\cite{rosi_fast_2013} to target an out-of-equilibrium critical quantum spin liquid of the Rokhsar-Kivelson type~\cite{rokhsar_superconductivity_1988}. 
We observe the key features of such a state, including the absence of local order, many-body coherences between nearly equal-amplitude dimer configurations over up to 18 sites, and universal correlations consistent with predictions from field theory~\cite{fradkin_field_2013}.
Together, 
these results pave the way for using dynamical control in analog-digital quantum simulators to study complex quantum many-body systems.
\end{abstract}

\maketitle

Emergent phenomena in interacting many-body systems lie at the frontier of multiple subfields across the physical sciences~\cite{anderson_more_1972, girvin_modern_2019,eisert_quantum_2015}. 
Understanding these phenomena is challenging, as numerical simulations rapidly become intractable 
in strongly correlated systems,
while direct microscopic probes are typically inaccessible in real-world systems and materials. 
Quantum processors offer a route to large-scale, programmable simulations of many-body systems and their dynamics.
In particular, 
analog quantum simulators, which aim to directly realize a quantum Hamiltonian~\cite{lloyd_universal_1996}, have already shed light on diverse phenomena ranging from quantum phase transitions~\cite{greiner_quantum_2002,scholl_programmable_2021} and quantum magnetism~\cite{deng_effective_2005,xu_neutral-atom_2025}, to non-equilibrium dynamics~\cite{ andersen_thermalization_2025} and thermalization~\cite{kaufman_quantum_2016, schreiber_observation_2015}. 
To access new regimes of many-body physics,
recent efforts have focused on tailoring the native dynamics~\cite{de_leseleuc_observation_2019, gonzalez-cuadra_observation_2025, young_observing_2024} or engineering the necessary interactions 
through periodic driving
~\cite{periwal_programmable_2021, miller_two-axis_2024, morong_engineering_2023}.
In parallel, the advent of gate-based digital simulation is rapidly opening doors to new physical models and complex observables~\cite{cochran_visualizing_2025, evered_probing_2025,will_probing_2025}, although large gate counts and circuit depths pose a challenge for 
near-term implementations at large scales~\cite{beverland_assessing_2022}.

\begin{figure*}
\includegraphics[width=2\columnwidth]{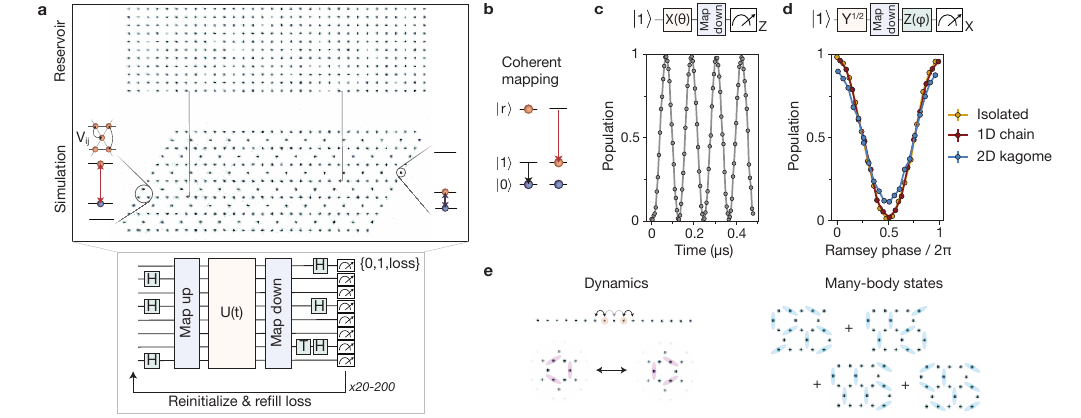}
\caption{\textbf{Analog-digital atomic processor for probing many-body quantum matter.} \textbf{a,} Analog evolution is performed in the interacting Rydberg qubit manifold and the hyperfine spin qubit is used for digital processing. Non-destructive readout of each atom obtains spin and loss information, after which lost sites are refilled from the reservoir and the spin state is reinitialized. The circuit repeats until the reservoir is depleted. The geometry of the simulation zone is programmed for each experiment, shown here for a 271-site kagome lattice. 
\textbf{b,} Two $\pi$-pulses are used to map between the qubit manifolds, shown for mapping down from the Rydberg to the hyperfine qubit. 
\textbf{c,} 
Rabi oscillations on the Rydberg qubit, measured after mapping down.
The contrast is 97.8(1)\% and loss is $\approx 1.8\%$ (not shown),
dominated by loss when turning off the tweezer traps and rearrangement infidelity as the reservoir depletes. 
Here and later, all data are locally postselected on no atom loss. \textbf{d,} Ramsey fringe showing coherence after mapping down $\ket{+}$. Atoms are located at next-nearest-neighbor sites for the 1D chain and 2D kagome lattice, emulating a blockaded many-body state. The contrast is reduced in 2D where effects of long-range Rydberg interaction tails are larger. 
\textbf{e,} Many-body states and dynamics can be engineered and probed using this zoned analog-digital architecture.}
\label{fig1}
\end{figure*}

Hybrid approaches that combine continuous analog evolution with coherent digital computation offer a promising route to overcoming these challenges~\cite{preskill_quantum_2018}, leveraging the efficiency of analog dynamics while retaining the universal control afforded by quantum circuits~\cite{senoo_high-fidelity_2025, andersen_thermalization_2025, brydges_probing_2019}.
Here we realize an analog-digital quantum simulator in a Rydberg atom array based on coherent mapping between the interacting Rydberg and digital hyperfine qubit manifolds. 
We use this architecture 
to first explore Hamiltonian engineering via periodic driving, employing programmable initial states to study 
six-body ring exchange and two-body hopping as examples.
These represent key ingredients for studying dynamics and excitations of complex phases, and notably underpin the realization of dynamical lattice gauge theories which are difficult to simulate classically~\cite{halimeh_quantum_2025}.

We then apply our hybrid approach to two-dimensional systems, focusing on the $U(1)$ quantum spin liquid (QSL)~\cite{fradkin_short_1990,fradkin_field_2013,rokhsar_superconductivity_1988,moessner_quantum_2008,read_valence-bond_1989}, an exotic state whose experimental observation remains elusive.
Featuring long-range entanglement and no local order, it is a paradigmatic example of a strongly-correlated system described by an emergent gauge theory similar to electromagnetism~\cite{fradkin_short_1990, fradkin_field_2013} and is closely linked to the resonating valence bond theory introduced by Anderson~\cite{fazekas_ground_1974}. 
As compared with the recently studied gapped $\mathbb{Z}_2$ QSLs~\cite{semeghini_probing_2021, satzinger_realizing_2021}, $U(1)$ QSLs are more challenging to realize due to their unstable, gapless nature~\cite{fradkin_field_2013}.
Enabled by the increased experimental cycle rate, we target an out-of-equilibrium $U(1)$ QSL
in the honeycomb quantum dimer model~\cite{sahay_quantum_2023} using analog state preparation optimized through a closed-loop quantum-classical procedure.
We observe microscopic ordering characteristic of a Rokhsar-Kivelson state~\cite{rokhsar_superconductivity_1988}, correlations consistent with field theory predictions, and further employ digital gates to directly measure many-body coherences.
\subsection*{Fast analog-digital quantum simulations}

In our experiments, programmable arrays of $^{87}$Rb atoms trapped in optical tweezers are used to implement both analog and digital evolution~\cite{evered_probing_2025, bluvstein_fault-tolerant_2025}. 
Analog evolution is implemented via excitation of the $\ket{1}\leftrightarrow \ket{r}$ Rydberg qubit with global laser pulses, such that the  
many-body unitary evolution $U(t)$ is governed by the Hamiltonian
\begin{equation}\label{eq:Hamiltonian}
    \frac{H}{\hbar} = \frac{\Omega(t)}{2}\sum_i\sigma^x_i - \sum_i n_i (\Delta(t) - \delta_i(t)) + \sum_{i<j} V_{ij}n_in_j
\end{equation}
where $\Omega$ is the Rabi frequency, $\Delta$ is the global detuning on the Rydberg transition, $\delta_i$ are site-resolved detunings, $\sigma_i^x = \ket{1_i}\bra{r_i} + \ket{r_i}\bra{1_i}$ and $n_i = \ket{r_i}\bra{r_i}$ is the Rydberg occupation number.
$V_{ij} = V_0/|\mathbf{x}_i - \mathbf{x}_j|^6$ denotes the van der Waals (vdW) interaction which prevents multiple Rydberg excitations within the blockade radius $R_b/a = (V_0/\Omega)^{1/6}$  for lattice spacing $a$; throughout we use $R_b/a \approx  1.3-1.4$, forbidding nearest-neighbor excitations. 

Digital processing is performed in the long-lived hyperfine clock qubit, $\ket{0} \leftrightarrow \ket{1}$, on which optical Raman transitions implement fully programmable single-qubit gates.
Fast Raman and Rydberg $\pi$-pulses coherently transfer the quantum state between the two qubit manifolds (Fig.~1b).
We illustrate this method for single qubits
by performing Rydberg Rabi oscillations and mapping the resulting state down to the hyperfine qubit for readout
(Fig.~1c). We further benchmark the preservation of phase coherence during the mapping in Fig.~1d, where the slightly reduced Ramsey contrast is due to crosstalk from the long-range tails of the vdW interaction during the mapping time (Methods).

This hybrid approach allows for two key features. First, it enables state preparation and measurement beyond the Pauli $Z$-basis. 
Second, non-destructive readout of the hyperfine qubit~\cite{bluvstein_fault-tolerant_2025} enables error detection based on loss information, where atom loss is a dominant source of error in Rydberg simulations that conventionally results in a measurement error. This includes correlated loss events previously observed during analog Rydberg evolution~\cite{boulier_spontaneous_2017}. 
The mapping protocol converts state initialization errors and population in blockade-violating states into 
these 
detectable loss events (Methods).
Finally, since atoms are retained after readout, they can be reinitialized and reused, thereby increasing the experimental cycle rate by over an order of magnitude.
Together these features give access to new observables, provide native error-mitigation techniques, and allow for measurements with high sample complexity.

\subsection*{Dynamic Floquet Hamiltonian engineering}
\begin{figure}
\includegraphics[width=1\columnwidth]{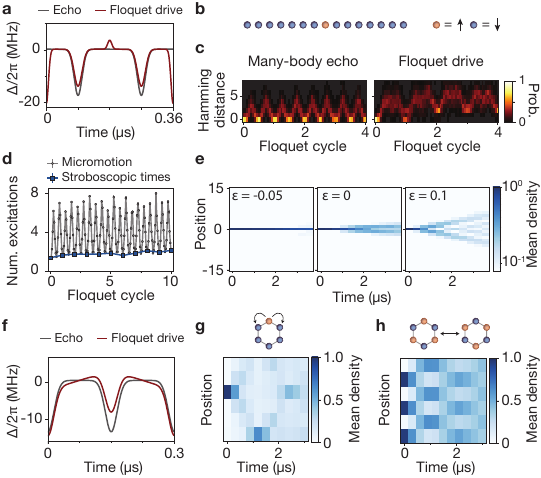}
\caption{\textbf{Hamiltonian engineering via periodic global detuning drives.} 
\textbf{a,} Example detuning profiles per Floquet cycle. Gray profile implements a many-body echo via $\pi$ phase jumps of the Rydberg drive, realized as smooth detuning pulses. Perturbations are added on top of this echo to engineer an effective Hamiltonian (red profile).
\textbf{b,} Schematic of a single Rydberg excitation initialized at the center of a 31-site chain, used as the initial state in \textbf{c-e}.
\textbf{c,} Dynamics of microstate populations, grouped by Hamming distance from the initial state in the $Z$-basis and computed over the central 19 sites of the chain.
Under the many-body echo, the state periodically revives, here twice per Floquet cycle (left).
The Floquet drive perturbs the dynamics, illustrated by the modified evolution of the populations, to yield an effective Floquet Hamiltonian in the Rydberg qubit manifold (right).
\textbf{d,} At stroboscopic times, the number of spin excitations is approximately conserved. 
\textbf{e,} Single-spin quantum walk under an effective blockaded XX Hamiltonian. The hopping rate is tuned via the drive parameter $\varepsilon$ (Methods).
\textbf{f,} Detuning profile per Floquet cycle used to engineer simultaneous two-body hopping and ring-exchange dynamics, and resulting dynamics on a six-site hexagonal plaquette starting from
\textbf{g,} a single spin and \textbf{h,} a N\'eel state. 
Data are averaged over 15 plaquettes evolved in parallel. The decay in contrast is primarily attributed to the sensitivity to static detuning offsets (Methods).
Explicit parameterizations for \textbf{a} and \textbf{f} and the associated parameters for each plot are given in Methods.}
\label{fig2}
\end{figure}

We first explore the engineering of interactions under nearest-neighbor Rydberg blockade
in small systems 
using periodic driving~\cite{ koyluoglu_floquet_2024,feldmeier_quantum_2024}. 
The key idea is to use a time-periodic global detuning drive to engineer a Floquet Hamiltonian, inspired by the dynamical control of quantum many-body scars~\cite{bluvstein_controlling_2021}.
In the simplest case, detuning $\pi$-pulses implement a many-body
echo, resulting in no stroboscopic dynamics~\cite{bluvstein_controlling_2021} (Fig.~2c). Perturbations away from this echo can then be used to control the evolution and generate the desired Floquet Hamiltonian, given by the sum of time-evolved operators over each period (Methods). 
Notably, although the intermediate dynamics generally involves many non-number-conserving states, we can engineer approximate number conservation at stroboscopic times (Fig.~2d).

Using this approach, we experimentally realize two types of effective Hamiltonians. First, we realize a hopping Hamiltonian with chemical potential $\mu$,
nearest-neighbor blockaded hopping energy $J$, 
and next-nearest neighbor interaction $U$,
which can be continuously tuned by varying the parameters of the periodic detuning profile $\Delta(t)$~\cite{koyluoglu_floquet_2024} (Methods). Fig.~2e demonstrates the observation of a single-particle quantum walk in a one-dimensional spin chain for increasing $J$, where the characteristic interference pattern is visible in the density profile. 
Importantly, this approach enables faster number-conserving hopping dynamics than can be reached at large static detuning (ED Fig.~3b).
These results demonstrate that the Floquet protocol works in practice and is well captured by the ansatz hopping Hamiltonian, even beyond the regime of small perturbations to the many-body echo (Methods).

Next, we demonstrate the ability to engineer higher-body interactions. Specifically, on hexagonal plaquettes we generate ring exchange dynamics by additionally engineering three-body interactions such that, at large $\mu$, the six-body ring exchange appears at second order~\cite{feldmeier_quantum_2024}.
To isolate the dynamics associated with the two-body and six-body dynamics, we prepare and map up initial states of a single spin excitation and a N\'eel state, respectively (Figs.~2f-h). 
Crucially, we observe that the hopping and ring-exchange dynamics occur on the same timescale, quantitatively different to perturbative dynamics generated at large static detuning at higher orders in $\Omega/\Delta$.
Such dynamics can also stabilize entangled many-body states (ED Fig.~4e).

Finally, we explore techniques for directly probing the resulting quantum dynamics.
Focusing on the hopping Hamiltonian, we initialize and evolve two Rydberg excitations separated by a single site (Figs.~3a,b), and observe a clear modification of the interference pattern for the two-spin dynamics compared to that of isolated single spins shown in Fig.~2e. This can be understood as the blockade constraint inducing an effective repulsive spin-spin interaction, which reduces the density within the overlapping light cones of the two spins.
This measurement therefore reveals the interacting nature of the Hamiltonian.
To gain further insight into the low-energy properties of the engineered Hamiltonian, we employ many-body spectroscopy which directly measures the single-particle dispersion~\cite{knap_probing_2013}. 
The key idea is to perform an interferometric measurement between a reference state with no excitations, $\ket{\Psi_\downarrow} = \ket{1}^{\otimes 31}$, and a probe state with a single excitation, $\sigma^x_0 \ket{\Psi_\downarrow}$.
We prepare the central spin of the chain in one of four superposition states in the hyperfine qubit, map this state up to Rydberg qubit, 
evolve
under the Floquet Hamiltonian for time $t$, and map back down for measurement in the $X$-basis (Fig.~3c). 
Using these measurements, we extract the Green's function $\mathcal{G}(r,t) = \bra{\Psi_\downarrow}\sigma^x_r(t) \sigma^x_0(0) \ket{\Psi_\downarrow}$ and plot its Fourier transform in Fig.~3d.
The fitted dispersion relation 
$\omega(k)/2\pi  =  0.407(2) - 0.281(2)\cos{ka}$
has the expected tight-binding form, and the real and imaginary parts of the spectral function are qualitatively consistent with Kramers-Kronig relations.
Furthermore, this measurement demonstrates simultaneous phase coherence of the hyperfine and Rydberg qubits throughout the entire evolution (Methods).

\begin{figure}
\includegraphics[width=1\columnwidth]{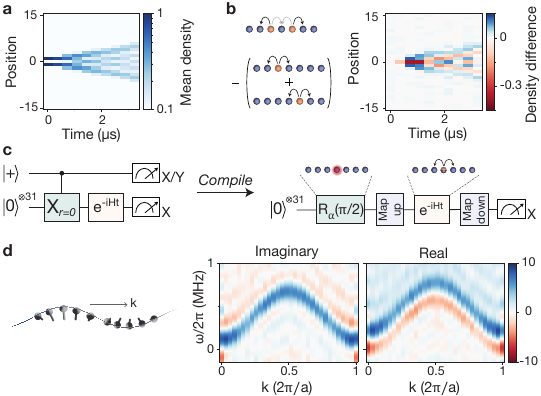}
\caption{\textbf{Dynamical probes of interactions and excitations.} \textbf{a, }Two-spin quantum walk under an effective blockaded XX Hamiltonian. \textbf{b, }Taking the difference with the summed density profiles of the two individual spins reveals an effective spin-spin repulsion. For clarity, we subtract the background evolution of the vacuum initial state. \textbf{c, }Protocol for many-body spectroscopy of a single spin excitation on the central site, $r = 0$. In this case, the ancilla in $\ket{+}$ is replaced by preparation of the central site in four bases, $\alpha \in \{\pm X,\pm Y\}$, and classical post-processing to obtain $\mathcal{G}(r,t)$ (Methods).
\textbf{d, }The Fourier transform, $\mathcal{G}(k,\omega)$, as a function of momentum, $k$, and frequency, $\omega$, across the 1D Brillouin zone. 
Left schematic illustrates the spin wave excitation.
The imaginary component is the spectral function, whose peak is fit to obtain the single-particle dispersion. From this we extract $(\mu+4U) / 2\pi = 0.407(2)$\,MHz and $J/2\pi=-0.140(1)$\,MHz, and independently verify these coefficients using Hamiltonian learning of the dynamics of different initial states (Methods). Background evolution, appearing at $k = 0$, is subtracted.
}
\label{fig3}
\end{figure}

\subsection*{Non-equilibrium quantum spin liquid}

\begin{figure*}
\includegraphics[width=2\columnwidth]{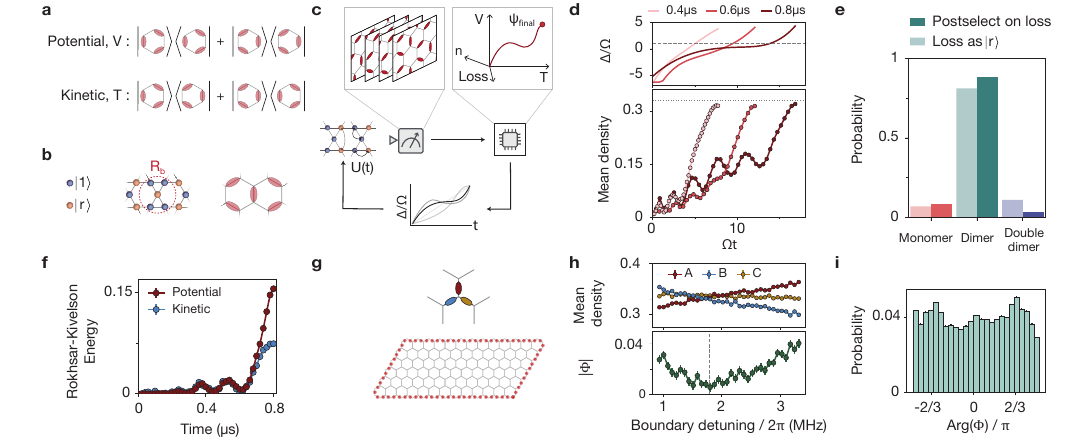}
\caption{\textbf{Dynamical preparation of U(1) quantum spin lakes.
} \textbf{a,} Potential and kinetic energy per hexagonal plaquette in the Rokhsar-Kivelson Hamiltonian. Potential energy counts the number of flippable plaquettes and kinetic energy captures the plaquette resonances. \textbf{b,} Mapping of Rydberg atoms on the vertices of the kagome lattice to a QDM on the honeycomb lattice. The main data in this experiment uses $R_b/a\approx1.32$. \textbf{c,} A multi-objective Bayesian optimizer is used in a closed loop with experiment. Each experimental run returns $X$ and $Z$ basis snapshots of the prepared state from which observables are computed; the RK energies and mean Rydberg density, $n$, are maximized, whilst atom loss is minimized. The preparation protocol is updated automatically and a new run is started. 
\textbf{d,} (Top) Best-performing detuning sweeps obtained from three overnight optimization runs with eight free parameters. The total time was allowed to vary to produce the longest sweep and was fixed for the shorter sweeps. See ED Fig.~5b for a summary of parameters. Dashed line marks the ground state phase transition. (Bottom) Corresponding evolution of loss-postselected mean Rydberg density. Oscillations are due to imperfect adiabaticity starting from all qubits in $\ket{1}$. Dotted line indicates 1/3 filling, corresponding to maximal packing.
\textbf{e,} Final monomer, dimer, and blockade-violating double-dimer populations per vertex for the 0.8$\,\mu$s drive.
\textbf{f,} Evolution of the kinetic and potential RK energies. The reduction of $\langle T\rangle$ relative to $\langle V\rangle$ is larger at later times as the Rydberg populations, and thus vdW tail energies, increase.
\textbf{g,} (Top) Schematic of the three kagome sublattices (A,B,C) per vertex. (Bottom) Local light shifts are applied to the 74 boundary atoms to compensate for their reduced mean-field vdW interaction energy. 
\textbf{h,} The minimum in the nematic order parameter, $\Phi$, corresponds to uniform bulk Rydberg population on the three sublattices. 
Dashed line marks the detuning strength used for all subsequent data. 
\textbf{i,} Angular distribution of $\Phi$ reveals the emergent global rotational symmetry. Three sharp peaks at $0,\pm2\pi/3$ are expected in a nematic phase.
}
\label{fig4}
\end{figure*}

We next employ these tools for preparing and probing complex many-body states. Specifically, 
we study the quantum dimer model (QDM) on the honeycomb lattice, which is predicted to host a deconfined QSL with an emergent $U(1)$ gauge symmetry arising from its bipartite nature~\cite{moessner_quantum_2008,fradkin_field_2013}.
The QSL state is the equal-amplitude and equal-phase superposition of all perfect dimer coverings and is found at the critical point of the QDM with the Hamiltonian $H_{\mathrm{RK}} = -tT + vV$, where $T$ is the kinetic energy and $V$ is the potential energy, defined in Fig.~4a.
$T$ captures the quantum coherences of the state, favoring plaquettes which can resonate via ring exchange, while $V$ penalizes such `flippable' plaquettes. 
Known as the Rokhsar-Kivelson (RK) Hamiltonian~\cite{rokhsar_superconductivity_1988}, this model features a critical point, $v=t$, known as the RK point.

To realize the QDM, atoms are placed on the vertices of the kagome lattice and encode the presence of a dimer on the bonds of the medial honeycomb lattice (Fig.~4b).
Rydberg blockade enforces that each vertex is touched by at most one dimer, with vertices without a dimer referred to as monomers. 
While at large $\Delta$ the ground state of the Rydberg Hamiltonian, Eq.~\eqref{eq:Hamiltonian}, on this lattice is a nematic phase~\cite{samajdar_quantum_2021}, it was recently pointed out that, out-of-equilibrium, finite-sized regions of QSL order known as quantum spin `lakes'
can emerge~\cite{sahay_quantum_2023,giudici_dynamical_2022}.
We therefore start by optimizing state preparation protocols in this out-of-equilibrium regime, further motivated by theoretical observations that tailored protocols can mitigate the impact of, for example, van der Waals tails and decoherence \cite{gjonbalaj_shortcuts_2025,vu_optimizing_2025,zeng_adiabatic_2025}.
More specifically, 
we employ a closed-loop optimization of different detuning sweeps, using the measured observables from the quantum evolution to update the sweep parameters and target the desired RK state (i.e., the ground state at the RK point). 
We use a Bayesian optimizer to simultaneously maximize $\langle T\rangle$, $\langle V\rangle$, and mean Rydberg density whilst minimizing atom loss (Methods), and find preparation sweeps for three evolution times that all approach the maximum filling of $1/3$ (Fig.~4c,d).

Two characteristic features emerge from the optimized detuning sweeps. First, longer sweeps improve performance by slowing down at early times, indicating the initial evolution is quasi-adiabatic. Second, the sweeps speed up before the phase transition ---in contrast to ground state preparation--- and reach approximately the same final sweep rate for all protocols.
Intuitively, this can be understood as follows:
the quasi-adiabatic evolution prepares an equal phase superposition of states and the final rapid sweep implements a dynamical projection onto the maximally-packed dimer coverings. 
During this final stage, the sweep must be sufficiently slow to adiabatically remove monomer excitations, 
while not so slow as to resolve any energy differences between dimer coverings and create phase excitations. 
As a result, there exists an optimum in the final sweep rate that balances these competing energy scales (Methods), thus explaining why this rate is approximately the same across the different optimized sweeps.

Focusing on the slowest sweep, we find a high final dimer population alongside finite monomer and double-dimer densities (Fig.~4e), arising from excitations as well as dressing of the low-energy maximally-packed dimer manifold at finite $\Delta/\Omega$ (Methods). The RK kinetic and potential energies grow in step with these dimer populations (Fig.~4f); we explore these energies more below. 
Notably, we find that local loss postselection improves the state fidelity, and closely reproduces the results obtained when postselecting on the absence of loss across the entire lattice (ED Fig.~2c,d).

\subsection*{Signatures of a Rokhsar-Kivelson state}

\begin{figure*}
\includegraphics[width=2\columnwidth]{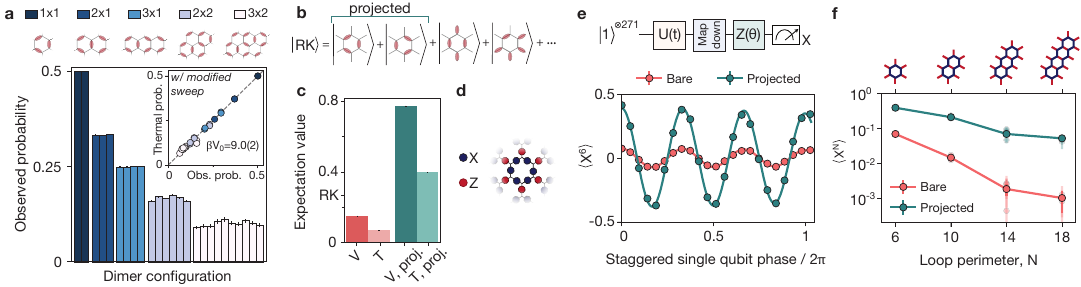}
\caption{\textbf{Superposition of dimer coverings.} 
\textbf{a,} Probability distribution of dimer configurations in isolated subsystems. The probabilities within each subsystem are nearly equal after projecting surrounding bonds onto no dimer  ($\ket{0}$ in the hyperfine qubit) and restricting to perfect dimer coverings without defects, i.e., monomers and double-dimers.
Inset shows a non-uniform distribution, prepared with a weakly modified preparation sweep (Methods), that is fit to a Boltzmann distribution with inverse temperature $\beta$. 
\textbf{b,} Schematic of the RK state. Two local configurations are isolated by projecting bonds around each single-plaquette subsystem onto no dimer. 
\textbf{c,} Bare and projected potential and kinetic energies, averaged over all plaquettes in the lattice. The RK prediction is $\langle T \rangle = \langle V \rangle \approx 0.29$ from classical Monte Carlo simulations (Methods). Defects reduce this expectation value in experiment, and the relative reduction of the kinetic energy is largely attributed to measurement error.
\textbf{d,} Local measurement bases for the projected kinetic energy.
\textbf{e,} Parity oscillations on single plaquettes using local Raman to apply $Z$ rotations on every other site after mapping down. $U(t)$ represents the analog state preparation.
\textbf{f,} Bare and projected X loops of increasing perimeter. Solid points are averaged over different measured geometries, shown individually as small points.}
\label{fig5}
\end{figure*}

A key property of the RK state is the absence of local order, implying that seeding of order by the lattice boundaries must be suppressed.
Such boundary effects imbalance the population on the three kagome sublattices (A,B,C), which can be quantified by the nematic order parameter, $\Phi = \sum_j \left(n_j^A + n_j^B e^{2\pi i /3} + n_j^C e^{-2\pi i /3}\right)$, 
where the sum runs over vertices (Fig.~4g). 
By applying static local light shifts to the boundary atoms, we prepare a uniform density across all three sublattices (Fig.~4h) and, moreover, find that the distribution of the phase of the nematic order parameter is close to uniform, revealing the emergent rotational symmetry of the state and distinguishing it from a nematic phase (Fig.~4i).

To further probe the microscopic order, we next examine the probability distribution of dimer coverings. By choosing small subsystems and projecting the surrounding bonds onto not having a dimer in post-processing, we isolate the local dimer covering from the surrounding lattice.
Within such subsystems of up to 27 atoms plus 12 projected bonds, 
we observe that the probability distribution across perfect dimer coverings is almost flat (Fig.~5a).
To investigate the robustness of this uniformity, we slightly perturb the state preparation profile and find that even small modifications result in an unequal distribution (ED Fig.~7a,b).
As an example, in the inset of Fig.~5a, we fit such an instance to a thermal distribution at finite temperature, where the classical energy of each subsystem is equal to its internal van der Waals interaction energy, finding excellent agreement.
In contrast, for the original sweep profile, the uniform distribution corresponds to an approximately infinite-temperature state that exactly coincides with the equal-amplitude prediction for the RK state.
We further perform the same analysis for different projected boundary conditions around each subsystem and again find good
agreement with the RK prediction (ED Fig.~7c).

We next turn to the many-body coherences between dimer covering.
Closed X loop operators, of which the kinetic energy operator is the smallest example, map different dimer coverings onto each other such that the expectation probes the coherences of the state. The expectation value of these loops, even in the ideal QSL, is not unity; that is, it is not a stabilizer state. 
However, projecting all the bonds surrounding the chosen X loop onto not having a dimer isolates the states with finite expectation value such that $\langle V \rangle_{\mathrm{proj.}} = \langle T \rangle_{\mathrm{proj.}} = +1$ for the RK state (Fig.~5b).
For the potential energy, we find an $\approx5\times$ increase in the expectation value after projection (Fig.~5c).
To measure the projected kinetic energy, after mapping down we apply local $Y^{1/2}$ gates to the sites in one of three plaquette sublattices (Fig.~5d).
Although we find the kinetic energy is reduced relative to the potential energy, we attribute this predominately to microstate-dependent phase accumulation during the mapping down that acts as an off-diagonal measurement error (Methods).

Intuitively, the X loops are local GHZ-like states on the loop perimeter embedded within the many-body state. To test this we perform parity oscillations, again on one plaquette sublattice at a time (Fig.~5e). 
Using local hyperfine $Z$ rotations on alternate sites along the loop, we measure coherent oscillations in $\langle X^6 \rangle$ with a many-body phase given by $3\times$ the applied single-qubit phase, as expected for the state $\frac{1}{\sqrt{2}}(\ket{101010} + \ket{010101})$. 
We similarly find that the loop expectation value is insensitive to the global readout phase.

Extending to larger loops of up to 33 sites in total, where the loop perimeter is always measured in the $X$-basis and surrounding atoms in the $Z$-basis for projection, we observe a slow decay in the projected expectation value with loop perimeter and a faster decay in the bare loop (Fig.~5f).
This decaying ratio $\langle X^N \rangle/\langle X^N \rangle_{\mathrm{proj.}}$ reflects the exponentially increasing number of dimer configurations over larger subsystems~\cite{moessner_quantum_2008}.
The perimeter-law decay of the projected X loops, which are all +1 in the ideal QSL, is due to virtual `magnetic' excitations, i.e., phases between dimer configurations generated at the endpoint of open Z strings, alongside incoherent errors~\cite{semeghini_probing_2021, verresen_prediction_2021}.

\subsection*{Emergent universal correlations}

\begin{figure}
\includegraphics[width=1\columnwidth]{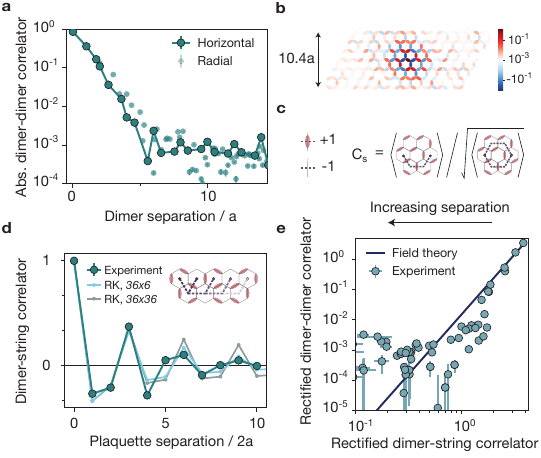}
\caption{\textbf{Correlations in the emergent U(1) gauge theory.}
\textbf{a,} Dimer-dimer connected correlator. 
Solid points show the absolute value of correlations restricted to dimers within plaquettes in the same row.
Correlations decay then plateau, indicative of weak long-range order, with additional non-universal structure arising from the lattice geometry.
Small points are radially averaged over the lattice after sign rectification.
\textbf{b,} Spatial plot showing the signed correlations with respect to the central site.
\textbf{c,} Definition of the dimer-string correlator between plaquette centers with FM normalization.
\textbf{d,} Dimer-string correlator, restricted to plaquettes in the same row (inset). 
The corresponding expectation value for the RK state is shown for two lattice dimensions (in units of plaquettes) obtained from Monte Carlo simulations. Monte Carlo simulations use periodic boundary conditions and the results are averaged over all four topological sectors (Methods). 
\textbf{e,} Relation between magnitudes of rectified dimer-dimer and dimer-string correlators. Field theory predicts a ratio of 4 on logarithmic axes. Each marker corresponds to the correlators at a given vertex separation. 
}
\label{fig6}
\end{figure}

To probe the emergent $U(1)$ gauge theory, we next study the correlations of the state. We focus on two types of correlators ---dimer-dimer and dimer-string--- which are predicted by the field theory to decay as $1/r^{2}$ and $1/r^{1/2}$, respectively~\cite{fradkin_field_2013}.
Remarkably, at the RK point, such $Z$-basis correlators on a lattice are known analytically~\cite{fisher_statistical_1963} and exactly coincide with the predictions of continuum field theory at long distances, after accounting for the lattice-dependent substructure (Methods).
We first investigate the dimer-dimer correlations $C_d(|\mathbf{x}_i - \mathbf{x}_j|) = \langle \sigma^z_i\sigma^z_k\rangle - \langle \sigma^z_i\rangle \langle \sigma^z_j\rangle$ where $\sigma^z_i = 2n_i - 1$, finding an exponential decay of the magnitude (Fig.~6a) and an antiferromagnetic sign structure (Fig.~6b). 
The plateau at large separations suggests the presence of weak long-range order induced by the open boundary conditions, which explicitly break the $\mathbb{Z}_3$ symmetry of the honeycomb lattice.
We then turn to the dimer-string correlator, $C_s(|\mathbf{x}_i - \mathbf{x}_j|) = \prod_{\ell\in \Gamma_{ij}}\sigma_\ell^z$, that measures the dimer parity along an open string $\Gamma_{ij}$ between two plaquettes, illustrated in Fig.~6c.
Similar to the definition of the Fredenhagen-Marcu (FM) order parameter for topological order~\cite{fredenhagen_charged_1983}, we normalize by the closed loop parity, mitigating the effect of dressing of the string by monomers and incoherent errors that both result in a perimeter-law decay of closed loops~\cite{semeghini_probing_2021,verresen_prediction_2021} (ED Fig.~8e).
The resulting expectation value displays a characteristic sign structure and magnitude decaying to zero for longer strings (Fig.~6d), once again ruling out the presence of crystalline ordering.
We further measure normalized open X strings that probe for monomers created at the string endpoints and find that the expectation value is small and decays with increasing string length (ED Fig.~8g), consistent with a deconfined gauge theory~\cite{gregor_diagnosing_2010,verresen_prediction_2021}.

To reveal the universal physics, we analyze rectified dimer-dimer and dimer-string correlators, where the rectification 
maps these both onto vertex-vertex correlators described by the field theory (Methods and ED Fig.~9).
In Fig.~6e we compare the logarithms of the two rectified correlators and find good agreement with the universal ratio of 4 predicted by the field theory, which is expected to hold even at finite system size (Methods).
We attribute the exponential decay of the correlators to finite-size effects and the presence of monomers.
Both arguments from field theory and classical Monte Carlo simulations of the RK state are consistent with such an exponential decay on a finite-size lattice, converging towards algebraic correlations with increasing size (Methods).
Interestingly, we find the decay of the correlators is largely independent of lattice height, with a difference emerging only in the magnitude of the plateau in the dimer-dimer correlations (ED Fig.~10).
Our results are therefore consistent with the preparation of quantum spin lakes that appear locally like the RK state,
with a length scale set by the monomer excitation density
\cite{sahay_quantum_2023}.
Furthermore, we find close agreement with unitary time-dependent matrix product state simulations of the experimental protocol (ED Figs.~8,9), indicating that incoherent errors do not appreciably affect these results.

\subsection*{Discussion and outlook}
These observations reveal the key signatures of a non-equilibrium $U(1)$ Rokhsar-Kivelson state and provide insights into the complex many-body physics of this state.
This is especially notable given 
the instability of the $U(1)$ QSL in equilibrium and its gapless nature, which together render it inaccessible via either adiabatic sweeps or finite-depth 
digital circuits.
In this out-of-equilibrium regime, we find that microscopic properties are sensitive to the specific protocol used for state preparation, highlighting the value of experimental optimization, informed by theory (Fig.~4). 
Rather remarkably, the key features of the resulting state are well-described by the equilibrium theoretical predictions of the Rokhsar-Kivelson Hamiltonian at its critical point (Figs.~5,6).
Quantifying the deviations from these predictions at longer length scales and developing a detailed understanding of the interplay between finite monomer density, system size, and open boundary conditions constitute interesting directions for future explorations.

These results can be extended along several directions. 
By using local measurements, an approximate equilibrium parent Hamiltonian can be obtained using Hamiltonian learning techniques~\cite{qi_determining_2019}, 
potentially enabling studies of the deformations from the ideal Rokhsar-Kivelson model.
By engineering such Hamiltonians using, e.g. Floquet techniques, one could then study low-energy excitations 
relevant to understanding emergent gauge dynamics in resonating valence bond states.
This and other Hamiltonian engineering approaches could make use of variational quantum circuits~\cite{cerezo_variational_2021, katz_hybrid_2025}, possibly combined with closed-loop optimization incorporating control over analog evolution, digital gates, and coherent atom rearrangement. Furthermore, such methods can be used for inverse quantum simulation~\cite{kokail_inverse_2026} to design systems or materials with desired properties.

More broadly, these experiments demonstrate the potential of a hybrid analog-digital approach to quantum simulation. Digital controls enable local operations, error detection, and fast experimental cycle rates, while analog evolution efficiently implements many-body interactions.
We find coherent mapping to be a powerful tool and envision several pathways to further improve its performance (Methods).
Error mitigation based on loss information improves effective fidelities within existing technical capabilities.
Using atomic motion to toggle between interacting and non-interacting geometries (Fig.~2,3) enables measurement of two-time correlation functions~\cite{lu_algorithms_2021} or entanglement witnesses~\cite{vidal_computable_2002}, and could be used for circuit-based state preparation~\cite{cochran_visualizing_2025,satzinger_realizing_2021,evered_probing_2025} or to employ state purification techniques~\cite{childs_streaming_2025}.
Extending to interleaved layers of digital and analog evolution may enable new opportunities for quantum simulation or sensing~\cite{katz_hybrid_2025}.
Many of these techniques can be applied to a range of platforms~\cite{katz_hybrid_2025, andersen_thermalization_2025,senoo_high-fidelity_2025}, greatly expanding the breadth of accessible many-body physics in practical quantum simulators.\\

\noindent \textit{Note:} During the completion of this work, we became aware
of related work realizing $U(1)$ quantum spin lakes using ultracold atoms in an optical lattice~\cite{Karch2026}.

\clearpage
\newpage

\bibliographystyle{naturemag_arxiv2.bst}
\bibliography{references.bib}

\begin{thebibliography}{100}
\expandafter\ifx\csname url\endcsname\relax
  \def\url#1{\texttt{#1}}\fi
\expandafter\ifx\csname urlprefix\endcsname\relax\def\urlprefix{URL }\fi
\providecommand{\bibinfo}[2]{#2}
\providecommand{\eprint}[2][]{\url{#2}}

\bibitem{anderson_more_1972}
\bibinfo{author}{Anderson, P.~W.}
\newblock \bibinfo{title}{More {Is} {Different}}.
\newblock \emph{\bibinfo{journal}{Science}} \textbf{\bibinfo{volume}{177}}, \bibinfo{pages}{393--396} (\bibinfo{year}{1972}).

\bibitem{girvin_modern_2019}
\bibinfo{author}{Girvin, S.~M.} \& \bibinfo{author}{Yang, K.}
\newblock \emph{\bibinfo{title}{Modern {Condensed} {Matter} {Physics}}} (\bibinfo{publisher}{Cambridge University Press}, \bibinfo{year}{2019}).

\bibitem{eisert_quantum_2015}
\bibinfo{author}{Eisert, J.}, \bibinfo{author}{Friesdorf, M.} \& \bibinfo{author}{Gogolin, C.}
\newblock \bibinfo{title}{Quantum many-body systems out of equilibrium}.
\newblock \emph{\bibinfo{journal}{Nature Physics}} \textbf{\bibinfo{volume}{11}}, \bibinfo{pages}{124--130} (\bibinfo{year}{2015}).

\bibitem{preskill_quantum_2018}
\bibinfo{author}{Preskill, J.}
\newblock \bibinfo{title}{Quantum {Computing} in the {NISQ} era and beyond}.
\newblock \emph{\bibinfo{journal}{Quantum}} \textbf{\bibinfo{volume}{2}}, \bibinfo{pages}{79} (\bibinfo{year}{2018}).

\bibitem{bluvstein_fault-tolerant_2025}
\bibinfo{author}{Bluvstein, D.} \emph{et~al.}
\newblock \bibinfo{title}{A fault-tolerant neutral-atom architecture for universal quantum computation}.
\newblock \emph{\bibinfo{journal}{Nature}} \textbf{\bibinfo{volume}{649}}, \bibinfo{pages}{39--46} (\bibinfo{year}{2026}).

\bibitem{goldman_periodically_2014}
\bibinfo{author}{Goldman, N.} \& \bibinfo{author}{Dalibard, J.}
\newblock \bibinfo{title}{Periodically {Driven} {Quantum} {Systems}: {Effective} {Hamiltonians} and {Engineered} {Gauge} {Fields}}.
\newblock \emph{\bibinfo{journal}{Physical Review X}} \textbf{\bibinfo{volume}{4}}, \bibinfo{pages}{031027} (\bibinfo{year}{2014}).

\bibitem{knap_probing_2013}
\bibinfo{author}{Knap, M.} \emph{et~al.}
\newblock \bibinfo{title}{Probing {Real}-{Space} and {Time}-{Resolved} {Correlation} {Functions} with {Many}-{Body} {Ramsey} {Interferometry}}.
\newblock \emph{\bibinfo{journal}{Physical Review Letters}} \textbf{\bibinfo{volume}{111}}, \bibinfo{pages}{147205} (\bibinfo{year}{2013}).

\bibitem{rosi_fast_2013}
\bibinfo{author}{Rosi, S.} \emph{et~al.}
\newblock \bibinfo{title}{Fast closed-loop optimal control of ultracold atoms in an optical lattice}.
\newblock \emph{\bibinfo{journal}{Physical Review A}} \textbf{\bibinfo{volume}{88}}, \bibinfo{pages}{021601} (\bibinfo{year}{2013}).

\bibitem{rokhsar_superconductivity_1988}
\bibinfo{author}{Rokhsar, D.~S.} \& \bibinfo{author}{Kivelson, S.~A.}
\newblock \bibinfo{title}{Superconductivity and the {Quantum} {Hard}-{Core} {Dimer} {Gas}}.
\newblock \emph{\bibinfo{journal}{Physical Review Letters}} \textbf{\bibinfo{volume}{61}}, \bibinfo{pages}{2376--2379} (\bibinfo{year}{1988}).

\bibitem{fradkin_field_2013}
\bibinfo{author}{Fradkin, E.}
\newblock \emph{\bibinfo{title}{Field {Theories} of {Condensed} {Matter} {Physics}}} (\bibinfo{publisher}{Cambridge University Press}, \bibinfo{address}{Cambridge}, \bibinfo{year}{2013}).

\bibitem{lloyd_universal_1996}
\bibinfo{author}{Lloyd, S.}
\newblock \bibinfo{title}{Universal {Quantum} {Simulators}}.
\newblock \emph{\bibinfo{journal}{Science}} \textbf{\bibinfo{volume}{273}}, \bibinfo{pages}{1073--1078} (\bibinfo{year}{1996}).

\bibitem{greiner_quantum_2002}
\bibinfo{author}{Greiner, M.}, \bibinfo{author}{Mandel, O.}, \bibinfo{author}{Esslinger, T.}, \bibinfo{author}{Hänsch, T.~W.} \& \bibinfo{author}{Bloch, I.}
\newblock \bibinfo{title}{Quantum phase transition from a superfluid to a {Mott} insulator in a gas of ultracold atoms}.
\newblock \emph{\bibinfo{journal}{Nature}} \textbf{\bibinfo{volume}{415}}, \bibinfo{pages}{39--44} (\bibinfo{year}{2002}).

\bibitem{scholl_programmable_2021}
\bibinfo{author}{Scholl, P.} \emph{et~al.}
\newblock \bibinfo{title}{Programmable quantum simulation of {2D} antiferromagnets with hundreds of {Rydberg} atoms}.
\newblock \emph{\bibinfo{journal}{Nature}} \textbf{\bibinfo{volume}{595}}, \bibinfo{pages}{233--238} (\bibinfo{year}{2021}).

\bibitem{deng_effective_2005}
\bibinfo{author}{Deng, X.-L.}, \bibinfo{author}{Porras, D.} \& \bibinfo{author}{Cirac, J.~I.}
\newblock \bibinfo{title}{Effective spin quantum phases in systems of trapped ions}.
\newblock \emph{\bibinfo{journal}{Physical Review A}} \textbf{\bibinfo{volume}{72}}, \bibinfo{pages}{063407} (\bibinfo{year}{2005}).

\bibitem{xu_neutral-atom_2025}
\bibinfo{author}{Xu, M.} \emph{et~al.}
\newblock \bibinfo{title}{A neutral-atom {Hubbard} quantum simulator in the cryogenic regime}.
\newblock \emph{\bibinfo{journal}{Nature}} \textbf{\bibinfo{volume}{642}}, \bibinfo{pages}{909--915} (\bibinfo{year}{2025}).

\bibitem{andersen_thermalization_2025}
\bibinfo{author}{Andersen, T.~I.} \emph{et~al.}
\newblock \bibinfo{title}{Thermalization and criticality on an analogue–digital quantum simulator}.
\newblock \emph{\bibinfo{journal}{Nature}} \textbf{\bibinfo{volume}{638}}, \bibinfo{pages}{79--85} (\bibinfo{year}{2025}).

\bibitem{kaufman_quantum_2016}
\bibinfo{author}{Kaufman, A.~M.} \emph{et~al.}
\newblock \bibinfo{title}{Quantum thermalization through entanglement in an isolated many-body system}.
\newblock \emph{\bibinfo{journal}{Science}} \textbf{\bibinfo{volume}{353}}, \bibinfo{pages}{794--800} (\bibinfo{year}{2016}).

\bibitem{schreiber_observation_2015}
\bibinfo{author}{Schreiber, M.} \emph{et~al.}
\newblock \bibinfo{title}{Observation of many-body localization of interacting fermions in a quasirandom optical lattice}.
\newblock \emph{\bibinfo{journal}{Science}} \textbf{\bibinfo{volume}{349}}, \bibinfo{pages}{842--845} (\bibinfo{year}{2015}).

\bibitem{de_leseleuc_observation_2019}
\bibinfo{author}{de~Léséleuc, S.} \emph{et~al.}
\newblock \bibinfo{title}{Observation of a symmetry-protected topological phase of interacting bosons with {Rydberg} atoms}.
\newblock \emph{\bibinfo{journal}{Science}} \textbf{\bibinfo{volume}{365}}, \bibinfo{pages}{775--780} (\bibinfo{year}{2019}).

\bibitem{gonzalez-cuadra_observation_2025}
\bibinfo{author}{Gonzalez-Cuadra, D.} \emph{et~al.}
\newblock \bibinfo{title}{Observation of string breaking on a (2 + 1){D} {Rydberg} quantum simulator}.
\newblock \emph{\bibinfo{journal}{Nature}} \textbf{\bibinfo{volume}{642}}, \bibinfo{pages}{321--326} (\bibinfo{year}{2025}).

\bibitem{young_observing_2024}
\bibinfo{author}{Young, D.~J.} \emph{et~al.}
\newblock \bibinfo{title}{Observing dynamical phases of {BCS} superconductors in a cavity {QED} simulator}.
\newblock \emph{\bibinfo{journal}{Nature}} \textbf{\bibinfo{volume}{625}}, \bibinfo{pages}{679--684} (\bibinfo{year}{2024}).

\bibitem{periwal_programmable_2021}
\bibinfo{author}{Periwal, A.} \emph{et~al.}
\newblock \bibinfo{title}{Programmable interactions and emergent geometry in an array of atom clouds}.
\newblock \emph{\bibinfo{journal}{Nature}} \textbf{\bibinfo{volume}{600}}, \bibinfo{pages}{630--635} (\bibinfo{year}{2021}).

\bibitem{miller_two-axis_2024}
\bibinfo{author}{Miller, C.} \emph{et~al.}
\newblock \bibinfo{title}{Two-axis twisting using {Floquet}-engineered {XYZ} spin models with polar molecules}.
\newblock \emph{\bibinfo{journal}{Nature}} \textbf{\bibinfo{volume}{633}}, \bibinfo{pages}{332--337} (\bibinfo{year}{2024}).

\bibitem{morong_engineering_2023}
\bibinfo{author}{Morong, W.} \emph{et~al.}
\newblock \bibinfo{title}{Engineering {Dynamically} {Decoupled} {Quantum} {Simulations} with {Trapped} {Ions}}.
\newblock \emph{\bibinfo{journal}{PRX Quantum}} \textbf{\bibinfo{volume}{4}}, \bibinfo{pages}{010334} (\bibinfo{year}{2023}).

\bibitem{cochran_visualizing_2025}
\bibinfo{author}{Cochran, T.~A.} \emph{et~al.}
\newblock \bibinfo{title}{Visualizing dynamics of charges and strings in (2 + 1){D} lattice gauge theories}.
\newblock \emph{\bibinfo{journal}{Nature}} \textbf{\bibinfo{volume}{642}}, \bibinfo{pages}{315--320} (\bibinfo{year}{2025}).

\bibitem{evered_probing_2025}
\bibinfo{author}{Evered, S.~J.} \emph{et~al.}
\newblock \bibinfo{title}{Probing the {Kitaev} honeycomb model on a neutral-atom quantum computer}.
\newblock \emph{\bibinfo{journal}{Nature}} \textbf{\bibinfo{volume}{645}}, \bibinfo{pages}{341--347} (\bibinfo{year}{2025}).

\bibitem{will_probing_2025}
\bibinfo{author}{Will, M.} \emph{et~al.}
\newblock \bibinfo{title}{Probing non-equilibrium topological order on a quantum processor}.
\newblock \emph{\bibinfo{journal}{Nature}} \textbf{\bibinfo{volume}{645}}, \bibinfo{pages}{348--353} (\bibinfo{year}{2025}).

\bibitem{beverland_assessing_2022}
\bibinfo{author}{Beverland, M.~E.} \emph{et~al.}
\newblock \bibinfo{title}{Assessing requirements to scale to practical quantum advantage}, \bibinfo{note}{arXiv:2211.07629} (\bibinfo{year}{2022}).

\bibitem{senoo_high-fidelity_2025}
\bibinfo{author}{Senoo, A.} \emph{et~al.}
\newblock \bibinfo{title}{High-fidelity entanglement and coherent multi-qubit mapping in an atom array}, \bibinfo{note}{arXiv:2506.13632} (\bibinfo{year}{2025}).

\bibitem{brydges_probing_2019}
\bibinfo{author}{Brydges, T.} \emph{et~al.}
\newblock \bibinfo{title}{Probing {Rényi} entanglement entropy via randomized measurements}.
\newblock \emph{\bibinfo{journal}{Science}} \textbf{\bibinfo{volume}{364}}, \bibinfo{pages}{260--263} (\bibinfo{year}{2019}).

\bibitem{halimeh_quantum_2025}
\bibinfo{author}{Halimeh, J.~C.}, \bibinfo{author}{Mueller, N.}, \bibinfo{author}{Knolle, J.}, \bibinfo{author}{Papić, Z.} \& \bibinfo{author}{Davoudi, Z.}
\newblock \bibinfo{title}{Quantum simulation of out-of-equilibrium dynamics in gauge theories}, \bibinfo{note}{arXiv:2509.03586} (\bibinfo{year}{2025}).

\bibitem{fradkin_short_1990}
\bibinfo{author}{Fradkin, E.} \& \bibinfo{author}{Kivelson, S.}
\newblock \bibinfo{title}{Short range resonating valence bond theories and superconductivity}.
\newblock \emph{\bibinfo{journal}{Modern Physics Letters B}} \textbf{\bibinfo{volume}{4}}, \bibinfo{pages}{225--232} (\bibinfo{year}{1990}).

\bibitem{moessner_quantum_2008}
\bibinfo{author}{Moessner, R.} \& \bibinfo{author}{Raman, K.~S.}
\newblock \bibinfo{title}{Quantum dimer models}, \bibinfo{note}{arXiv:0809.3051} (\bibinfo{year}{2008}).

\bibitem{read_valence-bond_1989}
\bibinfo{author}{Read, N.} \& \bibinfo{author}{Sachdev, S.}
\newblock \bibinfo{title}{Valence-bond and spin-peierls ground states of low-dimensional quantum antiferromagnets}.
\newblock \emph{\bibinfo{journal}{Phys. Rev. Lett.}} \textbf{\bibinfo{volume}{62}}, \bibinfo{pages}{1694--1697} (\bibinfo{year}{1989}).

\bibitem{fazekas_ground_1974}
\bibinfo{author}{Fazekas, P.} \& \bibinfo{author}{Anderson, P.~W.}
\newblock \bibinfo{title}{On the ground state properties of the anisotropic triangular antiferromagnet}.
\newblock \emph{\bibinfo{journal}{The Philosophical Magazine: A Journal of Theoretical Experimental and Applied Physics}} \textbf{\bibinfo{volume}{30}}, \bibinfo{pages}{423--440} (\bibinfo{year}{1974}).

\bibitem{semeghini_probing_2021}
\bibinfo{author}{Semeghini, G.} \emph{et~al.}
\newblock \bibinfo{title}{Probing topological spin liquids on a programmable quantum simulator}.
\newblock \emph{\bibinfo{journal}{Science}} \textbf{\bibinfo{volume}{374}}, \bibinfo{pages}{1242--1247} (\bibinfo{year}{2021}).

\bibitem{satzinger_realizing_2021}
\bibinfo{author}{Satzinger, K.~J.} \emph{et~al.}
\newblock \bibinfo{title}{Realizing topologically ordered states on a quantum processor}.
\newblock \emph{\bibinfo{journal}{Science}} \textbf{\bibinfo{volume}{374}}, \bibinfo{pages}{1237--1241} (\bibinfo{year}{2021}).

\bibitem{sahay_quantum_2023}
\bibinfo{author}{Sahay, R.}, \bibinfo{author}{Vishwanath, A.} \& \bibinfo{author}{Verresen, R.}
\newblock \bibinfo{title}{Quantum {Spin} {Puddles} and {Lakes}: {NISQ}-{Era} {Spin} {Liquids} from {Non}-{Equilibrium} {Dynamics}}, \bibinfo{note}{arXiv:2211.01381} (\bibinfo{year}{2023}).

\bibitem{boulier_spontaneous_2017}
\bibinfo{author}{Boulier, T.} \emph{et~al.}
\newblock \bibinfo{title}{Spontaneous avalanche dephasing in large {Rydberg} ensembles}.
\newblock \emph{\bibinfo{journal}{Physical Review A}} \textbf{\bibinfo{volume}{96}}, \bibinfo{pages}{053409} (\bibinfo{year}{2017}).

\bibitem{koyluoglu_floquet_2024}
\bibinfo{author}{Koyluoglu, N.~U.}, \bibinfo{author}{Maskara, N.}, \bibinfo{author}{Feldmeier, J.} \& \bibinfo{author}{Lukin, M.~D.}
\newblock \bibinfo{title}{Floquet {Engineering} of {Interactions} and {Entanglement} in {Periodically} {Driven} {Rydberg} {Chains}}.
\newblock \emph{\bibinfo{journal}{Physical Review Letters}} \textbf{\bibinfo{volume}{135}}, \bibinfo{pages}{113603} (\bibinfo{year}{2025}).

\bibitem{feldmeier_quantum_2024}
\bibinfo{author}{Feldmeier, J.}, \bibinfo{author}{Maskara, N.}, \bibinfo{author}{Köylüoğlu, N.~U.} \& \bibinfo{author}{Lukin, M.~D.}
\newblock \bibinfo{title}{Quantum simulation of dynamical gauge theories in periodically driven {Rydberg} atom arrays}, \bibinfo{note}{arXiv:2408.02733} (\bibinfo{year}{2024}).

\bibitem{bluvstein_controlling_2021}
\bibinfo{author}{Bluvstein, D.} \emph{et~al.}
\newblock \bibinfo{title}{Controlling quantum many-body dynamics in driven {Rydberg} atom arrays}.
\newblock \emph{\bibinfo{journal}{Science}} \textbf{\bibinfo{volume}{371}}, \bibinfo{pages}{1355--1359} (\bibinfo{year}{2021}).

\bibitem{samajdar_quantum_2021}
\bibinfo{author}{Samajdar, R.}, \bibinfo{author}{Ho, W.~W.}, \bibinfo{author}{Pichler, H.}, \bibinfo{author}{Lukin, M.~D.} \& \bibinfo{author}{Sachdev, S.}
\newblock \bibinfo{title}{Quantum phases of {Rydberg} atoms on a kagome lattice}.
\newblock \emph{\bibinfo{journal}{PNAS}} \textbf{\bibinfo{volume}{118}}, \bibinfo{pages}{e2015785118} (\bibinfo{year}{2021}).

\bibitem{giudici_dynamical_2022}
\bibinfo{author}{Giudici, G.}, \bibinfo{author}{Lukin, M.~D.} \& \bibinfo{author}{Pichler, H.}
\newblock \bibinfo{title}{Dynamical {Preparation} of {Quantum} {Spin} {Liquids} in {Rydberg} {Atom} {Arrays}}.
\newblock \emph{\bibinfo{journal}{Physical Review Letters}} \textbf{\bibinfo{volume}{129}}, \bibinfo{pages}{090401} (\bibinfo{year}{2022}).

\bibitem{gjonbalaj_shortcuts_2025}
\bibinfo{author}{Gjonbalaj, N.~O.}, \bibinfo{author}{Sahay, R.} \& \bibinfo{author}{Yelin, S.~F.}
\newblock \bibinfo{title}{Shortcuts to {Analog} {Preparation} of {Non}-{Equilibrium} {Quantum} {Lakes}}, \bibinfo{note}{arXiv:2502.03518} (\bibinfo{year}{2025}).

\bibitem{vu_optimizing_2025}
\bibinfo{author}{Vu, D.} \emph{et~al.}
\newblock \bibinfo{title}{Optimizing the dynamical preparation of quantum spin lakes on the ruby lattice}, \bibinfo{note}{arXiv:2512.09040} (\bibinfo{year}{2025}).

\bibitem{zeng_adiabatic_2025}
\bibinfo{author}{Zeng, Z.} \emph{et~al.}
\newblock \bibinfo{title}{Adiabatic echo protocols for robust quantum many-body state preparation}, \bibinfo{note}{arXiv:2506.12138} (\bibinfo{year}{2025}).

\bibitem{verresen_prediction_2021}
\bibinfo{author}{Verresen, R.}, \bibinfo{author}{Lukin, M.~D.} \& \bibinfo{author}{Vishwanath, A.}
\newblock \bibinfo{title}{Prediction of {Toric} {Code} {Topological} {Order} from {Rydberg} {Blockade}}.
\newblock \emph{\bibinfo{journal}{Physical Review X}} \textbf{\bibinfo{volume}{11}}, \bibinfo{pages}{031005} (\bibinfo{year}{2021}).

\bibitem{fisher_statistical_1963}
\bibinfo{author}{Fisher, M.~E.} \& \bibinfo{author}{Stephenson, J.}
\newblock \bibinfo{title}{Statistical {Mechanics} of {Dimers} on a {Plane} {Lattice}. {II}. {Dimer} {Correlations} and {Monomers}}.
\newblock \emph{\bibinfo{journal}{Physical Review}} \textbf{\bibinfo{volume}{132}}, \bibinfo{pages}{1411--1431} (\bibinfo{year}{1963}).

\bibitem{fredenhagen_charged_1983}
\bibinfo{author}{Fredenhagen, K.} \& \bibinfo{author}{Marcu, M.}
\newblock \bibinfo{title}{Charged states in z2 gauge theories}.
\newblock \emph{\bibinfo{journal}{Communications in Mathematical Physics}} \textbf{\bibinfo{volume}{92}}, \bibinfo{pages}{81--119} (\bibinfo{year}{1983}).

\bibitem{gregor_diagnosing_2010}
\bibinfo{author}{Gregor, K.}, \bibinfo{author}{Huse, D.~A.}, \bibinfo{author}{Moessner, R.} \& \bibinfo{author}{Sondhi, S.~L.}
\newblock \bibinfo{title}{Diagnosing {Deconfinement} and {Topological} {Order}}, \bibinfo{note}{arXiv:1011.4187} (\bibinfo{year}{2010}).

\bibitem{qi_determining_2019}
\bibinfo{author}{Qi, X.-L.} \& \bibinfo{author}{Ranard, D.}
\newblock \bibinfo{title}{Determining a local {Hamiltonian} from a single eigenstate}.
\newblock \emph{\bibinfo{journal}{Quantum}} \textbf{\bibinfo{volume}{3}}, \bibinfo{pages}{159} (\bibinfo{year}{2019}).

\bibitem{cerezo_variational_2021}
\bibinfo{author}{Cerezo, M.} \emph{et~al.}
\newblock \bibinfo{title}{Variational quantum algorithms}.
\newblock \emph{\bibinfo{journal}{Nature Reviews Physics}} \textbf{\bibinfo{volume}{3}}, \bibinfo{pages}{625--644} (\bibinfo{year}{2021}).

\bibitem{katz_hybrid_2025}
\bibinfo{author}{Katz, O.} \emph{et~al.}
\newblock \bibinfo{title}{Hybrid digital-analog protocols for simulating quantum multi-body interactions}, \bibinfo{note}{arXiv:2512.21385} (\bibinfo{year}{2025}).

\bibitem{kokail_inverse_2026}
\bibinfo{author}{Kokail, C.} \emph{et~al.}
\newblock \bibinfo{title}{Inverse {Quantum} {Simulation} for {Quantum} {Material} {Design}}, \bibinfo{note}{arXiv:2601.12239} (\bibinfo{year}{2026}).

\bibitem{lu_algorithms_2021}
\bibinfo{author}{Lu, S.}, \bibinfo{author}{Bañuls, M.~C.} \& \bibinfo{author}{Cirac, J.~I.}
\newblock \bibinfo{title}{Algorithms for {Quantum} {Simulation} at {Finite} {Energies}}.
\newblock \emph{\bibinfo{journal}{PRX Quantum}} \textbf{\bibinfo{volume}{2}}, \bibinfo{pages}{020321} (\bibinfo{year}{2021}).

\bibitem{vidal_computable_2002}
\bibinfo{author}{Vidal, G.} \& \bibinfo{author}{Werner, R.~F.}
\newblock \bibinfo{title}{Computable measure of entanglement}.
\newblock \emph{\bibinfo{journal}{Physical Review A}} \textbf{\bibinfo{volume}{65}}, \bibinfo{pages}{032314} (\bibinfo{year}{2002}).

\bibitem{childs_streaming_2025}
\bibinfo{author}{Childs, A.~M.} \emph{et~al.}
\newblock \bibinfo{title}{Streaming quantum state purification}.
\newblock \emph{\bibinfo{journal}{Quantum}} \textbf{\bibinfo{volume}{9}}, \bibinfo{pages}{1603} (\bibinfo{year}{2025}).

\bibitem{Karch2026}
\bibinfo{author}{Karch, {\u{S}}.} \emph{et~al.}
\newblock \bibinfo{title}{in preparation} (\bibinfo{year}{2026}).

\bibitem{manovitz_quantum_2025}
\bibinfo{author}{Manovitz, T.} \emph{et~al.}
\newblock \bibinfo{title}{Quantum coarsening and collective dynamics on a programmable simulator}.
\newblock \emph{\bibinfo{journal}{Nature}} \textbf{\bibinfo{volume}{638}}, \bibinfo{pages}{86--92} (\bibinfo{year}{2025}).

\bibitem{brown_gray-molasses_2019}
\bibinfo{author}{Brown, M.}, \bibinfo{author}{Thiele, T.}, \bibinfo{author}{Kiehl, C.}, \bibinfo{author}{Hsu, T.-W.} \& \bibinfo{author}{Regal, C.}
\newblock \bibinfo{title}{Gray-{Molasses} {Optical}-{Tweezer} {Loading}: {Controlling} {Collisions} for {Scaling} {Atom}-{Array} {Assembly}}.
\newblock \emph{\bibinfo{journal}{Physical Review X}} \textbf{\bibinfo{volume}{9}}, \bibinfo{pages}{011057} (\bibinfo{year}{2019}).

\bibitem{barredo_atom-by-atom_2016}
\bibinfo{author}{Barredo, D.}, \bibinfo{author}{de~Léséleuc, S.}, \bibinfo{author}{Lienhard, V.}, \bibinfo{author}{Lahaye, T.} \& \bibinfo{author}{Browaeys, A.}
\newblock \bibinfo{title}{An atom-by-atom assembler of defect-free arbitrary two-dimensional atomic arrays}.
\newblock \emph{\bibinfo{journal}{Science}} \textbf{\bibinfo{volume}{354}}, \bibinfo{pages}{1021--1023} (\bibinfo{year}{2016}).

\bibitem{ebadi_quantum_2021}
\bibinfo{author}{Ebadi, S.} \emph{et~al.}
\newblock \bibinfo{title}{Quantum phases of matter on a 256-atom programmable quantum simulator}.
\newblock \emph{\bibinfo{journal}{Nature}} \textbf{\bibinfo{volume}{595}}, \bibinfo{pages}{227--232} (\bibinfo{year}{2021}).

\bibitem{levine_dispersive_2022}
\bibinfo{author}{Levine, H.} \emph{et~al.}
\newblock \bibinfo{title}{Dispersive optical systems for scalable {Raman} driving of hyperfine qubits}.
\newblock \emph{\bibinfo{journal}{Physical Review A}} \textbf{\bibinfo{volume}{105}}, \bibinfo{pages}{032618} (\bibinfo{year}{2022}).

\bibitem{bluvstein_logical_2024}
\bibinfo{author}{Bluvstein, D.} \emph{et~al.}
\newblock \bibinfo{title}{Logical quantum processor based on reconfigurable atom arrays}.
\newblock \emph{\bibinfo{journal}{Nature}} \textbf{\bibinfo{volume}{626}}, \bibinfo{pages}{58--65} (\bibinfo{year}{2024}).

\bibitem{levine_parallel_2019}
\bibinfo{author}{Levine, H.} \emph{et~al.}
\newblock \bibinfo{title}{Parallel {Implementation} of {High}-{Fidelity} {Multiqubit} {Gates} with {Neutral} {Atoms}}.
\newblock \emph{\bibinfo{journal}{Physical Review Letters}} \textbf{\bibinfo{volume}{123}}, \bibinfo{pages}{170503} (\bibinfo{year}{2019}).

\bibitem{bluvstein_quantum_2022}
\bibinfo{author}{Bluvstein, D.} \emph{et~al.}
\newblock \bibinfo{title}{A quantum processor based on coherent transport of entangled atom arrays}.
\newblock \emph{\bibinfo{journal}{Nature}} \textbf{\bibinfo{volume}{604}}, \bibinfo{pages}{451--456} (\bibinfo{year}{2022}).

\bibitem{evered_high-fidelity_2023}
\bibinfo{author}{Evered, S.~J.} \emph{et~al.}
\newblock \bibinfo{title}{High-fidelity parallel entangling gates on a neutral-atom quantum computer}.
\newblock \emph{\bibinfo{journal}{Nature}} \textbf{\bibinfo{volume}{622}}, \bibinfo{pages}{268--272} (\bibinfo{year}{2023}).

\bibitem{scholl_erasure_2023}
\bibinfo{author}{Scholl, P.} \emph{et~al.}
\newblock \bibinfo{title}{Erasure conversion in a high-fidelity {Rydberg} quantum simulator}.
\newblock \emph{\bibinfo{journal}{Nature}} \textbf{\bibinfo{volume}{622}}, \bibinfo{pages}{273--278} (\bibinfo{year}{2023}).

\bibitem{wu_erasure_2022}
\bibinfo{author}{Wu, Y.}, \bibinfo{author}{Kolkowitz, S.}, \bibinfo{author}{Puri, S.} \& \bibinfo{author}{Thompson, J.~D.}
\newblock \bibinfo{title}{Erasure conversion for fault-tolerant quantum computing in alkaline earth {Rydberg} atom arrays}.
\newblock \emph{\bibinfo{journal}{Nature Communications}} \textbf{\bibinfo{volume}{13}}, \bibinfo{pages}{4657} (\bibinfo{year}{2022}).

\bibitem{ma_high-fidelity_2023}
\bibinfo{author}{Ma, S.} \emph{et~al.}
\newblock \bibinfo{title}{High-fidelity gates and mid-circuit erasure conversion in an atomic qubit}.
\newblock \emph{\bibinfo{journal}{Nature}} \textbf{\bibinfo{volume}{622}}, \bibinfo{pages}{279--284} (\bibinfo{year}{2023}).

\bibitem{festa_blackbody-radiation-induced_2022}
\bibinfo{author}{Festa, L.}
\newblock \bibinfo{title}{Blackbody-radiation-induced facilitated excitation of {Rydberg} atoms in optical tweezers}.
\newblock \emph{\bibinfo{journal}{Physical Review A}} \textbf{\bibinfo{volume}{105}} (\bibinfo{year}{2022}).

\bibitem{goldschmidt_anomalous_2016}
\bibinfo{author}{Goldschmidt, E.} \emph{et~al.}
\newblock \bibinfo{title}{Anomalous {Broadening} in {Driven} {Dissipative} {Rydberg} {Systems}}.
\newblock \emph{\bibinfo{journal}{Physical Review Letters}} \textbf{\bibinfo{volume}{116}}, \bibinfo{pages}{113001} (\bibinfo{year}{2016}).

\bibitem{cao_autoionization-enhanced_2025}
\bibinfo{author}{Cao, A.}, \bibinfo{author}{Yelin, T.~L.}, \bibinfo{author}{Eckner, W.~J.}, \bibinfo{author}{Oppong, N.~D.} \& \bibinfo{author}{Kaufman, A.~M.}
\newblock \bibinfo{title}{Autoionization-enhanced {Rydberg} dressing by fast contaminant removal}.
\newblock \emph{\bibinfo{journal}{Physical Review Letters}} \textbf{\bibinfo{volume}{134}}, \bibinfo{pages}{133201} (\bibinfo{year}{2025}).

\bibitem{abanin_effective_2017}
\bibinfo{author}{Abanin, D.~A.}, \bibinfo{author}{De~Roeck, W.}, \bibinfo{author}{Ho, W.~W.} \& \bibinfo{author}{Huveneers, F.}
\newblock \bibinfo{title}{Effective {Hamiltonians}, prethermalization, and slow energy absorption in periodically driven many-body systems}.
\newblock \emph{\bibinfo{journal}{Physical Review B}} \textbf{\bibinfo{volume}{95}}, \bibinfo{pages}{014112} (\bibinfo{year}{2017}).

\bibitem{scholl_microwave_2022}
\bibinfo{author}{Scholl, P.} \emph{et~al.}
\newblock \bibinfo{title}{Microwave {Engineering} of {Programmable} {X}{X}{Z} {Hamiltonians} in {Arrays} of {Rydberg} {Atoms}}.
\newblock \emph{\bibinfo{journal}{PRX Quantum}} \textbf{\bibinfo{volume}{3}}, \bibinfo{pages}{020303} (\bibinfo{year}{2022}).

\bibitem{geier_floquet_2021}
\bibinfo{author}{Geier, S.} \emph{et~al.}
\newblock \bibinfo{title}{Floquet {Hamiltonian} {Engineering} of an {Isolated} {Many}-{Body} {Spin} {System}}.
\newblock \emph{\bibinfo{journal}{Science}} \textbf{\bibinfo{volume}{374}}, \bibinfo{pages}{1149--1152} (\bibinfo{year}{2021}).

\bibitem{potirniche_floquet_2017}
\bibinfo{author}{Potirniche, I.-D.}, \bibinfo{author}{Potter, A.~C.}, \bibinfo{author}{Schleier-Smith, M.}, \bibinfo{author}{Vishwanath, A.} \& \bibinfo{author}{Yao, N.~Y.}
\newblock \bibinfo{title}{Floquet symmetry-protected topological phases in cold atomic systems}.
\newblock \emph{\bibinfo{journal}{Physical Review Letters}} \textbf{\bibinfo{volume}{119}}, \bibinfo{pages}{123601} (\bibinfo{year}{2017}).

\bibitem{fendley_competing_2004}
\bibinfo{author}{Fendley, P.}, \bibinfo{author}{Sengupta, K.} \& \bibinfo{author}{Sachdev, S.}
\newblock \bibinfo{title}{Competing density-wave orders in a one-dimensional hard-boson model}.
\newblock \emph{\bibinfo{journal}{Physical Review B}} \textbf{\bibinfo{volume}{69}}, \bibinfo{pages}{075106} (\bibinfo{year}{2004}).

\bibitem{maskara_discrete_2021}
\bibinfo{author}{Maskara, N.} \emph{et~al.}
\newblock \bibinfo{title}{Discrete {Time}-{Crystalline} {Order} {Enabled} by {Quantum} {Many}-{Body} {Scars}: {Entanglement} {Steering} via {Periodic} {Driving}}.
\newblock \emph{\bibinfo{journal}{Physical Review Letters}} \textbf{\bibinfo{volume}{127}}, \bibinfo{pages}{090602} (\bibinfo{year}{2021}).

\bibitem{else_prethermal_2017}
\bibinfo{author}{Else, D.~V.}, \bibinfo{author}{Bauer, B.} \& \bibinfo{author}{Nayak, C.}
\newblock \bibinfo{title}{Prethermal {Phases} of {Matter} {Protected} by {Time}-{Translation} {Symmetry}}.
\newblock \emph{\bibinfo{journal}{Physical Review X}} \textbf{\bibinfo{volume}{7}}, \bibinfo{pages}{011026} (\bibinfo{year}{2017}).

\bibitem{abanin_rigorous_2017}
\bibinfo{author}{Abanin, D.}, \bibinfo{author}{De~Roeck, W.}, \bibinfo{author}{Ho, W.~W.} \& \bibinfo{author}{Huveneers, F.}
\newblock \bibinfo{title}{A {Rigorous} {Theory} of {Many}-{Body} {Prethermalization} for {Periodically} {Driven} and {Closed} {Quantum} {Systems}}.
\newblock \emph{\bibinfo{journal}{Communications in Mathematical Physics}} \textbf{\bibinfo{volume}{354}}, \bibinfo{pages}{809--827} (\bibinfo{year}{2017}).

\bibitem{pastori_characterization_2022}
\bibinfo{author}{Pastori, L.}, \bibinfo{author}{Olsacher, T.}, \bibinfo{author}{Kokail, C.} \& \bibinfo{author}{Zoller, P.}
\newblock \bibinfo{title}{Characterization and {Verification} of {Trotterized} {Digital} {Quantum} {Simulation} via {Hamiltonian} and {Liouvillian} {Learning}}.
\newblock \emph{\bibinfo{journal}{PRX Quantum}} \textbf{\bibinfo{volume}{3}}, \bibinfo{pages}{030324} (\bibinfo{year}{2022}).

\bibitem{gu_practical_2024}
\bibinfo{author}{Gu, A.}, \bibinfo{author}{Cincio, L.} \& \bibinfo{author}{Coles, P.~J.}
\newblock \bibinfo{title}{Practical {Hamiltonian} learning with unitary dynamics and {Gibbs} states}.
\newblock \emph{\bibinfo{journal}{Nature Communications}} \textbf{\bibinfo{volume}{15}}, \bibinfo{pages}{312} (\bibinfo{year}{2024}).

\bibitem{wilde_scalably_2022}
\bibinfo{author}{Wilde, F.} \emph{et~al.}
\newblock \bibinfo{title}{Scalably learning quantum many-body {Hamiltonians} from dynamical data}, \bibinfo{note}{arXiv:2209.14328} (\bibinfo{year}{2022}).

\bibitem{maskara_programmable_2025}
\bibinfo{author}{Maskara, N.} \emph{et~al.}
\newblock \bibinfo{title}{Programmable simulations of molecules and materials with reconfigurable quantum processors}.
\newblock \emph{\bibinfo{journal}{Nature Physics}} \textbf{\bibinfo{volume}{21}}, \bibinfo{pages}{289--297} (\bibinfo{year}{2025}).

\bibitem{verresen_stable_2019}
\bibinfo{author}{Verresen, R.}, \bibinfo{author}{Vishwanath, A.} \& \bibinfo{author}{Pollmann, F.}
\newblock \bibinfo{title}{Stable {Luttinger} liquids and emergent {U(1)} symmetry in constrained quantum chains}, \bibinfo{note}{arXiv:1903.09179} (\bibinfo{year}{2019}).

\bibitem{alcaraz_exactly_1999}
\bibinfo{author}{Alcaraz, F.~C.} \& \bibinfo{author}{Bariev, R.~Z.}
\newblock \bibinfo{title}{An {Exactly} {Solvable} {Constrained} {XXZ} {Chain}}, \bibinfo{note}{arXiv:9904042} (\bibinfo{year}{1999}).

\bibitem{white_quantum_2026}
\bibinfo{author}{White, R.} \emph{et~al.}
\newblock \bibinfo{title}{Quantum {Cellular} {Automata} on a {Dual}-{Species} {Rydberg} {Processor}}, \bibinfo{note}{arXiv:2601.16257} (\bibinfo{year}{2026}).

\bibitem{cesa_engineering_2026}
\bibinfo{author}{Cesa, F.} \emph{et~al.}
\newblock \bibinfo{title}{Engineering discrete local dynamics in globally driven dual-species atom arrays}, \bibinfo{note}{arXiv:2601.16961} (\bibinfo{year}{2026}).

\bibitem{kasteleyn_statistics_1961}
\bibinfo{author}{Kasteleyn, P.~W.}
\newblock \bibinfo{title}{The statistics of dimers on a lattice: {I}. {The} number of dimer arrangements on a quadratic lattice}.
\newblock \emph{\bibinfo{journal}{Physica}} \textbf{\bibinfo{volume}{27}}, \bibinfo{pages}{1209--1225} (\bibinfo{year}{1961}).

\bibitem{kasteleyn_dimer_1963}
\bibinfo{author}{Kasteleyn, P.~W.}
\newblock \bibinfo{title}{Dimer {Statistics} and {Phase} {Transitions}}.
\newblock \emph{\bibinfo{journal}{Journal of Mathematical Physics}} \textbf{\bibinfo{volume}{4}}, \bibinfo{pages}{287--293} (\bibinfo{year}{1963}).

\bibitem{temperley_dimer_1961}
\bibinfo{author}{Temperley, H. N.~V.} \& \bibinfo{author}{Fisher, M.~E.}
\newblock \bibinfo{title}{Dimer problem in statistical mechanics-an exact result}.
\newblock \emph{\bibinfo{journal}{The Philosophical Magazine: A Journal of Theoretical Experimental and Applied Physics}} \textbf{\bibinfo{volume}{6}}, \bibinfo{pages}{1061--1063} (\bibinfo{year}{1961}).

\bibitem{lieb_solution_1967}
\bibinfo{author}{Lieb, E.~H.}
\newblock \bibinfo{title}{Solution of the {Dimer} {Problem} by the {Transfer} {Matrix} {Method}}.
\newblock \emph{\bibinfo{journal}{Journal of Mathematical Physics}} \textbf{\bibinfo{volume}{8}}, \bibinfo{pages}{2339--2341} (\bibinfo{year}{1967}).

\bibitem{wilkins_derivation_2023}
\bibinfo{author}{Wilkins, N.} \& \bibinfo{author}{Powell, S.}
\newblock \bibinfo{title}{Derivation of field theory for the classical dimer model using bosonization}.
\newblock \emph{\bibinfo{journal}{Physical Review E}} \textbf{\bibinfo{volume}{107}}, \bibinfo{pages}{054126} (\bibinfo{year}{2023}).

\bibitem{zhou_emergent_2021}
\bibinfo{author}{Zhou, Z.}, \bibinfo{author}{Yan, Z.}, \bibinfo{author}{Liu, C.}, \bibinfo{author}{Chen, Y.} \& \bibinfo{author}{Zhang, X.-F.}
\newblock \bibinfo{title}{Emergent {Rokhsar}-{Kivelson} point in realistic quantum {Ising} models}, \bibinfo{note}{arXiv:2106.05518} (\bibinfo{year}{2021}).

\bibitem{ardonne_topological_2004}
\bibinfo{author}{Ardonne, E.}, \bibinfo{author}{Fendley, P.} \& \bibinfo{author}{Fradkin, E.}
\newblock \bibinfo{title}{Topological order and conformal quantum critical points}.
\newblock \emph{\bibinfo{journal}{Annals of Physics}} \textbf{\bibinfo{volume}{310}}, \bibinfo{pages}{493--551} (\bibinfo{year}{2004}).

\bibitem{polyakov_gauge_1987}
\bibinfo{author}{Polyakov, A.~M.}
\newblock \emph{\bibinfo{title}{Gauge {Fields} and {Strings}}} (\bibinfo{publisher}{Taylor \& Francis}, \bibinfo{year}{1987}).

\bibitem{zurek_dynamics_2005}
\bibinfo{author}{Zurek, W.~H.}, \bibinfo{author}{Dorner, U.} \& \bibinfo{author}{Zoller, P.}
\newblock \bibinfo{title}{Dynamics of a {Quantum} {Phase} {Transition}}.
\newblock \emph{\bibinfo{journal}{Physical Review Letters}} \textbf{\bibinfo{volume}{95}}, \bibinfo{pages}{105701} (\bibinfo{year}{2005}).

\bibitem{henley_coulomb_2010}
\bibinfo{author}{Henley, C.~L.}
\newblock \bibinfo{title}{The ``{Coulomb} phase''' in frustrated systems}.
\newblock \emph{\bibinfo{journal}{Annual Review of Condensed Matter Physics}} \textbf{\bibinfo{volume}{1}}, \bibinfo{pages}{179--210} (\bibinfo{year}{2010}).

\bibitem{olson_ax_2025}
\bibinfo{author}{Olson, M.} \emph{et~al.}
\newblock \bibinfo{title}{Ax: {A} {Platform} for {Adaptive} {Experimentation}} (\bibinfo{year}{2025}).

\bibitem{kim_large-scale_2019}
\bibinfo{author}{Kim, D.} \emph{et~al.}
\newblock \bibinfo{title}{Large-{Scale} {Uniform} {Optical} {Focus} {Array} {Generation} with a {Phase} {Spatial} {Light} {Modulator}}.
\newblock \emph{\bibinfo{journal}{Optics Letters}} \textbf{\bibinfo{volume}{44}}, \bibinfo{pages}{3178} (\bibinfo{year}{2019}).

\bibitem{chew_ultraprecise_2024}
\bibinfo{author}{Chew, Y.~T.} \emph{et~al.}
\newblock \bibinfo{title}{Ultraprecise holographic optical tweezer array}.
\newblock \emph{\bibinfo{journal}{Physical Review A}} \textbf{\bibinfo{volume}{110}}, \bibinfo{pages}{053518} (\bibinfo{year}{2024}).

\bibitem{kais_review_2014}
\bibinfo{author}{Lidar, D.~A.}
\newblock \bibinfo{title}{Review of {Decoherence}‐{Free} {Subspaces}, {Noiseless} {Subsystems}, and {Dynamical} {Decoupling}}.
\newblock In \emph{\bibinfo{booktitle}{Advances in {Chemical} {Physics}}}, \bibinfo{pages}{295--354} (\bibinfo{publisher}{Wiley}, \bibinfo{year}{2014}).

\bibitem{zaletel_time-evolving_2015}
\bibinfo{author}{Zaletel, M.~P.}, \bibinfo{author}{Mong, R. S.~K.}, \bibinfo{author}{Karrasch, C.}, \bibinfo{author}{Moore, J.~E.} \& \bibinfo{author}{Pollmann, F.}
\newblock \bibinfo{title}{Time-evolving a matrix product state with long-ranged interactions}.
\newblock \emph{\bibinfo{journal}{Physical Review B}} \textbf{\bibinfo{volume}{91}}, \bibinfo{pages}{165112} (\bibinfo{year}{2015}).

\bibitem{hauschild_efficient_2018}
\bibinfo{author}{Hauschild, J.} \& \bibinfo{author}{Pollmann, F.}
\newblock \bibinfo{title}{Efficient numerical simulations with {Tensor} {Networks}: {Tensor} {Network} {Python} ({TeNPy})}.
\newblock \emph{\bibinfo{journal}{SciPost Physics Lecture notes}} \bibinfo{pages}{005} (\bibinfo{year}{2018}).

\bibitem{alet_classical_2006}
\bibinfo{author}{Alet, F.}, \bibinfo{author}{Ikhlef, Y.}, \bibinfo{author}{Jacobsen, J.~L.}, \bibinfo{author}{Misguich, G.} \& \bibinfo{author}{Pasquier, V.}
\newblock \bibinfo{title}{Classical dimers with aligning interactions on the square lattice}.
\newblock \emph{\bibinfo{journal}{Physical Review E}} \textbf{\bibinfo{volume}{74}}, \bibinfo{pages}{041124} (\bibinfo{year}{2006}).

\bibitem{sandvik_correlations_2006}
\bibinfo{author}{Sandvik, A.~W.} \& \bibinfo{author}{Moessner, R.}
\newblock \bibinfo{title}{Correlations and confinement in nonplanar two-dimensional dimer models}.
\newblock \emph{\bibinfo{journal}{Physical Review B}} \textbf{\bibinfo{volume}{73}}, \bibinfo{pages}{144504} (\bibinfo{year}{2006}).

\bibitem{moessner_phase_2001}
\bibinfo{author}{Moessner, R.}, \bibinfo{author}{Sondhi, S.~L.} \& \bibinfo{author}{Chandra, P.}
\newblock \bibinfo{title}{The phase diagram of the hexagonal lattice quantum dimer model}.
\newblock \emph{\bibinfo{journal}{Physical Review B}} \textbf{\bibinfo{volume}{64}}, \bibinfo{pages}{144416} (\bibinfo{year}{2001}).

\end{thebibliography}

\clearpage
\newpage 

\section*{Methods}

\small \noindent\textbf{Experimental system} \\
Our experimental apparatus is described in detail in Refs.~\cite{manovitz_quantum_2025,evered_probing_2025,bluvstein_fault-tolerant_2025}, with several key modifications to enable these analog-digital experiments (ED Fig.~1). 
Individual $^{87}$Rb atoms are stochastically loaded into a static array of 852-nm optical tweezers generated by a spatial light modulator (SLM), with a loading efficiency of 75\% using $\Lambda$-enhanced gray molasses cooling on the D1 line~\cite{brown_gray-molasses_2019}. A second set of 852-nm optical tweezers is generated with a pair of crossed acousto-optic deflectors (AODs; DTSX-400, AA Opto-Electronic) and enable atoms to be rearranged into defect-free arrays~\cite{barredo_atom-by-atom_2016, scholl_programmable_2021,ebadi_quantum_2021}. We use fluorescence imaging with a 0.65-NA objective (Special Optics) onto a CMOS camera
(Hamamatsu ORCA-Quest C15550-20UP) for readout.

The hyperfine qubit is encoded in the clock states $\ket{0} = \ket{F=1,m_F=0}$ and $\ket{1} = \ket{F=2,m_F=0}$ and single-qubit operations on this qubit are implemented via two-photon Raman excitation~\cite{levine_dispersive_2022}. The Raman light is directed either through a global path for global single-qubit gates or through a pair of crossed AODs for fully-programmable local gates~\cite{bluvstein_fault-tolerant_2025}.
In this work, we operate the Raman laser at a decreased intermediate-state detuning of +53\,GHz in order to reach a large two-photon Rabi frequency of $\sim8$\,MHz, which is necessary for coherent mapping (see ``Coherent mapping''). 
Owing to the large differential light shift at this detuning ($\sim1$\,MHz), we choose to implement local $X$ rotations indirectly via local $Z$ rotations combined with global $Y^{1/2}$ rotations~\cite{bluvstein_logical_2024, evered_probing_2025}.

For excitation of the Rydberg qubit, composed of $\ket{1} = \ket{5S_{1/2}, F=2, m_F =0}$ and $\ket{r} = \ket{nS_{1/2}, m_J=-1/2}$, we use a two-photon excitation scheme with 420-nm and 1013-nm Rydberg beams. Here $n$ denotes the Rydberg state principal quantum number and we use either $n=53$ or $n=70$, as discussed below. In contrast to our previous analog quantum simulation experiments~\cite{manovitz_quantum_2025,ebadi_quantum_2021, semeghini_probing_2021}, which used excitation from the stretched states $\ket{5S_{1/2}, F=2, m_F = \pm2}$, here we drive transitions from the clock state. This choice comes with two downsides: the two-photon Rabi frequency is reduced by $\sqrt2$, for the same laser power and scattering rate~\cite{levine_parallel_2019}, and the clock state can couple to both Rydberg magnetic sublevels, $m_J = \pm 1/2$. 
To suppress the latter effect, we first increase our magnetic field to 11.3\,G such that the Zeeman splitting of 32\,MHz is more than $2\times$ larger than the maximum detuning $\Delta$ used in this work. 
Second, we choose to excite to the lower-energy $m_J=-1/2$ state to ensure that all blockade-violating states ---including those involving atoms in both $m_J$ sublevels--- lie at higher energy than the desired blockaded subspace. In the inverse configuration, cancellation of the van der Waals interaction and Zeeman splitting, which are of comparable magnitude, can spuriously bring such blockade-violating states near to resonance.
Finally, where possible, we avoid large positive $\Delta$. This is relevant for the Floquet engineering drives where we have the freedom to choose negative detuning $\pi$-pulses.

Our analog-digital approach to quantum simulation is based on coherently transferring the quantum state of each atom between the digital, long-lived hyperfine qubit and the interacting Rydberg qubit.
Similar methods have been demonstrated previously in Refs.~\cite{bluvstein_quantum_2022,senoo_high-fidelity_2025}, and here we extend this approach to larger two-dimensional arrays.
An important addition in this work is the use of non-destructive readout of the hyperfine qubit via spin-to-position conversion~\cite{bluvstein_fault-tolerant_2025}. This enables loss detection and, since most atoms are retained during the readout, the atoms can subsequently be reinitialized and reused. 
The full experimental sequence for coherent mapping and qubit reuse is illustrated in ED Fig.~1b,c and Supplementary Video 1, and involves the following steps:
\begin{enumerate}
  \item Global state initialization of $\ket{0}$ using Raman-assisted optical pumping~\cite{levine_parallel_2019}.
  \item Local and global single-qubit gates to prepare a product state of arbitrary local superpositions of $\ket{0}$ and $\ket{1}$.
  \item A fast Rydberg $\pi$-pulse to transfer population from $\ket{1}$ to $\ket{r}.$
  \item A fast Raman $\pi$-pulse to transfer population from $\ket{0}$ to $\ket{1}.$
  \item Analog Rydberg evolution by driving the $\ket{1} \leftrightarrow \ket{r}$ transition with time-dependent Rabi frequency $\Omega$ and detuning $\Delta$.
  \item A fast Raman $\pi$-pulse to transfer population from $\ket{1}$ to $\ket{0}$.
  \item A fast Rydberg $\pi$-pulse to transfer population from $\ket{r}$ to $\ket{1}$.
  \item Local and global single-qubit gates to measure in arbitrary local bases. 
  \item Loss-resolved non-destructive qubit readout.
  \item Refilling of missing sites with atoms from the reservoir.
\end{enumerate}
This cycle is repeated until the reservoir is depleted, typically allowing between 20 and 200 cycles (in 2D and 1D arrays, respectively) before reloading a new magneto-optical trap to refill the array. Details of mid-circuit rearrangement from the reservoir are given in Ref.~\cite{bluvstein_fault-tolerant_2025}, with the difference that here we use global imaging, allowing the reservoir occupations to be updated after each cycle rather than keeping track of used atoms in software. 
This is especially important in this work as loss of reservoir atoms is non-negligible when the SLM tweezers are turned off for up to 4\,$\mu$s during the analog evolution; this is required since the Rydberg state is anti-trapped by 852-nm light. Future work would benefit from independent control of the reservoir traps and the rest of the array.
\\

\small \noindent\textbf{Details of experiment configuration}\\
To enable these experiments, we employ a range of configurations, including different array geometries and Rydberg excitation parameters.
We use a zoned architecture in which the static SLM traps are arranged into a simulation zone and a reservoir zone. The reservoir, fixed throughout this work to be 11 rows tall and 32 columns wide, serves as a storage for atoms that are used to refill missing sites at the end of each experimental cycle (see above).
The geometry of the simulation zone is programmed individually for each experiment, and it is where all operations ---including single-qubit gates, Rydberg evolution and qubit readout--- are performed.
In addition, the experiments in this work are carried out using two different choices of Rydberg state, $n$, and intermediate state detuning. These are chosen to balance trade-offs associated with scattering, Rydberg state lifetime, interaction energies, and excitation matrix elements.

For the 1D experiments in Figs.~1--3, we operate at $n=53$, enabling us to use an intermediate-state detuning of $-6.3$\,GHz to minimize scattering error, as well as suppressing decoherence from electric field fluctuations, at the cost of reduced Rydberg lifetime.
For efficient use of laser power, we use narrow Gaussian Rydberg beams (waist $\approx35$\,$\mu$m). These parameters are also compatible with high-fidelity two-qubit gates~\cite{evered_high-fidelity_2023}.
For the analog Rydberg dynamics, the atoms are spaced by $a = 3.6\,\mu$m ($V_0 = 14.2$\,MHz) and we use a Rydberg Rabi frequency of 2.3\,MHz, except for in Fig.~1c.
Since the local Raman tweezer waist of 2.5\,$\mu$m is too large to address individual sites at this close spacing, we use moveable AOD tweezers and increase the atom spacing to $a = 6.0\,\mu$m for the state initialization and measurement (ED Fig.~3a). 
Dynamical decoupling of the hyperfine qubit is used to preserve coherence during this reconfiguration.
Two additional rows of static SLM traps in the simulation zone are used for spin-to-position conversion.

For all remaining experiments performed in a 2D region, we instead operate at $n = 70$, motivated by the larger spacing used for Rydberg dynamics, here $a = 6.0\,\mu$m ($V_0 = 18.1$\,MHz). This spacing is more comfortable for both mid-circuit rearrangement and for resolving sites separated by only $a/2$ during imaging, a consequence of the spin-to-position conversion which effectively doubles the density of the array.
The simulation zone is thus $62\,\mu$m tall and $185\,\mu$m wide; the exact geometries are shown Fig.~1a (kagome lattice) and ED Fig.~4a (15 hexagonal plaquettes).
To uniformly address this zone, we use tophat Rydberg intensity profiles which we homogenize to 2-3\% peak-to-peak variation using the automated calibration described in Ref.~\cite{evered_probing_2025}. This is crucial to ensure homogeneous Rydberg laser intensities for coherent mapping, which enforces a more stringent requirement on homogeneity than our previous analog simulation experiments (see below).
Given limited laser power, here we operate at an intermediate-state detuning of +1\,GHz.
\\

\small \noindent\textbf{Coherent mapping} \\
There are several sources of infidelity that must be considered when performing coherent mapping between the hyperfine and Rydberg manifolds, which we group into population-transfer errors and phase errors.\\

\noindent \textit{Population-transfer errors.}\\
A principal limitation is the long-range vdW interaction tails which detune the Rydberg $\pi$-pulse away from resonance, reducing its fidelity. While we compensate for the mean-field interaction shift, the exact local resonance frequency differs between microstates and cannot be compensated. 
The dominant contribution comes from next-nearest neighbor sites which have a tail energy of 220\,kHz in the 1D chain and 670\,kHz in the 2D kagome lattice. 
To mitigate this effect, we increase the Rydberg Rabi frequency for the mapping pulses to 9.7\,MHz (1D) or 12.6\,MHz (2D) by increasing the 420-nm intensity by $15-20\times$ relative to the analog dynamics.
For these parameters, numerical simulations estimate an upper bound on the Rydberg $\pi$-pulse infidelity of $\approx1.3\%$ in 2D when all possible microstates are equally populated (for an otherwise perfect pulse without vdW tails). 
Importantly, this infidelity leaves the population in $\ket{r}$, which is subsequently lost. These errors are therefore detected with loss-resolved readout and do not lead to a measurement error. 

This phenomenon of loss-conversion of errors by the mapping protocol occurs for two other error sources. 
First, blockade-violating states (i.e., states with nearest-neighbor atoms both in $\ket{r}$) are strongly shifted off-resonance and are thus similarly left in the $\ket{r}$ state and lost; we estimate a $\approx 40\%$ loss probability per qubit in these cases. 
Second, any population erroneously in $\ket{0}$ prior to the mapping down is transferred to $\ket{r}$ and also lost; this can originate from imperfect initial state preparation as well as scattering errors and Rydberg decay during the analog dynamics.
Finally, when mapping up, we always to initialize all qubits in $\ket{0}$ and apply single-qubit gates to prepare $\ket
1$, ensuring that gate errors do not prepare blockade-violating states.\\ 

\noindent \textit{Phase errors.} \\
In addition to imperfect population transfer, both spurious coherent phase accumulation and incoherent dephasing reduce the fidelity of off-diagonal (i.e., $X$- and $Y$-basis) measurements. In particular, population in $\ket{r}$ accumulates a phase relative to the $\{\ket{0},\ket{1}\}$ ground-state manifold during the gap time for the Raman $\pi$-pulse (100\,ns unless stated otherwise). 
The origin of this phase includes:
\begin{itemize}
    \item \textit{Long-range Rydberg interaction tails.} A microstate-dependent phase is accrued due to the vdW interaction between atoms idling in $\ket{r}$.
    \item \textit{420-nm, 1013-nm and Raman light shifts.} During the Raman $\pi$-pulse, the Raman and 1013-nm beams both impart phase shifts on the ground-Rydberg transition of $\sim \pi$. During the Rydberg $\pi$-pulse the 420-nm beam additionally induces a phase of $\sim 2\pi$ on the hyperfine qubit; this is particularly large when blue-detuned by +1\,GHz as the light shift has opposite sign for $\ket{0}$ and $\ket{1}$. We calibrate and compensate these global phases, but residual inhomogeneity of the Rydberg beam intensities results in a spatially-varying phase. We find this to be negligible here, but if necessary this could also be calibrated and locally corrected.
    \item \textit{Dephasing.} Doppler shifts and fluctuations of the 1013-nm light shift on the ground-Rydberg transition result in a random phase of $\sim 2\pi \times 0.02$ radians for 2D experiments ($T_2^*\approx 1-2\,\mu$s at $n=70$, including 1013-nm spatial inhomogeneity). The 420-nm light is not on during the 100\,ns gap and so does not contribute.
\end{itemize}

\noindent \textit{Mitigation and future improvements.} \\
To mitigate these effects, we minimize the time between the analog evolution and the coherent mapping steps. This is accomplished using fast 70\,ns $\pi$-pulses for both the Raman and the Rydberg (square and Gaussian pulse shapes, respectively). The 100\,ns gap ensures timing jitter between the Raman and Rydberg pulses (typically $<$10\,ns) does not lead to pulse overlap. 
We further synchronize all the arbitrary waveform generators (Spectrum Instrumentation) used to generate the Raman and Rydberg pulses to the same 10\,MHz frequency standard (SRS FS725) to prevent drift in the pulse alignment over the repeated cycles.

In Figure~1, we benchmark the coherent mapping for both populations and coherent superpositions and find that it works well in the 1D chain, consistent with the above error sources and known state preparation and readout fidelities~\cite{bluvstein_fault-tolerant_2025, evered_high-fidelity_2023}. In 2D, the larger vdW tail energy is the main limitation for mapping coherent superpositions.
In ED Fig.~1d we measure the coherence after mapping down as a function of interaction energy and gap time, confirming that the reduction in contrast in experiment is in agreement with unitary numerical simulations including vdW tails.

Despite these limitations, we find coherent mapping to be an invaluable tool for probing complex many-body quantum states in one and two dimensions.
In particular, directly measuring coherences of a many-body state in the \{$\ket{1},\ket{r}$\} manifold is otherwise impractical: to apply a single-qubit $\pi/2$-pulse would require increasing the Rydberg Rabi frequency by an additional one to two orders of magnitude.
Two possible approaches to further improve this technique in 2D are (1) reducing the magnitude of the vdW interaction strength relative to other energy scales or (2) reconstructing the desired off-diagonal observable using additional local measurements. 
For (1), a comparable fidelity to the 1D chain is reached at $\approx 4\times$ reduced tail energy for the kagome lattice. This would require a 25\% increase in the lattice spacing, with a corresponding increase in Rydberg laser power to maintain the same intensity, and an $\approx 4\times$ slower analog evolution to maintain the same $R_b/a$.
For (2), one can infer an expectation value $\langle O\rangle$ without any phase error by measuring the evolved operator $U_{\mathrm{vdW}} O U^\dagger_{\mathrm{vdW}}$, where $U_{\mathrm{vdW}}$ is the known unitary evolution during the gap time. In general this will be a sum of Pauli strings requiring local measurements in mixed bases, although symmetries may reduce the number of measurement settings. Moreover, rather than assuming a known vdW interaction strength, in certain cases one can use parity oscillations to directly extract and compensate for the accumulated phase by conditioning on the Rydberg populations of neighboring qubits in post-processing. The remaining limitation of this approach is that the Rydberg $\pi$-pulse still experiences a microstate-dependent detuning shift.\\

\small \noindent\textbf{Loss in analog simulations} \\
Atom loss is one of the main error sources in analog Rydberg simulation, arising predominantly from blackbody-induced decay of the Rydberg state, loss from turning off the optical tweezers during the Rydberg evolution, and scattering from the intermediate state in the two-photon excitation scheme. Furthermore, other error channels can be converted to loss errors via the coherent mapping protocol (see above) or other erasure-conversion schemes~\cite{scholl_erasure_2023, wu_erasure_2022, ma_high-fidelity_2023}. 
To mitigate the impact of loss, we use loss-resolved readout to directly detect which atoms are missing. This identifies true losses as well as vacancies from imperfect rearrangement. In these experiments, we find that postselection using this loss information is an efficient approach that improves effective fidelities~\cite{senoo_high-fidelity_2025, scholl_erasure_2023}: only the qubits within the operator being measured need to be postselected on atom presence, such that the sample overhead scales with operator weight rather than system size (ED Fig.~2c,d).

We further use the ability to detect loss to characterize correlations between loss events in Rydberg simulations. In particular, blackbody decay of Rydberg \textit{S} states to opposite-parity Rydberg \textit{P} states is known to trigger the loss of many atoms, known as avalanche events~\cite{festa_blackbody-radiation-induced_2022, goldschmidt_anomalous_2016, cao_autoionization-enhanced_2025}.
Indeed, we find that the distribution of the number of lost atoms per snapshot deviates from a Poisson distribution for large number of losses (ED Fig.~2a), indicative of such large-scale correlated loss events. Furthermore, the two-point loss correlations decay as $1/r^3$ for radial separation $r$ (ED Fig.~2b), consistent with dipolar \textit{S-P} interactions that can tune otherwise unpopulated blockade-violating states into resonance.

For Floquet experiments in the 1D chain (Fig.~2a--e and Fig.~3), we retain $\approx$ 95\% of the data after loss postselection. In Figs.~2c,d we postselect on all 19 sites and keep 55\%.
For the hexagonal plaquettes, we postselect on all six sites, keeping 77\% of the data.
For the quantum spin liquid experiments in Figures~4--6, as representative examples, we retain 96\% for the mean density in Fig.~4d, 90\% for dimer populations in Fig.~4e, and 81\% for the six-body RK energies in Fig.~4f.\\

\small \noindent\textbf{Approach to Floquet Hamiltonian engineering} \\
Periodic driving can transform a system's native Hamiltonian into an effective Hamiltonian with qualitatively different dynamics~\cite{goldman_periodically_2014, abanin_effective_2017}, commonly achieved through modulation of the single-qubit Pauli frame~\cite{scholl_microwave_2022, geier_floquet_2021} or the interaction strengths~\cite{potirniche_floquet_2017}.
Here we use the Floquet engineering protocol developed in Refs.~\cite{koyluoglu_floquet_2024,feldmeier_quantum_2024}, summarized below, to realize effective Hamiltonians beyond the native Rydberg Hamiltonian ($H$, Eq.~\ref{eq:Hamiltonian}) using a global detuning drive with period $\tau$, $\Delta(t+n\tau)=\Delta(t)$ and a constant Rabi frequency, $\Omega$.
This approach relies on constrained dynamics generated by Rabi oscillations within the nearest-neighbor blockade approximation, where the many-body unitary evolution is governed by the `PXP' model~\cite{fendley_competing_2004}
\begin{align}\label{eq:PXP_ham}
    H/\hbar \approx \underbrace{\frac{\Omega}{2} \sum_{i} \mathcal{P} \sigma_i^x \mathcal{P}}_{H_{\text{PXP}}} - \Delta(t) N
\end{align}
with $N=\sum_i n_i$. The projector $\mathcal{P}=\prod_{\braket{i,j}} (1-n_i n_j)$ enforces the blockade constraint by removing configurations in which nearest-neighbor sites $\braket{i,j}$ are simultaneously excited to the Rydberg state. Within this constrained manifold, $\pi$-pulses of the global detuning effectively reverse the sign of the interacting PXP evolution~\cite{maskara_discrete_2021}, $e^{i \pi N} H_{\text{PXP}} e^{i \pi N} = - H_{\text{PXP}}$. 
This property enables a `many-body echo' by means of equally-spaced $\pi$-pulses located at $t_e + n \tau$ and $t_e + \tau/2 +n\tau$ within each Floquet cycle $n$, 
\begin{align*}
    \Delta_e(t\big|t_e,w_e) = -\pi\sum_{n\in \mathbb{N}_0} f\left(t\big|t_e+n\frac{\tau}{2},w_e\right)
\end{align*}
where $t_e \leq \tau/2$ and $f(t|\overline{t},w) = \frac{1}{w\sqrt{\pi}}e^{-(t-\overline{t})^2/w^2}$ describes narrow Gaussian pulses centered at $\overline{t}$ with width $w$ (Fig.~2a).
In the instantaneous pulse limit $w_e \to 0$, the PXP evolution is reversed during the interval between the two $\pi$-pulses: within each Floquet period, the system evolves forward for time $t_e$, backward for $\tau/2$, and forward again for $\tau/2-t_e$. The resulting net PXP evolution up to time $t$ can be expressed as $U(t)=e^{-iH_{\mathrm{PXP}}\, t_{\mathrm{eff}}(t)}$, with an effective time $t_{\mathrm{eff}}(t)$, 
\begin{align*}
    t_{\mathrm{eff}}(0\leq t \leq \tau) = 
    \begin{cases} 
		t & 0 \leq t < t_e \\
		-t + 2 t_e & t_e \leq t < t_e + \tau/2 \\
		t - \tau & t_e + \tau/2 \leq t < \tau
	\end{cases},
\end{align*}
and $t_{\mathrm{eff}}(t+\tau) = t_{\mathrm{eff}}(t)$.
In particular, the system exhibits periodic revivals as there is no net evolution at stroboscopic times, $t_{\mathrm{eff}}(n\tau) = 0$, and thus the Floquet unitary $U_F = U(\tau)=\mathbf{1}$.

By perturbing the many-body echo drive, $\Delta(t) = \Delta_e(t) + \Delta_p(t)$, we controllably generate a variety of higher-body interactions in the Floquet effective Hamiltonian. The effect of the global detuning perturbation $\Delta_p(t)$ can be captured in the interaction picture with respect to the many-body echo evolution~\cite{koyluoglu_floquet_2024,feldmeier_quantum_2024}, resulting in the Floquet unitary $U_{\mathrm{int}} = \mathcal{T}\exp{\left\{i\hbar\int_{0}^{\tau} dt \Delta_p(t) N(t_{\mathrm{eff}}(t))\right\}}$. Here, the total Rydberg occupation operator $N$ is conjugated by the effective interacting PXP evolution, 
\begin{align*}
    N(t_{\mathrm{eff}}) \equiv e^{it_{\mathrm{eff}} H_{\text{PXP}}} N e^{-it_{\mathrm{eff}} H_{\text{PXP}}}.
\end{align*}
The key insight is that $N(t_{\mathrm{eff}})$ acquires overlap with multi-body interactions due to operator spreading. The choice of perturbation $\Delta_p(t)$ then determines the weights of these terms in the engineered effective Floquet Hamiltonian $U_{\mathrm{int}}\equiv\exp{\left\{ -i \tau H_{\mathrm{eff}}\right\}}$. To leading order, 
\begin{equation} \label{eq:H_eff_leading_order}
    \frac{H_{\mathrm{eff}}^{(0)}}{\hbar} = -\int_0^\tau dt \Delta_p(t) N(t_{\mathrm{eff}}(t))
\end{equation} 
with higher-order corrections obtained systematically from a high-frequency Floquet-Magnus expansion~\cite{else_prethermal_2017, abanin_rigorous_2017}. 

Deviations of $\Delta_e(t)$ from ideal instantaneous $\pi$-pulses ($w_e\to 0$) can also be absorbed into the global detuning perturbation: $\tilde{\Delta}_p(t)=\Delta_p(t) + \bigl[ \Delta_e(t\big|t_e,w_e)-\Delta_e(t\big|t_e,0) \bigr]$, and therefore enter the effective Hamiltonian on the same footing. 
Furthermore, van der Waals interaction tails beyond the blockade radius, which are neglected in the PXP approximation, are likewise dressed by the effective interacting PXP evolution, thereby generating additional contributions to the effective Hamiltonian. \\

\small \noindent\textbf{Engineering hopping dynamics}\\
In Figures~2 and 3, we engineer an effective blockaded XX Hamiltonian using a global detuning drive, parameterized by \{$\varepsilon$,$\gamma$,$\theta$,$\Delta_0$\} with the form
\begin{align*}
	&\Delta(t) / 2\pi = (\varepsilon-0.5)\sum_{n\in N} f\left(t\big|\frac{\tau}{4}+n\frac{\tau}{2},w_e\right) \notag \\
    &+(\gamma - \theta) \sum_{n\in N} f\left(t\big|n\tau,w_p\right) + \theta \sum_{n\in N} f\left(t\big|\frac{\tau}{2} +n\tau,w_p\right)+\Delta_0 / 2\pi
\end{align*}
where $w_e = 16$\,ns and $w_p = 8$\,ns are chosen to be as narrow as possible whilst ensuring the magnitude of $\Delta(t)$ does not exceed $\approx 20$\,MHz. This minimizes accidental modulation of the 420-nm intensity, and hence $\Omega$, arising from the frequency-dependent deflection efficiency of the acousto-optic modulators (AOMs) used to control the detuning profile; we correct for this effect within the $\pm20$\,MHz range.
We use $\tau = 0.36\,\mu$s and $\Omega = 2\pi \times 2.3$\,MHz ($R_b/a = 1.35$), ensuring strong blockade such that the PXP approximation is valid.
Since the effective PXP evolution is bounded by ${\Omega t_{\mathrm{eff}}}/{2\pi}\leq {\Omega \tau}/{8\pi} \approx 0.2$, we can exploit a short-time expansion of the time-evolved global number operator $N(t_{\mathrm{eff}})$ in the blockaded subspace~\cite{koyluoglu_floquet_2024}, 
\begin{equation*}
\begin{split}
    N(t_{\mathrm{eff}}) &= N + \frac{\Omega t_{\mathrm{eff}}
    }{2} H_{\text{PYP}} \\ 
    &+ \frac{(\Omega t_{\mathrm{eff}})^2}{4}  \left(- H_{\text{PXYP}} + \frac{1}{4} H_{\text{ZIZ}} - 3 N\right)+ \mathcal{O}((\Omega \tau)^3).
\end{split}
\end{equation*}
Here, $H_{PYP} \equiv\sum_i \mathcal{P} \sigma_i^y \mathcal{P}$ enacts blockaded local spin-flips with a relative phase, $H_{PXYP} \equiv \frac{1}{2}\sum_i \mathcal{P} \left( \sigma_i^x \sigma_{i+1}^x + \sigma_i^y \sigma_{i+1}^y \right) \mathcal{P}$ generates blockaded nearest-neighbor spin exchange, and $H_{ZIZ} = \sum_i \sigma^z_i \sigma^z_{i+2}$ is a diagonal next-nearest-neighbor interaction. We define $\sigma_i^x = \ket{1_i}\bra{r_i} + \ket{r_i}\bra{1_i}$,  $\sigma_i^y = i \ket{1_i}\bra{r_i} - i \ket{r_i}\bra{1_i}$, and $\sigma^z_i=2n_i-1=\ket{r_i}\bra{r_i}-\ket{1_i}\bra{1_i}$. These operators form the basis for the following ansatz for the effective Floquet Hamiltonian:
\begin{align}\label{eqn:XY_effective_ham}
    H_F/\hbar &\approx -\mu\; N +U\; H_{\text{ZIZ}} - J\; H_{\text{PXYP}} \\
    &- g_X\; H_{\text{PXP}} - g_Y\; H_{\text{PYP}} \nonumber
\end{align}

In Fig.~2a, we illustrate the detuning profile for $\varepsilon = 0.1$, $\theta = 0.05$, and $\gamma = -0.25$. For all data in the main text, $\theta = 0$ and $\Delta_0 \approx 2\pi\times-0.5$\,MHz (up to a systematic offset in $\Delta$) are fixed.
For the many-body echo in Fig.~2c, we use $\varepsilon = -0.06$ and $\gamma = 0$.
For Fig.~2e we use, from left to right, $(\varepsilon,\gamma)$ equal to ($-0.05$,$-$0.05), (0.0,$-$0.15), and (0.1,$-$0.35), where $\gamma$ is chosen for good number conservation.
In Fig.~2c (right) and Figure~3, we target fast hopping dynamics and use $\varepsilon = 0.2$ and $\gamma = -0.4$.
Even though many of these parameters are not small, we find that the engineered dynamics are well-described by Eq.~\eqref{eqn:XY_effective_ham}, where the coefficients are renormalised by the finite pulse widths and van der Waals tails. 
For example, we find that the hopping energy scales linearly with $\varepsilon$ (ED Fig.~3b).

Corrections from higher-order terms in the short-time and Floquet-Magnus expansions may also introduce additional operators into the effective Hamiltonian, with coefficients beyond quadratic order in $\Omega\tau,\varepsilon,\gamma,\theta$~\cite{koyluoglu_floquet_2024}.
In particular, for drives with $\gamma>\pi/2$, we can redefine $\gamma'=\gamma+\pi$, such that the drive is recast as a perturbation around a different many-body echo containing additional $\pi$-pulses at $t=n\tau$. This redefinition effectively symmetrizes the drive, $t_{\mathrm{eff}}(\tau \leq t \leq 2\tau)=-t_{\mathrm{eff}}(0 \leq t \leq \tau)$, and thereby removes terms linear in $t_{\mathrm{eff}}$ from the leading-order effective Hamiltonian. Because $N(t_{\mathrm{eff}})=N(0)$ at $t=n\tau$, another effect of this transformation is a modification of the prefactor of the global detuning associated with $\gamma$ in the leading-order effective Hamiltonian.\\

\small \noindent\textbf{Verification of the effective Hamiltonian} \\
To independently verify the effective Hamiltonian governing the Floquet-engineered dynamics, we fit experimentally measured stroboscopic observables to exact classical simulations of the ansatz model~\cite{pastori_characterization_2022, gu_practical_2024, wilde_scalably_2022}.
Specifically, we numerically optimize the Hamiltonian coefficients to minimize the discrepancy between simulated and experimentally measured bitstring probability distributions on a contiguous 8-site subsystem, which fully determine all observables diagonal in the chosen measurement basis and supported on the block (i.e., all local $k\!\le\!8$-body correlators in that basis). To mitigate boundary effects, we analyze the central eight sites of the experimentally realized open chain, and compare to eight consecutive sites of a 12-site periodic chain in simulation.

We consider several initial states $\ket{\psi_0}$, Floquet periods $n$, and measurement unitaries $\mathcal{M}$ to constrain the Hamiltonian terms that govern the hopping, detuning, and creation of excitations. First, we separately prepare and Floquet evolve the vacuum (all-ground) state $\ket{\psi_0}=\ket{\Psi_\downarrow}$, a single central excitation $\ket{\psi_0}=\sigma^x_0 \ket{\Psi_\downarrow}$, and two next-nearest-neighbor excitations $\ket{\psi_0}=\sigma^x_{-1} \sigma^x_{+1} \ket{\Psi_\downarrow}$, followed by measurement in the $Z$-basis ($\mathcal{M}=\mathbb{I}$). To probe phase-sensitive dynamics, we additionally prepare four different superpositions of the vacuum and single-excitation states, $\ket{\psi_0}=\frac{1}{\sqrt{2}}\left( \ket{\Psi_\downarrow}\pm \sigma^{x,y}_0 \ket{\Psi_\downarrow} \right)$, and measure in the $X$-basis ($\mathcal{M}=e^{-i\frac{\pi}{4}\sum_j \sigma^y_{j}}$). Aggregating these datasets, we perform a multi-parameter minimization of the cost function
\begin{equation*}
\begin{aligned}
    C(\mathcal{S})=\sum_{\substack{\{(|\psi_0\rangle,\mathcal{M})\}\\\mathbf{s}\in \{0,1\}^{\otimes 8}}}  \sum_{n=0}^{n_{\max}-1}&\frac{1}{2^8\cdot n_{\max}}\Big[ p_{\mathrm{exp}}\left(\mathbf{s}\big|\ket{\psi_0},n,\mathcal{M}\right)\\
    - &\big|\bra{\mathbf{s}}\mathcal{M}e^{-i(n\tau)H|_{J,h,g}}\ket{\psi_0}\big|^2\Big]^2
\end{aligned}
\end{equation*}
where $\mathcal{S} = (J,\mu,U, g_X,g_Y)$ is the set of input parameters and  $p_{\mathrm{exp}}\left(\mathbf{s}\big|\ket{\psi_0},n,\mathcal{M}\right)$ is the experimentally measured probability of observing bitstring $\mathbf{s}$ at time $n\tau$ in each of the above configurations.

We apply this procedure to verify the effective Hamiltonian engineered in Figure~3, yielding a cost function $C_{\min}=1.34\cdot 10^{-4}$ with optimal coefficients $J/2\pi=-0.143\,\mathrm{MHz}$, $\mu/2\pi=0.411$\,MHz, $U/2\pi=0.023\,\mathrm{MHz}$, $g_X/2\pi=0.000\,\mathrm{MHz}$, and $g_Y/2\pi = 0.021 \,\mathrm{MHz}$ (ED Fig.~3c,d). We find the fitted $J$ and $(\mu+4U)=2\pi \times 0.502\,\mathrm{MHz}$ are close to those extracted from the single-particle dispersion measurement (Fig.~3d), corroborating the blockaded XX Hamiltonian as a good description of the Floquet-engineered dynamics. \\

\small \noindent\textbf{Many-body spectroscopy} \\
Here we outline the procedure for measuring the single-particle Green's function, $\mathcal{G}(r,t) = \bra{\Psi_\downarrow}\sigma^x_r(t) \sigma^x_0(0) \ket{\Psi_\downarrow}$, in Fig.~3d~\cite{maskara_programmable_2025, knap_probing_2013}.\\

\noindent \textit{Measurement protocol.}\\
We prepare one of four different superposition states $|\psi_{\pm x,y} \rangle = \left( \ket{\Psi_\downarrow}\pm \sigma^{x,y}_0 \ket{\Psi_\downarrow} \right)/\sqrt{2}$ where $\sigma_0^{x,y}$ acts on the central site, $r=0$, and all other spins are in the ground state. After evolving under the Floquet-engineered dynamics for a time $t=n\tau$, we measure all sites in the $X$-basis in parallel to obtain the site-resolved expectation values
\begin{equation*}
\begin{aligned}
    \braket{\sigma_r^x(t)}_{\pm x,y} &\equiv \bra{\psi_{\pm x,y}}\sigma_r^x(t)\ket{\psi_{\pm x,y}}\\ 
    &= \bra{\Psi_\downarrow} \Big[ \sigma^x_r(t) + \sigma^{x,y}_0(0) \sigma^x_r(t) \sigma^{x,y}_0(0) \\
    &\qquad \pm\sigma^{x}_r(t) \sigma^{x,y}_0(0)\pm \sigma^{x,y}_0(0) \sigma^x_r(t) \Big]  \ket{\Psi_\downarrow}.
\end{aligned}
\end{equation*}
Taking differences cancels the first two terms and isolates the real and imaginary parts of the target Green's function:

\begin{equation*}
\begin{aligned}
\Re \mathcal{G}(r,t) &= \frac{1}{2}\bra{\Psi_\downarrow}
\left[\sigma^x_r(t)\sigma^{x}_0(0)
      + \sigma^{x}_0(0) \sigma^x_r(t) \right]
\ket{\Psi_\downarrow} \\
&=\frac{\braket{\sigma_r^x(t)}_{+x}-\braket{\sigma_r^x(t)}_{-x}}{4}
\end{aligned}
\end{equation*}
\begin{equation*}
\begin{aligned}
    \Im{\mathcal{G}(r,t)} &= \frac{1}{2i}\bra{\Psi_\downarrow} \left[\sigma^x_r(t)\sigma^{x}_0(0) -\sigma^{x}_0(0) \sigma^x_r(t) \right] \ket{\Psi_\downarrow}\\
    &= \frac{1}{2}\bra{\Psi_\downarrow} \left[\sigma^x_r(t)\sigma^{y}_0(0) + \sigma^{y}_0(0) \sigma^x_r(t) \right] \ket{\Psi_\downarrow}\\ 
    &=\frac{\braket{\sigma_r^x(t)}_{+y}-\braket{\sigma_r^x(t)}_{-y}}{4} \\
\end{aligned}
\end{equation*}
where $\sigma^y_0\ket{\Psi_{\downarrow}}=-i\sigma^x_0\ket{\Psi_{\downarrow}}$ is used in the second line.
Finally, by performing a discrete Fourier transform in time and space, we obtain the density of states (Fig.~3d) as a function of momentum, $k$, and frequency, $\omega$, across the 1D Brillouin zone:
\begin{align*}
    \mathcal{G}(k,\omega) = \sum_{n,r}  e^{-in \omega\tau} e^{-ikr}\; \mathcal{G}(r,n\tau).
\end{align*}
For the hopping Hamiltonian, Eq~\eqref{eqn:XY_effective_ham}, with chemical potential $\mu$,
blockaded hopping energy $J$,
and next-nearest neighbor interaction $U$,
the peak of the imaginary part follows the single-particle dispersion $\omega(k)=\mu+4U+2J\cos{ka}$~\cite{maskara_programmable_2025,koyluoglu_floquet_2024}. \\

\noindent{\it Calibration of phases.}\\
Any relative phase accumulation between the two qubit manifolds must be compensated to ensure the prepared state, analog evolution, and measurement are all in the desired basis.
Using single qubits, we first map up $\ket{+}_{0,1} =(\ket{0}+\ket{1})/\sqrt{2}$ and calibrate the phase of a readout Rydberg $\pi/2$-pulse, corresponding to faithfully preparing $\ket{+}_{1,r} = (\ket{1}+\ket{r})/\sqrt{2}$. We then remove the Rydberg $\pi/2$-pulse, map down, and calibrate the phase of a readout Raman $\pi/2$-pulse, corresponding to returning to $\ket{+}_{0,1}$. Finally, we calibrate the 420-nm differential light shift on the hyperfine qubit imparted during the analog evolution. 
These phases are then used in the four measurements described above. 
We further apply an in-software compensation of the mean-field phase from vdW tails during the mapping down ($\approx 0.04\pi$), estimated from the measured populations with the $\pi/2$-pulse before readout removed. Miscalibration of these phases mixes the real and imaginary parts of the Green's function. \\

\small \noindent \textbf{Domain wall dynamics under Rydberg blockade}\\
The constrained XX model in the absence of spin flips, Eq.~\eqref{eqn:XY_effective_ham} with $g_{X,Y} = 0$, can be mapped onto the integrable XXZ Heisenberg chain for which spin-wave excitations are known to be interacting~\cite{verresen_stable_2019}, as demonstrated by the spin scattering in Fig.~3b.
In ED Fig.~3e-g, we further investigate the effect of these interactions on the dynamics of domain walls on top of the maximally-packed N\'eel state, $\ket{\mathbb{Z}_2}$. By allowing for a weak spin-flip term $g_{X,Y}$, domain wall pairs can be created and subsequently propagate under the hopping dynamics, washing out the initial N\'eel order.
Importantly, the band of domain wall excitations must be resonant with the N\'eel state for this process to occur. By sweeping the energy of the N\'eel state across the band, tuned via the chemical potential, we observe this generation of gapless domain walls (ED Fig.~3g).
This can be interpreted as an out-of-equilibrium probe of the underlying Luttinger liquid ground state found when $g_{X,Y}\rightarrow 0$. This Luttinger liquid phase is a consequence of the emergent $U(1)$ symmetry~\cite{verresen_stable_2019, alcaraz_exactly_1999,koyluoglu_floquet_2024} and, interestingly, is 
the one-dimensional counterpart of the $U(1)$ QSL~\cite{fradkin_field_2013}.\\

\small \noindent\textbf{Engineering ring-exchange dynamics}\\
Ring-exchange dynamics can be realized in our Floquet approach either by directly engineering a six-body interaction ($g_6\times \mathcal{P}\,(\prod_{i=1}^6\sigma^x_i)\,\mathcal{P}$) or indirectly via a second-order process generated from three-body interactions ($g_3 \times \sum_i \mathcal{P}\,\sigma_i^y \sigma_{i+1}^y \sigma_{i+2}^y\,\mathcal{P}$) together with a large static detuning. Here we employ the latter approach.
Since accessing these higher-weight operators requires leaving the short-time regime, we follow the numerical optimization of the detuning profile of Ref.~\cite{feldmeier_quantum_2024}, which generates profiles that produce large overlap of the effective Hamiltonian with the desired three-body interactions. 
Concretely, we approximate the profile by the closed-form expression
\begin{equation*}
\begin{aligned}
    &\Delta(t) / 2\pi = (\varepsilon_0-0.5)\sum_{n\in N} f\left(t\big|n\tau,w_e\right)+\\
    &(\varepsilon_1-0.5)\sum_{n\in N} f\left(t\big|\frac{\tau}{2}+n\tau,w_e\right) 
    +A \cos{\big(\frac{2\pi}{\tau}t\big)} + \Delta_0/2\pi
\end{aligned}
\end{equation*}
parameterized by $\{\varepsilon_0,\varepsilon_1,A,\Delta_0\}$.
We use $\tau = 0.30\,\mu$s and $\Omega = 2\pi\times 2.3$\,MHz ($R_b/a = 1.41$), such that the effective PXP evolution satisfies ${\Omega t_{\mathrm{eff}}}/{2\pi}\leq {\Omega \tau}/{4\pi} \approx 0.35$, providing sufficient evolution to build appreciable overlap with the target three-body operators~\cite{feldmeier_quantum_2024}.
For the data in Fig.~2f-h, we optimize the detuning profile by hand, targeting fast dynamics and good number conservation, and obtain $w_e= 20$\,ns, $\varepsilon_0=0.025$, $\varepsilon_1=0.13$, $A=-1.75$ and $\Delta_0 \approx 2\pi \times 0.65$\,MHz, where we observe that $A$ is the primary parameter that controls the speed of the dynamics (ED Fig.~4b).
For these parameter values, the leading-order effective Hamiltonian of Eq.~\eqref{eq:H_eff_leading_order} exhibits three-body interactions of strength $g_3 \approx -2\pi \times 0.3$\,MHz and an effective detuning $\mu = -2\pi \times 0.8$\,MHz. 
Neglecting all terms other than $g_3$ and $\mu$, this induces an effective ring-exchange of strength $g^{(\mathrm{eff})}_6 \approx 2\pi \times 0.22$\,MHz, which predicts plaquette oscillations of period $\pi/g^{(\mathrm{eff})}_6 \approx 2.3\,\mu$s. This time scale is consistent with the oscillations observed in the experiment (Fig.~2h, ED Fig.~4d).
Moreover, $H_{\mathrm{eff}}^{(0)}$ contains blockaded hopping of strength $J\approx -2\pi \times 0.23$\,MHz, comparable to the strength of the effective ring-exchange and consistent with the experimental result of Fig.~2g.
We note however that in this protocol, perturbations around the many-body echo are strong, both due to the detuning drive as well as the stronger tails of the vdW interaction on the chosen ring geometry. The first-order expression of Eq.~\eqref{eq:H_eff_leading_order} should therefore be viewed as a qualitative guide for the direct optimization of the dynamics on the simulator.

Finally, we find that inhomogeneity between different plaquettes contributes to the decaying contrast of the oscillations (ED Fig.~4c).
This decay is reproduced in numerical simulations that include Rydberg dephasing, $T_2^*\approx 1\,\mu$s, incorporating both shot-to-shot fluctuations and spatial inhomogeneity of the 420-nm and 1013-nm light shifts.
We similarly find that the engineered dynamics are sensitive to proper optimization of $\Delta_0$. This sensitivity is the main limitation to extending this Floquet approach to two dimensions, where we find the large micro-state dependent vdW tails reduce the fidelity of the engineered number conservation; the symmetry of the two N\'eel states on the single plaquette provides a special case where this effect is small.
This therefore motivates exploring protocols that are robust to such detuning shifts, as well as architectures such as dual-species processors with individually tunable inter- and intra-species interactions~\cite{white_quantum_2026, cesa_engineering_2026}.\\

\small \noindent\textbf{Overview of the $U(1)$ quantum spin liquid}\\
Here we provide a theoretical overview of the Rokhsar-Kivelson Hamiltonian and the field theory description of the $U(1)$ quantum spin liquid.\\

\noindent \textit{Rokhsar-Kivelson state and Hamiltonian.
}\\
The Rokhsar-Kivelson (RK) state is the equal-amplitude and equal-phase superposition of dimer configurations on the honeycomb lattice: 
\begin{equation} \label{eq-RKwavefunction}
    \ket{\Psi_{\mathsf{RK}}} \propto \sum_{\mathcal{D}} \ket{\mathcal{D}}
\end{equation}
where $\mathcal{D}$ is any configuration of dimers with exactly one dimer touching each vertex, i.e., satisfying the local dimer constraint.
In literature, such a wavefunction first appeared as the ground state of the RK dimer model \cite{rokhsar_superconductivity_1988}, $H_{\mathsf{RK}} = -tT + vV$, at its RK point ($t = v$).
Here, $T$ is the kinetic energy and $V$ is the potential energy, defined in Fig.~4a.
At this RK point, the Hamiltonian is positive semi-definite and can be written as a sum of (non-commuting) projectors.
The wavefunction of Eq.~\eqref{eq-RKwavefunction} is a zero energy state of the Hamiltonian and is hence one of its ground states.\\

\noindent{\it Predictions for diagonal correlations.} \\
The diagonal observables of the RK state can be rewritten as observables in a classical gas of dimers about which much is known~\cite{fradkin_field_2013, moessner_quantum_2008}. 
In particular, for any diagonal observable $\hat{\mathcal{Z}}$ in the dimer basis, $\langle \mathcal{D}|\mathcal{Z}|\mathcal{D}'\rangle=\mathcal{Z}(\mathcal{D})\delta_{\mathcal{D},\mathcal{D}'}$, hence
\begin{align}\label{eq:classical_dimer}
    \bra{\Psi_{\mathsf{RK}}} \hat{\mathcal{Z}} \ket{\Psi_{\mathsf{RK}}} = \frac{1}{\sum_{\mathcal{D}}1}\sum_{\mathcal{D}}\mathcal{Z}(\mathcal{D}).
\end{align}
where $\sum_{\mathcal{D}} 1$ is the infinite-temperature partition function of the classical dimer gas.
The correlation functions of the classical dimer gas are known exactly \cite{kasteleyn_statistics_1961, kasteleyn_dimer_1963, temperley_dimer_1961,lieb_solution_1967} and can be reproduced by the $(2+0)\text{D}$ 
compact-boson conformal field theory (CFT) \cite{fradkin_field_2013} whose action is
\begin{equation} \label{eq-compactboson}
    S = \frac{1}{8\pi K} \int \text{d}^2 x\, (\partial_\mu \varphi)^2
\end{equation}
where for the RK state $K = 1$ and $\varphi$ is a compact bosonic field.
This can be seen through a number of methods, including height-field representations of dimer configurations \cite{fradkin_field_2013}.
Most explicitly, the classical dimer gas is free-fermion solvable via a transfer matrix method~\cite{lieb_solution_1967} and its correlations on the lattice can be explicitly related to those of Eq.~\eqref{eq-compactboson} in the continuum~\cite{wilkins_derivation_2023}.

The relevant correlation functions of CFT are of $\partial \phii = \partial_x \phii - i \partial_y \phii$, which obey $\langle \partial \phii(x) \partial \phii(0) \rangle \sim 1/x^2$, and are further associated with vertex operators $e^{i m \varphi}$, which are local operators for $m \in \mathbb{Z}$.
These vertex operators have scaling dimensions given by $\Delta_m = m^2 K $, implying that correlations of these vertex operators decay as
\begin{equation}
    \langle e^{i m [\phii(x) - \phii(0)]} \rangle \sim \frac{1}{x^{2m^2 K}}
\end{equation}
For $m=1$, $e^{i \phii}$ is related to the nematic order parameter.
In particular, let $x$ label the $A$ sublattices of the honeycomb lattice, which forms the tripartite triangular lattice.
If $\sigma(x) = 0, 1, 2$ labels the sublattices of the triangular lattice, then
\begin{equation}\label{eq:CFT_2pt}
    \omega^{\sigma(x)} \sum_{\alpha} n_{\alpha}(x) \omega^{\alpha} \sim e^{i \phii(x)}  + \partial \phii(x)
\end{equation}
where $n_{\alpha}(x)$ is onsite density at the three edges $\alpha = 0, 1, 2$ surrounding a vertex $x$ and $\omega = e^{2\pi i/3}$ is the third root of unity.
As such we expect the correlations of the nematic order parameter to decay as $\sim 1/x^2$.

Moreover, when $m = 1/2$, $e^{i m \phii}$ is not a local operator and appears as a string on the lattice.
Indeed, we find that \cite{zhou_emergent_2021}
\begin{equation}\label{eq:CFT_string}
    \omega^{\sigma(x)} \sum_{\alpha, \beta} T_{\gamma_{\alpha, \beta}} \omega^{\alpha -  \beta} \sim e^{i [\phii(x) - \phii(0)]/2} + \cdots
\end{equation}
where the $\cdots$ are fields whose decay is faster than the leading  $e^{i [\phii(x) - \phii(0)]/2}$ term,  $\gamma_{\alpha \beta}$ is a string that starts at edge $\alpha$ around vertex $0$ and ends at edge $\beta$ around vertex $x$ and $T_{\gamma_{\alpha \beta}}$ is a product of $\sigma^z$ operators at these edges.
Consequently, we expect that the string correlation function decays as $\sim 1/x^{1/2}$.

These vertex operators of the CFT correspond to the rectified dimer-dimer and dimer-string correlation functions studied in experiment (see ``Calculation of correlation functions'' and ED Fig.~9).

$\ $
\\
\noindent{\it Connection to U(1) gauge theory.} \\
To establish the wavefunction of Eq.~\eqref{eq-RKwavefunction} as a microscopic realization of the $U(1)$ quantum spin liquid, we recast its properties in the language of the $U(1)$ gauge theory.
The local dimer constraint can be defined on any lattice and, for example, on the kagome lattice it leads to an emergent $\mathbb{Z}_2$ gauge theory description of significant interest in the study of topological quantum matter~\cite{moessner_quantum_2008, fradkin_field_2013,verresen_prediction_2021, semeghini_probing_2021}.
However, a crucial property of the honeycomb lattice is that it is \textit{bipartite}.
Consequently, it is possible to orient each dimer on the lattice to point from the $A$ to $B$ sublattice.

This directionality is captured by the
`electric field' degree of freedom, defined as $E_\alpha(x) = e^{i Q_0 \cdot x} n_{\alpha}(x)$, where $x$ is a vertex, $\alpha = 0, 1, 2$ labels the three edges surrounding the vertex, and $Q_0 = \frac{2\pi}{3a} \cdot  (1, \frac{1}{\sqrt{3}})$ is chosen such that the electric field is positive (goes out of) the $A$ sublattice and negative (goes into) the $B$ sublattice.
The dimer constraint can thus be expressed as a Gauss's law, $\nabla \cdot E(x) \ket{\Psi_{\mathsf{RK}}} = e^{iQ_0 \cdot x} \ket{\Psi_{\mathsf{RK}}}$ where $q(x) = \nabla \cdot E(x) \equiv \sum_{\alpha \in \text{d} x} E_{\alpha}(x)$ is the electric charge operator given by lattice 
divergence of the electric field.
The integral (sum) of $q(x)$ over an enclosed region counts the net number of dimers crossing the enclosing loop, with opposite dimer orientations contributing with opposite sign.
The  spectrum of this operator therefore lives in the integers, identifying this as a $U(1)$ gauge constraint---i.e. the sort of gauge constraint that appears in electromagnetism, where charges are quantized to the integers. 

The physics of the RK point can be more firmly connected to the deconfined phase of a $U(1)$ gauge theory through the quantum Liftshitz theory~\cite{fradkin_field_2013, ardonne_topological_2004}.
While the diagonal static correlators of the RK state are described by the compact-boson CFT of Eq.~\eqref{eq-compactboson}, the dynamical correlations in $(2+1)$D are described by the action 
\begin{align} \label{eq-RKelectromagnetism}
    S &= \int \text{d}^3  x\, \left[ \frac{1}{2} (\partial_t \phii)^2 +  \frac{1}{2} g (\nabla^2 \phii)^2 \right]\\
    &=  \int \text{d}^3 x \left[ \frac{1}{2} \mathbf{B}^2 + \frac{g}{2} (\nabla \times \mathbf{E})^2 \right].
\end{align}
where $g = 1/4\pi K$, $\mathbf{E} = \nabla \times (\phii \hat{z})$ and $\mathbf{B} = \partial_t (\phii \hat{z})$.
This action is to be contrasted with the usual action of the $U(1)$ gauge theory---equivalently, the action of electromagnetism---whose Lagrangian is $\propto (\mathbf{E}^2 - \mathbf{B}^2)$.
As it will turn out, the absence of the $\mathbf{E}^2$ term in the action is essential for the physics of the RK point.
In particular, it is known that standard electromagnetism in $(2+1)$D is generically confining (in the absence of additional fermionic fields) \cite{polyakov_gauge_1987}.
Consequently, standard electromagnetism does not have algebraic correlations, whereas the electromagnetic action of the RK point does.\\

\noindent{\it Instabilities in equilibrium.} \\
The missing leading $\mathbf{E}^2$ term in the `RK electromagnetic action' of Eq.~\eqref{eq-RKelectromagnetism}
shows that the RK point is fine-tuned.
In particular, to realize this action
each term with a lower or equal number of derivatives as the $(\nabla \times \mathbf{E})^2$ term must be fine-tuned to zero.
The only such terms that are consistent with time-reversal and spatial rotation symmetry and are not total derivatives are  $\mathbf{E}^2$ and $(\nabla \cdot \mathbf{E})^2$.
In an exact dimer model, the Gauss law $(\nabla \cdot \mathbf{E})$ is fixed in the Hilbert space and can be ignored, implying that the Eq.~\eqref{eq-RKelectromagnetism} requires one parameter of fine-tuning---tuning the coefficient on the $\mathbf{E}^2$ term to zero.
In an emergent gauge theory, such as those that appear in spin systems,  the coefficient on the $(\nabla \cdot \mathbf{E})^2$ term also needs to be fine-tuned.\\

\noindent{\textit{Quantum spin lakes.}} \\
Despite this instability in equilibrium, the RK state can be prepared robustly from non-equilibrium quantum dynamics~\cite{sahay_quantum_2023, giudici_dynamical_2022}.
The dynamics of interest keeps the system at low-energy (in contrast to, say, quench dynamics) and thus the physics is governed by near-ground-state energetics.
Here we consider the PXP approximation for simplicity and discuss the effect of vdW tails in ``Closed-loop optimization of state preparation''.

At large detuning $\Delta$, the low-energy states of our system includes all dimer coverings on the honeycomb lattice, which are split at sixth order in perturbation theory $\sim \Omega^6/\Delta^5$.
This leads to a nematic ground state wavefunction that spontaneously breaks rotation and translation symmetry.
As $\Delta$ is reduced, monomers---vertices with no adjacent dimers---will start to fluctuate in the system.
These are interpreted as electric charges in the language of an emergent gauge theory.
Eventually, monomers in the system will proliferate leading to a trivial `Higgs' phase with a low density of dimers; indeed, when $\Delta \to -\infty$, the ground state is $\ket{0}^{\otimes N}$.

Ground state behavior in hand, consider the following simplified dynamical protocol.
Starting in the ground state at $\Delta = -\Delta_0$ where $\Delta_0/\Omega \gg 1$, the wavefunction is (remarkably) well approximated by: 
\begin{equation}
    \ket{\psi(0)} \approx  \mathcal{P}\bigotimes_{i = 1}^{N} \left( \ket{0} + \varepsilon \ket{1}\right)_i
\end{equation}
where $\mathcal{P}$ denotes the blockade constraint, encoding the small quantum fluctuations induced by $\Omega$ at large negative $\Delta$.
If we adiabatically evolve the system from $-\Delta_0$ to $+\Delta_0$, i.e., at a rate that is much smaller than the $\Omega^6/\Delta_0^5$ splitting between dimer configurations, then the final state $\ket{\psi(T)}$ would well approximate the nematic ground state at $+\Delta_0$.
Conversely, for an instantaneous sweep, $\ket{\psi(T)} \approx \ket{\psi(0)}$.

Suppose we sweep at a rate that is much slower than $\Delta_0$, corresponding to adiabatic evolution relative to dimer/monomer excitations (i.e., tracking the ground-state filling), but much faster than $\Omega^6/\Delta_0^5$ such that the dynamics is diabatic with respect to this small energy splittings between dimer coverings.
A natural conjecture for the state of the wavefunction after performing this sweep is the wavefunction of the system is unchanged except that monomers are projected out.
\begin{equation}
    \ket{\psi(T)} \propto \mathcal{P}_{\text{Gauss}} \ket{\psi(0)} = \ket{\Psi_{\mathsf{RK}}}
\end{equation}
where $\mathcal{P}_{\text{Gauss}}$ projects the wavefunction into the space of dimer configurations.
This was shown to faithfully reproduce the result of the dynamics in this intermediate-rate `hemidiabatic' regime in Ref.~\cite{sahay_quantum_2023}.
Crucially, because the rate of the sweep is determined by rough energy scales, it does not require the same fine-tuning as required for the equilibrium realization of the RK point's correlations.

Since as $\Delta$ is swept from $-\Delta_0$ to $+\Delta_0$ the gap to the monomer/dimer excitations closes, it is not possible to truly be adiabatic with respect to these excitations.
As a result, the Gauss law is not perfectly projected out and instead we are left with a dilute density of monomer excitations atop $\ket{\Psi_{\mathsf{RK}}}$, per an analog of the Kibble-Zurek mechanism~\cite{zurek_dynamics_2005}.
The density of these dilute excitations sets an effective lake size $L_{\text{lake}}$ over which one expects the system to have correlations similar to the $\text{RK}$ wavefunction.
Crucially, $L_{\text{lake}}$ is an intensive quantity in the experiment and is not expected to depend on system size.

$\ $

\noindent{\textit{Finite size effects.}} \\
Considering the RK wavefunction on an infinite cylinder of width $L_y$, the diagonal correlations are expected to follow those of the $(2+0)$D compact-boson CFT, Eq.~\eqref{eq-compactboson}, which reduces to the $(1 + 1)$D compact-boson CFT placed on a ring of size $L_y$~\cite{henley_coulomb_2010, fradkin_field_2013}.
The finite size effects of these correlations are well known to follow
\begin{equation}\label{eq:finite_size_CFT}
    \langle e^{i m [\phii(x) - \phii(0)]} \rangle \sim \exp\left(- \frac{2\pi \Delta_m x}{L_y} \right)
\end{equation}
For comparing to the experiment, we expect that $L_y$ is replaced with $L_{\text{lake}}$, which sets the effective size of the RK wavefunction.
This leads to two interesting predictions:
\begin{enumerate}
    \item[I.] The observed monomer density in the experiment sets the scale over which the system matches the RK wavefunction.
    Consequently, postselecting on no monomers should improve consistency with the RK wavefunction (Fig.~5a).

    \item[II.] The length scale of the decay of the experiment---predicted to be set by $L_{\text{lake}}$---should be largely independent of the height of the lattice used in the experiment (ED Fig.~10e).
    
\end{enumerate}
We find that both of these predictions are borne out in the data.
Furthermore, from Eq.~\eqref{eq:finite_size_CFT} we expect that the ratio of the logarithms of the rectified dimer-dimer ($m=1$) and dimer-string ($m=1/2$) correlators is equal to $\Delta_{m=1} / \Delta_{m=1/2} = 4$ for any system size. We find good agreement with this prediction in both classical Monte Carlo simulations and experiment (Fig.~6e and ED Fig.~10).\\

\small \noindent\textbf{Closed-loop optimization of state preparation} \\
In Figure~4, we optimize detuning sweeps to target the $U(1)$ quantum spin liquid using a Bayesian optimizer, implemented with Ax~\cite{olson_ax_2025}. 
Our initial ansatz consisted of a sum of Gaussians for both $\Omega(t)$ and $\Delta(t)$ (not shown), inspired by pulsed counterdiabatic driving schemes~\cite{gjonbalaj_shortcuts_2025}. However, the large positive detunings required in this approach led to increased atom loss and blockade violations, and the optimization converged to smooth sweeps, thus motivating the parameterization shown in ED Fig.~5a,b. There are eight free parameters, $\{ \Omega,T, \Delta_{\mathrm{min}},\Delta_{\mathrm{max}},\Delta_{\mathrm{inf}},t_{\mathrm{inf}}, s, t_{\mathrm{slope}} \}$ and we impose loose bounds such that $\Delta(t)$ is physically motivated. For example, we enforce $\Delta_{\mathrm{min}}<0$ and $\Delta_{\mathrm{max}}>0$, but do not enforce monotonicity (i.e., $s$ can be negative).
We employ a multi-objective optimization over four metrics ($\langle T\rangle$,$\langle V\rangle$, mean Rydberg density, and atom loss) rather than a scalar cost function (ED Fig.~5c). In particular, we do not minimize the RK energy directly, since the different measurement error for $\langle T\rangle$ and $\langle V\rangle$ would bias the results.
We find that the observables are largely correlated across trials (ED Fig.~5d), supporting this approach.
For each optimization we run 50-80 trials, at the end of which we repeat the best 2-4 trials to select the final detuning sweep (Fig.~4d).

Using exact diagonalization to simulate the experimental drive on a $3\times 3$ unit cell kagome lattice, we find that the optimized detuning sweeps naturally target the hemidiabatic regime \cite{sahay_quantum_2023,gjonbalaj_shortcuts_2025} necessary to prepare quantum spin lakes.
In particular, the final leg of the 0.8\,$\mu$s drive sweeps through the ground state phase transition at a rate which is adiabatic relative to monomer excitations 
but sudden relative to the low-energy splitting between different dimer coverings induced by perturbative resonances and vdW tails (ED Fig.~6a,b).
Indeed, as this final sweep rate is increased, $\langle T \rangle$ and $\langle V \rangle$ decrease due to monomer proliferation; as the rate is decreased, $\langle T \rangle$ and $\langle V \rangle$ decrease due to relative phases accumulated between different dimer coverings, primarily due to the vdW interaction (ED Fig.~6c).\\

\small \noindent\textbf{Boundary engineering with local detunings} \\
To prevent seeding of Rydberg excitations from the array edges, we apply local light shifts to the boundary atoms using a programmable array of optical tweezers (1.5\,$\mu$m waist) generated with a SLM~\cite{manovitz_quantum_2025}. The light is sourced from a titanium:sapphire laser (M Squared) operated 1\,THz red-detuned of the D2 line. 
For the Floquet engineering experiments, we do not use any boundary detunings.
To generate the appropriate boundary detunings for the quantum spin liquid, we start with an empirical weighting of roughly three times larger intensity on sites near the lattice corners and use the weighted Gerchberg-Saxton (WGS) algorithm to compute the phase hologram displayed on the SLM~\cite{kim_large-scale_2019}.
Next, we measure the prepared Rydberg density distribution using the optimized detuning sweep and update the weights of each tweezer to target a uniform population over the boundary, repeating this roughly five times to converge to the final weights. When updating the weights, we re-run the WGS algorithm with the previous phase profile as a starting guess and, in the simulated fourier plane, update only the amplitudes of each wavevector component and not the phases, improving convergence~\cite{chew_ultraprecise_2024}.
Finally, we optimize the overall magnitude of the light shift by minimizing the nematic order parameter $\Phi$ (Fig.~4h). Vertices that are connected to boundary atoms are excluded from $\Phi$,
since perfect dimer coverings on a lattice with edges and corners are expected to have higher dimer populations on the boundary~\cite{fisher_statistical_1963}.\\

\small \noindent\textbf{Probability distribution of dimer configurations} \\
The number of allowed dimer configurations on the lattice grows exponentially with system size~\cite{moessner_quantum_2008} such that, to study the population distribution in practice, we restrict to subsystems with a sufficiently small number of possible states. \\

\noindent{\it Isolating small subsystems.}\\
We postselect for snapshots (per subsystem) where the atoms on the outer bonds connecting the subsystem to the surrounding lattice are in the desired state.
From these instances, we compute the population distribution over subsystem configurations that satisfy this imposed boundary condition, aggregating over all possible subsystems related by lattice translations.
This boundary-fixing procedure requires an exponential number of samples but, crucially, the scaling is governed by the subsystem size rather than the lattice size. 
In experiment, the overhead is increased by postselection to exclude monomers, double-dimers, and atom loss, as required for comparison to the RK state and to reduce the number of configurations.
In Fig.~5a, for no dimers around the subsystem, 
the percentage of kept data per subsystem is 7.2\%, 1.8\%, 0.4\%, 0.3\% and 0.04\% for the 1$\times$1, 2$\times$1, 3$\times$1, 2$\times$2 and 3$\times$2 subsystem sizes, respectively. These correspond to 6, 11, 16, 18, and 27 sites inside the subsystem, plus 6, 8, 10, 10, and 12 sites on the boundary. The increased data acquisition rate, enabled by qubit reuse, makes these overheads sufficiently practical.\\

\noindent\textit{Thermal distribution.} \\
The probability distribution of the state prepared using the modified sweep (ED Fig.~7a) is well-described by a thermal distribution $p_i \propto e^{-\beta E_i}$ where $p_i$ is the observed probability of configuration $i$, $\beta$ is the inverse temperature, and $E_i$ is the energy of the configuration. Probabilities are defined within a given subsystem size, $S$, so that $\sum_{i\in S} p_i = 1$, and $\beta$ is the same across all subsystems. 
The energy of each configuration is given by the sum of the vdW tail energies, excluding small corrections from Rydberg atoms outside the subsystem and perturbative ring exchange $\propto \Omega^6/\Delta^5$.
We fit the observed probabilities to these thermal probabilities with $\beta$ as the only free parameter, obtaining the fitted thermal distribution shown in Fig.~5a (inset) and ED Fig.~7b. \\

\small \noindent\textbf{Calculation of correlation functions} \\
\noindent \textit{Dimer-string correlator.} \\
The expectation value of Z loops in the RK state is $\pm1$, with the sign given by the emergent Gauss law~\cite{sahay_quantum_2023}.
For the dimer-string correlator, $C_s$, we restrict to loops that are (ideally) +1 in order to remove sign ambiguities, otherwise the two open strings whose product forms the closed loop have the opposite sign. 
We use symmetric loop geometries in which the two open strings have the same length and take the (signed) geometric mean of their expectation values for the numerator of $C_s$. 
Given this, we exclude strings that terminate on boundary plaquettes where a symmetric closed loop cannot always be formed.
\\

\noindent\textit{Rectified correlators.}\\
The rectified dimer-dimer and dimer-string correlators for the honeycomb lattice are shown schematically in ED Fig.~9a,d.
We follow the convention of Ref.~\cite{zhou_emergent_2021} where the three plaquette sublattices (a,b,c) determine the phases in the correlation functions.
Concretely, the rectified dimer-dimer correlator, corresponding in field theory to Eq.~\eqref{eq:CFT_2pt}, is
$\widetilde C_{d} = \langle \phi^*(\mathbf{r}_i)\phi(\mathbf{r}_j) \rangle$ where $\phi(\mathbf{r}_i) = \sigma^{z}_{i,\mathrm{bc}} + \omega \sigma^{z}_{i,\mathrm{ac}} +  \omega^2 \sigma^{z}_{i,\mathrm{ab}}$ is the sum over the three bonds touching vertex $i$, $\omega = e^{2\pi i/3}$, and the bonds are labeled by the sublattices of the adjacent plaquette pair.
The rectified dimer-string correlator, corresponding in field theory to Eq.~\eqref{eq:CFT_string}, is
$\widetilde C_{s} = \langle \psi^*(\mathbf{r}_i)\psi(\mathbf{r}_j) \rangle$ where $\psi(\mathbf{r}_i) = S^a_i + \omega S^b_i + \omega^2S_i^c$ is the sum over the three adjacent plaquettes and $\langle S_iS_j \rangle \equiv C_s(|\mathbf{r}_i - \mathbf{r}_j|)$ is the unrectified string correlator studied in Fig.~6d.
In a translationally invariant system $\widetilde C_{d,s}$ are real quantities; here we find the imaginary part is negligible for the $\widetilde C_{d}$ but finite for $\widetilde C_{s}$.
We compute the correlators between all pairs of vertices, again excluding the boundary layer. These are then averaged over endpoints with the same radial separation, resulting in the rectified correlators shown in Fig.~6e and ED Figs.~9,10.

For all correlation functions, error bars are calculated using bootstrap resampling with replacement over 20 resamples of the full dataset; increasing the number of resamples does not appreciably change the estimated uncertainties.\\

\small \noindent\textbf{Measuring many-body coherences.}\\
To understand the effect of off-diagonal measurement error, we use MPS simulations of the preparation sweep, finding $\langle T \rangle / \langle V \rangle = 0.92$ for the final state.
We then simulate the mapping down by evolving this state under the vdW tails interaction for 100\,ns, which reduces the ratio to $\langle T \rangle / \langle V \rangle= 0.76$ (ED Fig.~8b).
This is compared to $\langle T \rangle / \langle V \rangle = 0.47(1)$ in experiment, where an additional reduction is expected as the MPS numerics cannot account for the infidelity from the Rydberg $\pi$-pulse when mapping down in a three-level system (see also ED Fig.~1d). 
The decay of $\langle T \rangle$ as the mapping down gap time increases is several times faster than expected for single-qubit dephasing, consistent with the phase from vdW tail being the dominant error (ED Fig.~8d).

In addition, we use exact diagonalization on a smaller $3 \times 3$ unit cell lattice to simulate the effect of ground-Rydberg dephasing on the dynamics, including on-site Doppler dephasing and global detuning fluctuations, as well as radial and axial spread of the wavefunction corresponding to $\approx $10\,$\mu$K atom temperature.
These are modelled as local and global detuning shifts (randomly sampled from Gaussians of width $2\pi \times 0.043$\,MHz and $2\pi \times 0.2$\,MHz, respectively), and in-plane and out-of-plane positional disorder (randomly sampled from Gaussians of width $0.02a$ and $0.08a$, respectively). 
In ED Fig.~6d,e, we find that the RK energies are largely insensitive to these error sources and attribute this insensitivity to two factors.
First, hemidiabatic preparation is largely insensitive to the exact start and end of the parameter sweep,
since, by definition, it is too rapid to resolve such low-energy details of the spectrum~\cite{sahay_quantum_2023}.
As such, small global detuning shifts have little effect on the final fidelity of the prepared state.
Second, the kinetic energy operators live in a decoherence-free subspace~\cite{kais_review_2014} within the blockaded manifold, such that a global $Z$ rotation (i.e., a detuning shift) does not change the measured kinetic energy, as we observe in experiment.\\

\small \noindent\textbf{Numerical methods}\\
Here we provide details of the two numerical methods used to understand our results in Fig.~6 and ED Figs.~8--10.\\

\noindent{\it Tensor Network Simulations.} \\
Matrix product state (MPS) simulations of unitary dynamics are performed using the ``W-II'' matrix product operator method of Ref.~\cite{zaletel_time-evolving_2015} which is implemented in the Tensor Network Python (TeNPy) package~\cite{hauschild_efficient_2018}. 
Specifically, we simulate the quantum dynamics under the Rydberg Hamiltonian, Eq.~\eqref{eq:Hamiltonian}, for the standard 0.8\,$\mu$s preparation protocol used in the main text.
While the true vdW potential of the experiment is long-range, this is challenging to exactly simulate in MPS-based simulations as the bond dimension of the Hamiltonian's MPO gets prohibitively large.
Consequently, we study two approximations:
\begin{enumerate}
    \item \textit{PXP.} This enforces the nearest-neighbor blockade constraint but excludes long-range vdW tails, Eq.~\eqref{eq:PXP_ham}. 
    In practice, the blockade is enforced exactly for the three atoms in the upper triangles of the kagome lattice, thereby replacing these three atoms with an effective $4$-state qudit, and is enforced energetically between these triangles with a nearest neighbor interaction $V_{ij} = 100\, \Omega$.
    
    \item \textit{PXP plus tails.} In addition to enforcing nearest-neighbor blockade as above, we include long-range vdW tails up to a truncation distance of $R_{\text{trunc}} = \sqrt{7}a$ (i.e. up to the next-next-next-nearest neighbor contributions). 
    The truncation distance is chosen such that observables change negligibly upon slightly decreasing this distance. 
\end{enumerate}
In both cases, our numerics are performed on strip geometries that are infinite in the $x$-direction and finite in the $y$-direction. Following the experiment, we use a boundary detuning of $\delta_{\text{bdy}} = 0.52\, \Omega$ for sites along the top and bottom edges of the strip. We use a Trotter time step of $dt = 0.01\,  \Omega^{-1}$, which we find is sufficient to achieve convergence.
To simulate the kinetic and potential energies (ED Fig.~8), we use the `PXP plus tails' approximation with a three-plaquette-tall lattice and a bond dimension of $\chi = 256$.
To simulate the $Z$-basis correlation functions (ED Fig.~9), we instead use the `PXP' approximation to allow us to use a four-plaquette-tall lattice to compute these long-range observables. The bond dimension is $\chi = 350$. 
For such $Z$-basis observables, we find that neglecting vdW tails is a good approximation and the correlations closely agree with those on the smaller lattice that include vdW tails.\\

\noindent{\it Classical Monte Carlo.}\\
Following the mapping to a classical dimer gas, Eq.~\eqref{eq:classical_dimer}, observables for the RK state can be evaluated as classical averages over $\mathcal{N}$ dimer coverings $\mathcal{D}$ sampled uniformly by Monte Carlo.
In addition to $Z$-basis observables, expectation values of off-diagonal observables $\hat{\mathcal{X}}$ can be obtained from averaging over transition matrix elements, with contributions only from finitely many $\mathcal{D}'$ connected to $\mathcal{D}$ via allowed moves along the support of $\mathcal{X}$
\begin{align}
    \langle \hat{\mathcal{X}} \rangle = \frac{1}{\sum_{\mathcal{D}}1}\sum_{\mathcal{D},\mathcal{D}'}\langle \mathcal{D}|\mathcal{X}|\mathcal{D}'\rangle.
\end{align}
For instance, the kinetic energy of a single plaquette is equivalent to its potential energy. 
We evaluate both diagonal and off-diagonal observables via classical Monte-Carlo sampling of $\mathcal{D}$ with uniform weight using a worm-update Markov Chain, which performs nonlocal loop updates and yields short autocorrelation times~\cite{alet_classical_2006,sandvik_correlations_2006}. Each worm update temporarily introduces a pair of monomers, propagates one of the monomers through the lattice via local dimer pivots, and terminates when the monomers reunite to form a dimer. As a result, each worm update flips the dimer configuration along a closed loop. 

The worm update starts from a valid close-packed dimer covering $\mathcal{D}$ on the honeycomb lattice, where each vertex $v$ is incident to exactly one dimer oriented along one of three lattice directions $l \in \{1,2,3\}$. The update then proceeds as follows.
\begin{itemize}
    \item \emph{Initialize.} Choose a vertex $u_0$ uniformly at random to be the \emph{tail} of the worm. Let $(u_0,v_1)$ be the (unique) dimer incident on $u_0$, connecting it to the initial \emph{head} of the worm at neighboring vertex $v_1$. Remove $(u_0,v_1)$, creating two monomers at $u_0$ and $v_1$.
    \item \emph{Propagate head.} At step $i=1,2,\dots$, the \emph{head} is at $v_i$, and $(u_{i-1},v_i)$ denotes the recently removed dimer entering $v_i$ with orientation $l_i$.
    \begin{enumerate}
        \item Reinsert a dimer emanating from $v_i$ in a new orientation $k_i$, chosen with transition probability 
        \begin{align}
            P_{l_i\to k_i}= \begin{cases} 1/2 & l_i \neq k_i \\ 0 & l_i = k_i \end{cases} \text{\;\;so that } \sum_{k_i} P_{l_i \to k_i}=1,
        \end{align}
        which satisfies local detailed balance at infinite temperature and forbids immediate backtracking. The inserted dimer $(v_i, u_i)$ annihilates the monomer at $v_i$ and connects it to a neighboring vertex $u_i$. 
        \item If $u_i = u_0$, terminate (see below). Otherwise, $u_i$ is also incident to a (unique) pre-existing dimer $(u_i,v_{i+1})$. Remove $(u_i,v_{i+1})$ to restore the dimer constraint at $u_i$ and advance the \emph{head} to the connecting vertex $v_{i+1}$. 
    \end{enumerate}
    \item \emph{Terminate.} The worm closes the loop at the first step $i=\ell>1$ for which $u_\ell=u_0$. The final dimer insertion $(v_\ell,u_0)$ connects the head $v_\ell$ back to the tail $u_0$ and annihilates the monomer pair. This restores close packing and yields a new dimer covering $\mathcal{D}'$, which differs from $\mathcal{D}$ by dimers flipped along a closed loop of length $2\ell$.
\end{itemize}
\noindent Monte Carlo samples are obtained by iterating this worm update to generate a Markov chain $\mathcal{D} \to \mathcal{D}' \to \cdots$. 

In this work, we sample Monte Carlo trajectories initialized from one of the three columnar dimer coverings, which maximize the number of flippable plaquettes~\cite{moessner_phase_2001}. We consider periodic honeycomb lattices of size $L_x \times L_y$ in units of plaquette separation ($1/2a$), containing $3\cdot L_x\cdot L_y$ dimers. For each lattice, we average not only along a single trajectory but also across independent runs with identical initialization and different random seeds. For the $36 \times 6$ lattice, we generate 10 trajectories of 19,600,001 to 30,100,001 samples each, for a total of 229,300,010 samples. For the $36 \times 9$ lattice, we generate 10 trajectories of 11,600,001 to 19,700,001 samples each, for a total of 142,600,010 samples. For the $36 \times 12$ lattice, we generate 10 trajectories of 8,500,001 to 9,400,001 samples each, for a total of 90,300,010 samples. For the largest lattice ($36 \times 36$), we average over 10 trajectories of 2,800,001 to 3,400,001 samples each, for a total of 32,600,010 samples.
These data are used to compute RK state observables as a function of system size.

For the RK state on a torus, there are four topological sectors $(\pm1 , \pm 1)$, specified by the $Z$ string eigenvalues of the two non-contractible loops wrapping around the horizontal and vertical directions of the lattice.
For both the dimer-dimer and dimer-string correlators, we find that conditioning on a single topological sector enhances the estimates correlations relative to sector-averaged estimates, and reduces finite-size effects. We show this in ED Fig.~10d where the data shown in the $(+1,+1)$ topological sector were obtained by retaining 1,550,589 to 1,753,196 samples per trajectory across 10 trajectories, for a total of 16,805,690 samples.
Within a fixed sector, the correlators are symmetric about $L_x/2$ due to periodic boundary conditions. When averaging over sectors, the dimer-string correlator continues to decay beyond half-length as the non-contractible loops average to zero. For comparison to experiment, we average over all four sectors, as the sector is not well-defined for open boundary conditions.\\

\noindent\textbf{Data Availability}\\
The data that supports the findings of this study are available from the corresponding author on request.\\

\noindent\textbf{Acknowledgments}\\
We thank 
M. Aidelsburger, J.P. Bonilla Ataides, S. Choi, E. Demler, S. Ebadi, A. Gu, D. Kufel, V. Menon, S. Sachdev, G. Semeghini, M. Serbyn,  A. Vishwanath, and D. Vu for helpful discussions.
We acknowledge financial support from the US Department of Energy (DOE Quantum Systems Accelerator Center, contract number 7568717), the 
IARPA and the Army Research Office, under the Entangled Logical Qubits program (Cooperative Agreement Number W911NF-23-2-0219),
DARPA ONISQ program (grant number W911NF2010021) and MeasQuIT program (grant number HR0011-24-9-0359), the Center for Ultracold Atoms (an NSF Physics Frontier Center, grant number PHY-2317134), the National Science Foundation (grant numbers PHY-2012023, CCF-2313084, QuSEC grant OMA-2326787, and QLCI grant OMA-2120757), the Army Research Office MURI (grant number W911NF-20-1-0082),  Wellcome Leap Quantum for Bio program, and QuEra Computing.
S.J.E. acknowledges support from the National Defense Science and Engineering Graduate (NDSEG) fellowship.
M.X. and J.F. acknowledge support from the Harvard Quantum Initiative Postdoctoral Fellowship in Science and Engineering.
\\

\noindent\textbf{Author contributions} A.A.G., S.J.E., S.H.L., M.X., D.B., M.K., and T.M. contributed to the building of the experimental setup, performed the measurements, and analyzed the data. 
N.U.K., N.M., and  J.F. contributed to the development of the Floquet engineering protocol. N.U.K., R.S., and N.O.G. performed numerical simulations for the quantum spin liquid experiments and N.U.K., R.S., N.O.G., R.V., and J.F. contributed to theoretical analyses. 
All work was supervised by R.V., S.F.Y., J.F., M.G., V.V., and M.D.L. All authors contributed to the project vision, discussed the results, and contributed to the manuscript.
\\

\noindent\textbf{Competing interests} M.G., V.V., and M.D.L. are co-founders, M.G. V.V., and M.D.L are shareholders, S.F.Y.’s spouse is a co-founder and shareholder, V.V. is Chief Technology Officer, M.G. is a consultant, and M.D.L. is Chief Scientist of QuEra Computing.\\ 

\noindent\textbf{Correspondence and requests for materials} should be addressed to M.D.L.\\

\setcounter{figure}{0}
\newcounter{EDfig}
\renewcommand{\figurename}{Extended Data Fig.}

\begin{figure*}
\includegraphics[width=2\columnwidth]{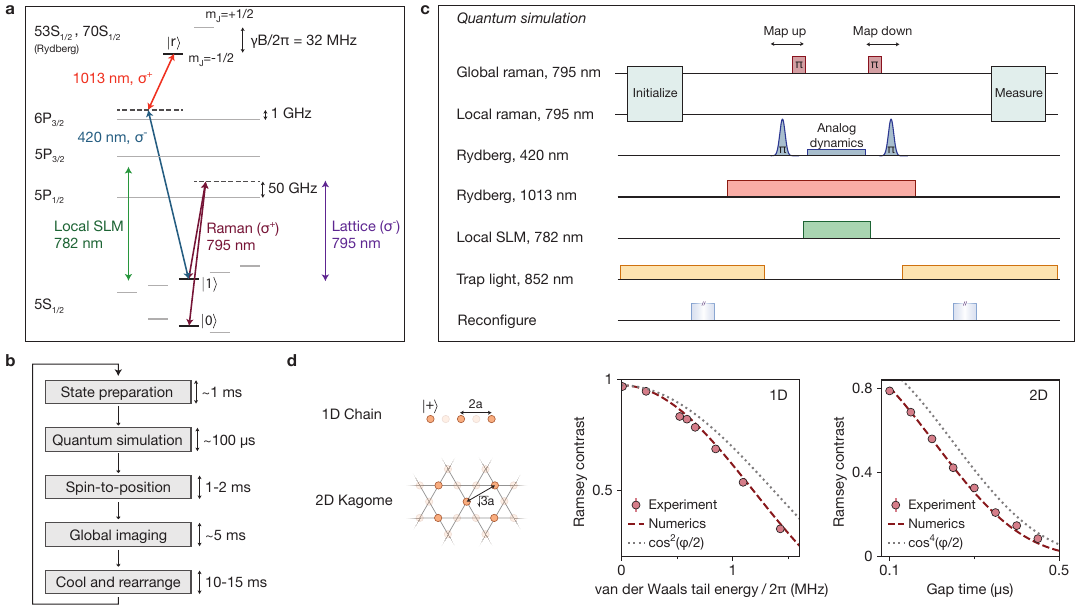}
\caption{\textbf{Analog-digital quantum simulations.} \textbf{a,} Level diagram showing the relevant atomic transitions in $^{87}$Rb. Raman transitions are used for single-qubit gates, the spin-dependent lattice is used for spin-to-position conversion, and the local SLM is used for local light shifts. The clock qubit $\ket{1} = \ket{F = 2, m_F=0}$ is excited to the lower-energy $m_J=-1/2$ Rydberg state via a two-photon transition. All Rydberg pair states lie higher in energy than the blockaded subspace.
We suppress coupling to the nearby $m_J = +1/2$ Rydberg state, which has 3$\times$ smaller matrix element, by increasing our DC magnetic field to 11.3\,G. 
The intermediate state detuning of +1 GHz and $n = 70$ Rydberg state is used for all 2D experiments. For the 1D chain in Figs.~2,3 we instead use $-6.3$ GHz and $n = 53$.
\textbf{b,} Schematic experiment timeline. The total loop takes 16.4\,ms for 1D experiments and 24\,ms for 2D experiments; in 2D,
spin-to-position conversion is performed in two groups due to limited laser power for the AOD tweezers, and image processing and rearrangement are slower for the larger lattice. 
\textbf{c,} Example pulse sequence used for analog-digital quantum simulation. Not all steps are used in all experiments. For the QSL experiments, the first Rydberg $\pi$-pulse is omitted when starting from the all-$\ket{0}$ initial state before mapping up. Gaussian Rydberg $\pi$-pulses are used to minimize coupling to the other $m_J$ state. The Rydberg Rabi frequency is reduced for the analog dynamics via the 420-nm intensity.
For 1D experiments at $n=53$ and performed in AOD tweezers, the atom spacing is reconfigured from $a = 6.0\,\mu$m for local Raman rotations to $3.6\,\mu$m for Rydberg dynamics. In 2D, $a = 6.0\,\mu$m is used throughout.
\textbf{d,} Characterization of coherent mapping. As in Fig.~1d, single qubits are arranged at next-nearest neighbor sites in 1D and 2D to emulate the vdW tails in a many-body state. Microstate-dependent phase accumulation from vdW tails occurs during the gap for the Raman $\pi$-pulse after the analog dynamics (here, a single-qubit $\pi/2$-pulse). We observe the decay in Ramsey contrast for increasing vdW tail energy, $E_{\mathrm{vdW}}$, by varying the lattice spacing in 1D (180\,ns gap time) and also for increasing gap time in 2D ($2\pi \times$670\,kHz tail energy at $\sqrt3 a$). Analytical formula includes only phase from vdW tails during the gap time $t$, given as $\cos^N(\varphi/2)$ for $N$ neighboring qubits and $\varphi = E_{\mathrm{vdW}}t$. Numerics simulate the entire measurement where the effect of vdW tails during the Rydberg pulses leads to a further reduction in contrast. 
Both curves are rescaled by the measured contrast of the hyperfine qubit without mapping and no other sources of error are included in the numerics.
}
\refstepcounter{EDfig}\label{fig:ED_systemdiagram}
\end{figure*}

\begin{figure*}
\includegraphics[width=2\columnwidth]{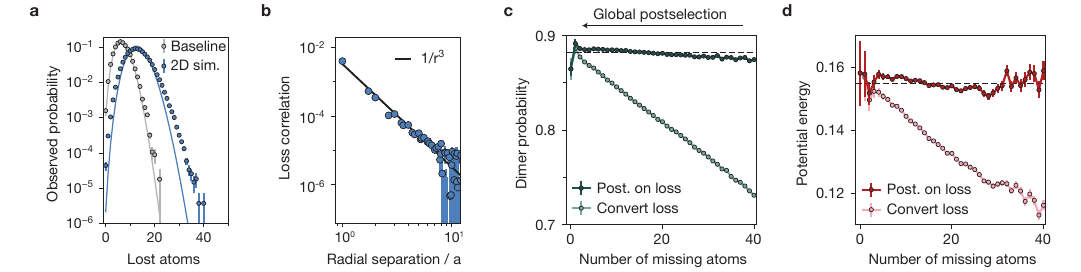}
\caption{\textbf{Loss detection in analog quantum simulations.} 
\textbf{a,} Distribution of losses per shot in QSL experiments. Baseline losses exclude the many-body dynamics and follow a Poisson distribution (solid line). The loss increases when performing analog dynamics and a non-Poissonian tail from correlated losses appears. There are 271 atoms in total. 
\textbf{b,} Spatial correlations of loss events. The two-point connected correlator follows a dipolar $1/r^3$ decay, suggesting the presence of Rydberg $S$-$P$ interactions resulting from an atom, originally in the $S$ state, undergoing blackbody decay to a nearby $P$ state. 
These long-range interactions facilitate so-called avalanche events~\cite{festa_blackbody-radiation-induced_2022, goldschmidt_anomalous_2016, boulier_spontaneous_2017} in which nearby atoms come into resonance with unwanted transitions and are also lost, explaining the tail in \textbf{a}. A background offset of $\approx 2.5\times10^{-5}$ is subtracted. Data in \textbf{a,b} are postselected on the atom being present in the prior image to remove correlations from mid-circuit rearrangement which are created when the reservoir starts to deplete. 
\textbf{c,d} Comparison of local and global loss postselection methods. 
Data are binned according to the number of missing atoms in each snapshot, including both losses and imperfect initial lattice filling.
Postselection on no lost qubits within the local operator approximates the expectation value obtained from shots with a vanishing number of missing atoms across the entire 271-site lattice (dark curves). Randomly converting loss to qubit state $\ket{0}$ or $\ket{1}$ in post-processing strongly affects the calculated expectation value (light curves). We illustrate this for the \textbf{c,} three-body dimer populations and \textbf{d,} six-body potential energy from the QSL experiments.
Dashed lines show the means of the entire dataset with local postselection.
There are 285,948 shots in total and 6 shots have zero missing atoms.
\\
\\
}
\refstepcounter{EDfig}
\end{figure*}

\begin{figure*}
\includegraphics[width=2\columnwidth]{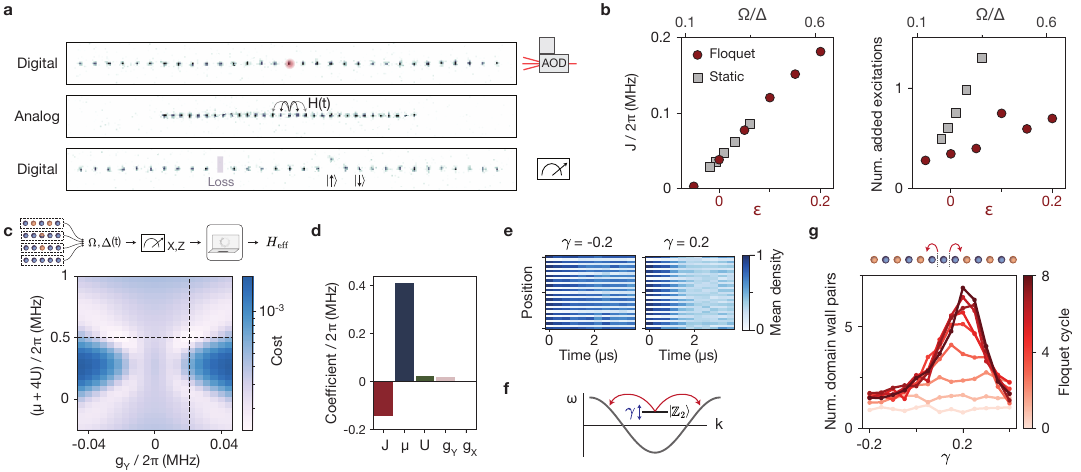}
\caption{\textbf{Engineering hopping dynamics via Floquet driving.}
\textbf{a,} Interleaved analog and digital operation. 
Moveable tweezers are used to reconfigure the chain between non-interacting and interacting geometries. Local Raman is used for state preparation and non-destructive readout obtains spin and loss information.
\textbf{b,} Comparison of Floquet and perturbative dynamics at large static detuning. Starting from a single excitation in a 31-site chain, a quantum walk is realized via Floquet driving (red) or perturbative second-order hopping (gray), tuned by $\varepsilon$ or $\Omega/\Delta$, respectively. For the same hopping energy (left), Floquet driving exhibits better number conservation (right). Floquet drive parameters are $\gamma + 2\varepsilon = -0.15$ and $\theta = 0$, optimized for number conservation. The effective dynamics at $\varepsilon=0.1$ has anomalously bad number conservation, possibly from an improperly calibrated resonant detuning for this data.
\textbf{c,} Verification of the effective Hamiltonian. The measured dynamics, starting from several initial states and measuring in different bases, are fit to numerical simulation of the ansatz blockaded XX Hamiltonian.
Plot shows a two-dimensional slice of the cost function, where the dashed lines mark the learned coefficients. 
\textbf{d,} Extracted coefficients of the effective Hamiltonian. The spin-flip terms $g_{X,Y}$ are small, consistent with approximate number conservation.
\textbf{e,} Domain wall excitations. Increasing $g_{X,Y}$ allows for the creation of domain wall pairs when a spin is flipped on top of the N\'eel $\ket{\mathbb{Z}_2}$ state, and subsequent hopping dynamics of these seeded domain walls washes out the initial order.
We use an odd-length chain with $\ket{r}$ initialized at the ends such that domain walls do not propagate from the boundaries.
\textbf{f,} This process occurs when the chemical potential, tuned via $\gamma$, brings the $\ket{\mathbb{Z}_2}$ state into resonance with the band of domain wall excitations. To a good approximation, $\gamma$ does not couple to the other terms in the effective Hamiltonian~\cite{koyluoglu_floquet_2024}.
\textbf{g,} A resonance appears when scanning the chemical potential across the band. This approximately coincides with the presence of a Luttinger liquid phase in the ground state phase diagram~\cite{koyluoglu_floquet_2024, alcaraz_exactly_1999, verresen_stable_2019}.
}
\refstepcounter{EDfig}
\end{figure*}

\begin{figure*}
\includegraphics[width=2\columnwidth]{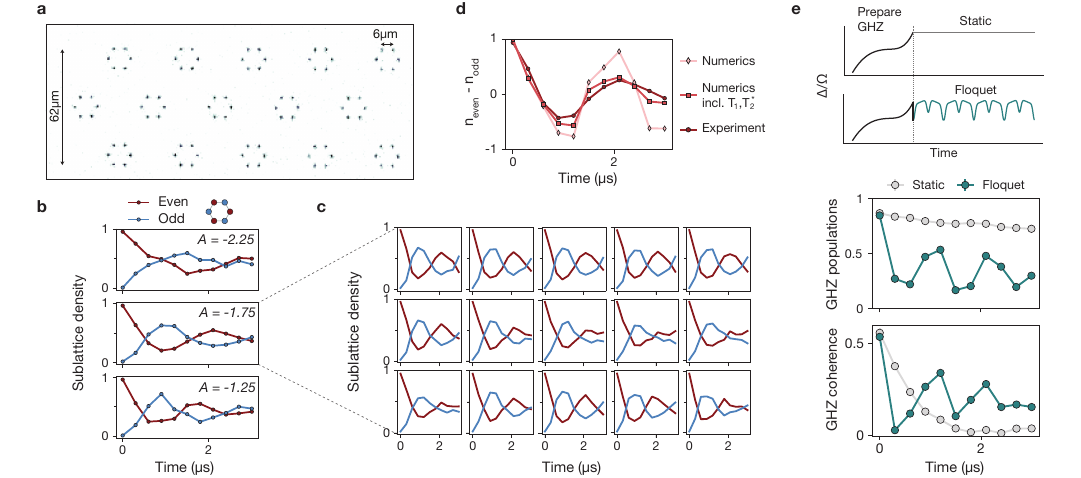}
\caption{\textbf{Engineering ring exchange via Floquet driving.} 
\textbf{a,} Atom image of the array of 15 isolated hexagonal plaquettes. 
\textbf{b,} Continuously tuneable six-body interactions. The speed of the ring-exchange dynamics is varied via the parameter $A$ of the Floquet drive, shown here via the populations on the two sublattices. $A=-1.75$ uses the same data as in Fig.~2h. 
\textbf{c,} Homogeneity of the engineered Hamiltonian. Sublattice dynamics are shown for individual plaquettes, where the several-percent-scale inhomogeneity of the 420-nm and 1013-nm light shifts result in the spatial variations of the static detuning offset. 
\textbf{d,} Numerical simulations of the engineered Floquet dynamics. Good agreement is found when including the measured $T_2^*\approx 1\,\mu s$ in numerics and further multiplying by an empirical exponential decay, exp[-$\langle n \rangle t/ T_1$], where $\langle n \rangle \approx 2.5$ is the mean Rydberg density and the Rydberg lifetime is $T_1 \approx 30 \,\mu s$, dominated by 1013-nm scattering. The $T_2^*$ is dominated by inhomogeneity between plaquettes and shot-to-shot fluctuations of the Rydberg beam light shifts and therefore is modelled as a Gaussian-distributed detuning shift with zero mean and $2\pi \times $200\,kHz standard deviation.
\textbf{e,} Stabilization of entangled states. The coherence time of a quasi-adiabatically prepared GHZ state, $ \approx \left(\ket{010101} + \ket{101010}\right)/\sqrt{2}$, can be extended by Floquet driving, as compared to holding the state at large static detuning, $\Delta/\Omega = 5.9$. We attribute this to the combined effect of an energy gap generated by the ring-exchange interaction and the reduced incoherent error owing to the lower average Rydberg population between stroboscopic times.
The engineered Hamiltonian contains non-number-conserving three-body interactions which admix the state $\ket{100100}$ and its cyclic permutations~\cite{feldmeier_quantum_2024}, resulting in oscillations in both the populations and coherences. Decay of populations at static detuning is consistent with Rydberg $T_1$.
}
\refstepcounter{EDfig}
\end{figure*}

\begin{figure*}
\includegraphics[width=2\columnwidth]{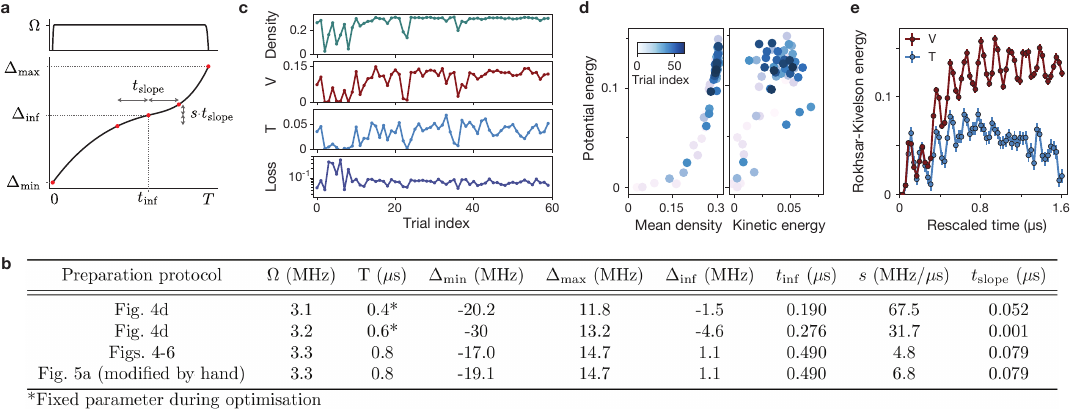}
\caption{\textbf{Experimental optimization of QSL state preparation.}
\textbf{a,} Parameterization of $\Omega(t)$ and $\Delta(t)$ profiles. The detuning sweep is a cubic interpolation between five points (red markers) defined by seven parameters. The Rabi frequency is constant with a 20\,ns turn-off time.
\textbf{b,} Summary table of parameters used in the main text, found using a Bayesian optimizer. Detuning sweep is later clipped between $\pm$20\,MHz. Final row is a by-hand modification of the optimized profile in row 3.
\textbf{c,} Results of one round of closed-loop optimization. Metrics used in the multi-objective optimization are shown as a function of trial number. Each trial takes roughly 7 minutes to obtain 2000 $Z$-basis and 4000 $X$-basis shots. At the end of the round, we repeat the best 2--4 trials to select the final sweep. For some rounds (not shown), we also included minimization of the magnitude of open length-5 X strings but found this had little effect and was within error bars for all reasonable sweeps.
\textbf{d,} The measured RK potential energy is positively correlated with both the mean density and RK kinetic energy across different preparation sweeps.
\textbf{e,} Sensitivity to sweep parameters. By rescaling the time of the 0.8\,$\mu$s sweep (row 3 in \textbf{b}), we find the RK energies can change significantly in this out-of-equilibrium regime. The local maxima of $\langle T\rangle$ and $\langle V \rangle$ coincide. For slower sweeps, $\langle T\rangle$ decreases faster than $\langle V\rangle$ which we explore numerically in ED Figure~6.
}
\refstepcounter{EDfig}
\end{figure*}

\begin{figure*}
\includegraphics[width=2\columnwidth]{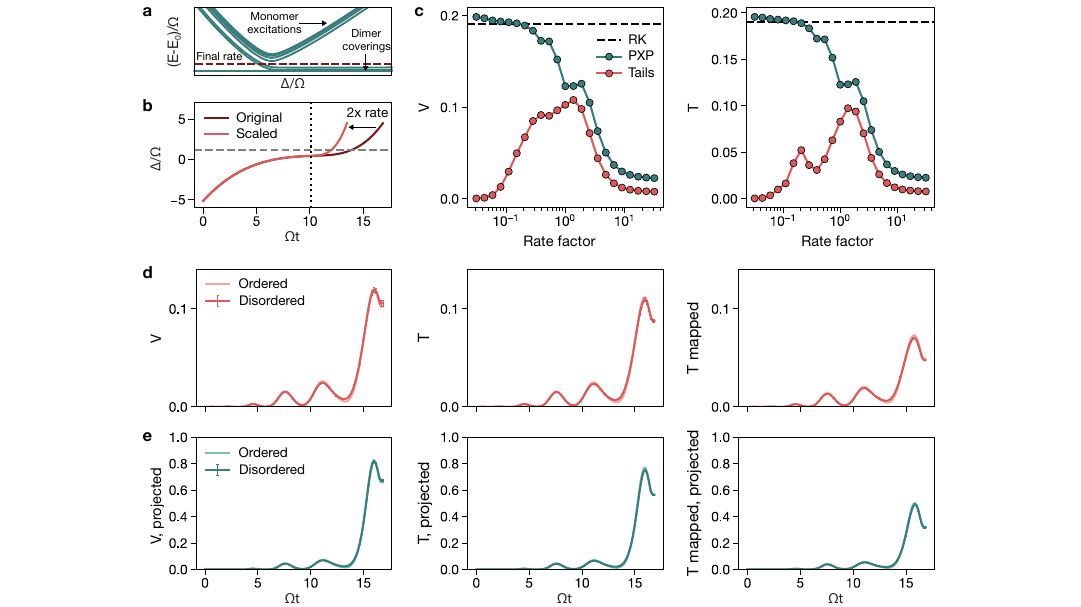}
\caption{\textbf{Preparation of quantum spin lakes in numerical simulations.}
\textbf{a,} Schematic of the energy spectrum of the Rydberg Hamiltonian on the kagome lattice as a function of detuning.
For $\Delta \gg \Omega$, monomer excitations are strongly gapped whereas vdW tails and perturbative couplings lead to a small splitting between dimer coverings.
Hemidiabatic preparation~\cite{sahay_quantum_2023, gjonbalaj_shortcuts_2025} corresponds to sweeping through the phase transition at a rate (dashed line) adiabatic relative to monomer excitations but sudden relative to the low-energy splitting.
\textbf{b,} To confirm that the final sweep rate in Fig.~\ref{fig4}d corresponds to the hemidiabatic regime, we modify the 0.8$\mu$s detuning profile by accelerating the sweep through the phase transition (gray dashed line) by a multiplicative rate factor after the inflection point (dotted line).
\textbf{c,} Evidence for hemidiabatic preparation of quantum spin lakes. 
We use exact diagonalization to simulate a $3\times3$ unit cell lattice with periodic boundary conditions along each lattice vector.
Starting from the ground state at $t=0$, we evolve under the modified sweep from \textbf{b} and plot the final potential and kinetic energies for both the PXP approximation and the full simulation of the long-range vdW interaction.
For large rate factors, we see that accelerating the sweep decreases these RK energies due to the large number of monomers in the final state.
For small rate factors, although the PXP model approaches the RK state ($\langle T \rangle = \langle V \rangle \approx 0.19$ for this system size), the inclusion of vdW tails changes the relative phases and amplitudes between dimer configurations which reduces both $T$ and $V$.
Rather, the optimal rate with tails corresponds to the hemidiabatic regime where the sweep is slow enough to remove monomers but fast enough that vdW tails dephasing is limited.
As expected, the experimentally optimized sweep corresponds to rates near the maxima of $T$ and $V$.
\textbf{d,} Effect of disorder on unprojected energies.
To account for sources of disorder in the experiment, we simulate the 0.8\,$\mu$s sweep starting from the empty product state $\ket{1}^{\otimes N}$ 100 times, randomly sampling on-site detuning shifts, global detuning shifts, in-plane positional shifts, and out-of-plane positional shifts from Gaussians with widths of $2\pi \times 0.043$\,MHz, $2\pi \times 0.2$\,MHz, $0.02a$, and $0.08a$, respectively.
We plot $V$, $T$, and $T$ after $100$\,ns of idle evolution under the vdW interaction to simulate the gap for the Raman pulse during mapping down.
We attribute the dip in these energies at the end of the sweep to finite-size effects which differ from the experiment.
\textbf{e,} Effect of same disorder on projected energies, where bonds around the loop are projected onto no dimer (see Fig.~5c).
}
\refstepcounter{EDfig}
\end{figure*}

\begin{figure*}
\includegraphics[width=2\columnwidth]{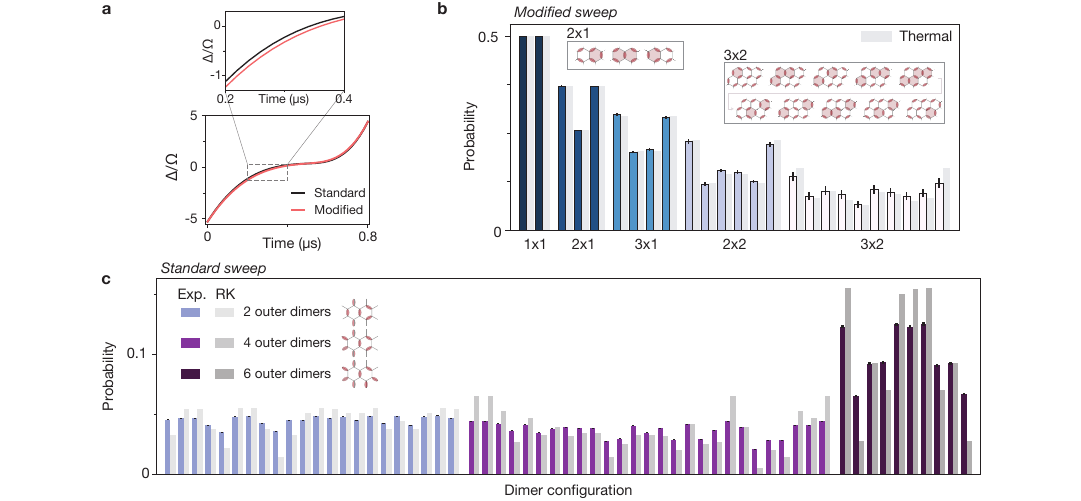}
\caption{\textbf{Dimer configurations in the Rokhsar-Kivelson state.} \textbf{a,} Comparison between the standard and weakly modified detuning sweeps used in the main text. Zoom-in highlights the scale of the modification. \textbf{b,} Probability distribution of dimer configurations in isolated subsystems using the modified sweep. The experimental distribution follows a thermal distribution with a fitted inverse temperature $\beta = 9.0(2) / V_0$, where $V_0$ is the nearest-neighbor interaction strength (light gray bars). The same data is used for the inset of Fig.~5a.
The dimer configurations are explicitly shown for two subsystem sizes. The vdW tail energy is larger for configurations with more flippable plaquettes within the subsystem, shaded in red, which explains the reduced probability of these configurations. \textbf{c,} Probability distribution within a 2$\times$1 subsystem for varying number of dimers surrounding the subsystem, using the standard 0.8\,$\mu$s sweep. 
The uniform distribution in Fig.~5a corresponds to 0 outer dimers and the single configuration with 8 outer dimers is omitted. 
The outer dimers restrict the state of the lattice surrounding the subsystem to be in one of a finite number of allowed dimer configuration which satisfy the boundary conditions, such that even the perfect RK state has a non-uniform probability distribution within subsystems in this analysis.
The RK distribution is shown for a $12\times 6$ lattice with periodic boundary conditions obtained from Monte Carlo simulations; we find a qualitatively similar distribution for other lattice dimensions although the precise amplitudes differ slightly (not shown). We find experiment is in good agreement with this qualitative distribution.
\\
\\
\\
\\
\\
\\}
\refstepcounter{EDfig}
\end{figure*}

\begin{figure*}
\includegraphics[width=2\columnwidth]{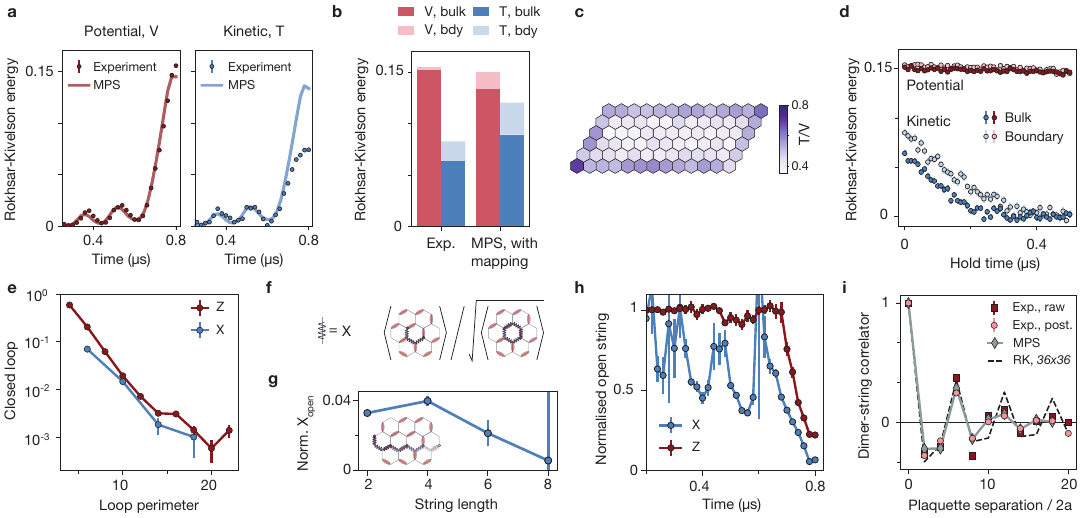}
\caption{\textbf{Loops and strings in the quantum spin liquid.}
\textbf{a,} Time-dependent matrix product state (MPS) simulations of $\langle V \rangle$ and $\langle T\rangle$, compared to experiment. For the final state, MPS predicts $\langle T\rangle/\langle V\rangle = 0.92$. The lattice used for the numerics is three plaquettes tall and infinite in the horizontal direction, with the van der Waals interaction truncated at $\sqrt{7}a$ and a bond dimension of $\chi=256$. \textbf{b,} Including the phase accumulation from vdW tails during the 100\,ns mapping down gap reduces the expectation value in MPS numerics to $\langle T\rangle/\langle V\rangle = 0.76$, compared to $\langle T\rangle/\langle V\rangle = 0.47(1)$ in experiment; numerics cannot account for the Rydberg $\pi$-pulse in the three-level system.
In experiment, plaquettes on the boundary (bdy) have higher $\langle T \rangle$ than in the bulk, where van der Waals tails are larger, but similar $\langle V \rangle$.
For this plot only, the four corner plaquettes are excluded.
\textbf{c,} Spatial plot shows the energy ratio in experiment. 
\textbf{d,} Evolution of experimental RK energies after state preparation, where the Rabi drive is turned off during the hold time. Decay of $\langle V \rangle$ is consistent with Rydberg $T_1$ error, while $\langle T \rangle$ decays faster as the different dimer configurations accumulate relative phases but the populations remain approximately unchanged.
\textbf{e,} Perimeter-law decay of closed X and Z loops. Data for X loops are the same as in Fig.~5f.
\textbf{f,} Normalization of open X strings. See Fig.~6c for the corresponding definition for Z strings.
\textbf{g,} Open X strings are small and decay to zero with increasing string length, consistent with the state being described by a deconfined field theory~\cite{semeghini_probing_2021, verresen_prediction_2021}.
\textbf{h,} Dynamics of open strings.
X strings, which create monomers at their endpoints, are finite when there is a high density of monomers. Z strings are trivially +1 for the initial product state. 
We show open strings with length 3 which are normalized by the length-6 closed loops that enclose a single plaquette. 
\textbf{i,} Effect of defects on the dimer-string correlator. Postselection on no monomers or double-dimers inside the closed loop (such that it is $+1$) slightly improves agreement between experiment and the RK prediction from Monte Carlo simulations.
}
\refstepcounter{EDfig}
\end{figure*}

\begin{figure*}
\includegraphics[width=2\columnwidth]{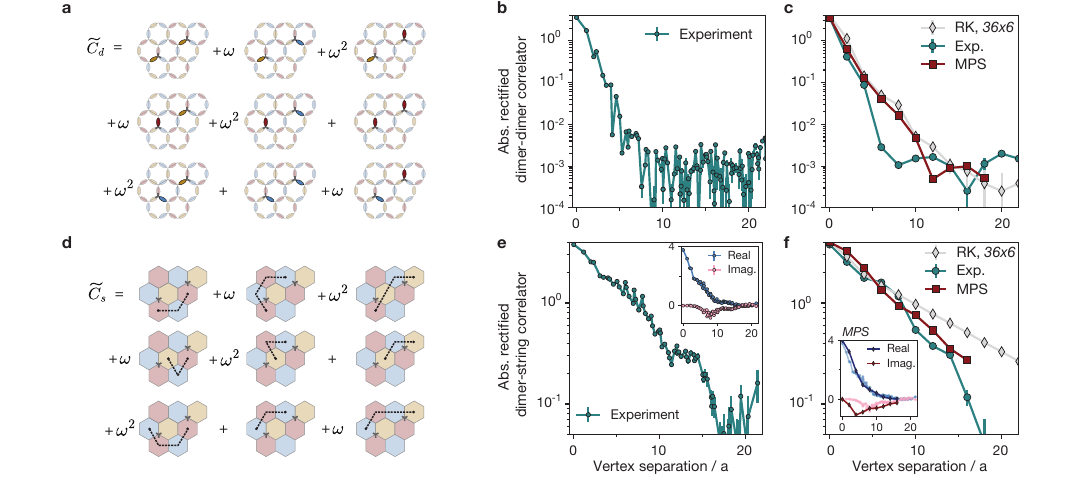}
\caption{\textbf{Diagonal correlators in the emergent U(1) gauge theory.}
 \textbf{a,} Schematic of the rectified dimer-dimer correlator between two vertices.  
The sum involves nine (unconnected) dimer-dimer correlation functions, as highlighted, and the colors indicate the relevant phases, determined by the sublattices (a,b,c) of the neighboring plaquette pair. Here, $\omega = e^{2\pi i/3}$.
\textbf{b,} Rectified dimer-dimer correlator in experiment.
\textbf{c,} Comparison to MPS simulations of the experimental preparation protocol and to Monte Carlo simulations of the RK state on a $36\times 6$ lattice with periodic boundary conditions. Inset shows real and imaginary components in MPS.
MPS numerics use the PXP approximation, a bond dimension of $\chi=350$, and a four-plaquette-tall lattice that is infinite in the horizontal direction.
\textbf{d,} Schematic of rectified dimer-string correlator between two vertices.
The sum involves nine unrectified dimer-string correlators between plaquette centers and the colors indicate the plaquette sublattices that define the phases.
\textbf{e,} Rectified dimer-string correlator in experiment.
While $\widetilde C_s$ is a real quantity for a translationally invariant system, here we find an appreciable imaginary component (inset).
\textbf{f,} Comparison of the dimer-string correlator to MPS and Monte Carlo simulations. Inset shows real and imaginary parts in MPS numerics, with the experimental data shown as faint markers.
In \textbf{c} and \textbf{f} the experimental correlations are shown along the horizontal direction only to compare to numerics, while in \textbf{b} and \textbf{e} they are radially averaged over the lattice, excluding vertices on the boundary.
}
\refstepcounter{EDfig}
\end{figure*}

\begin{figure*}
\includegraphics[width=2\columnwidth]{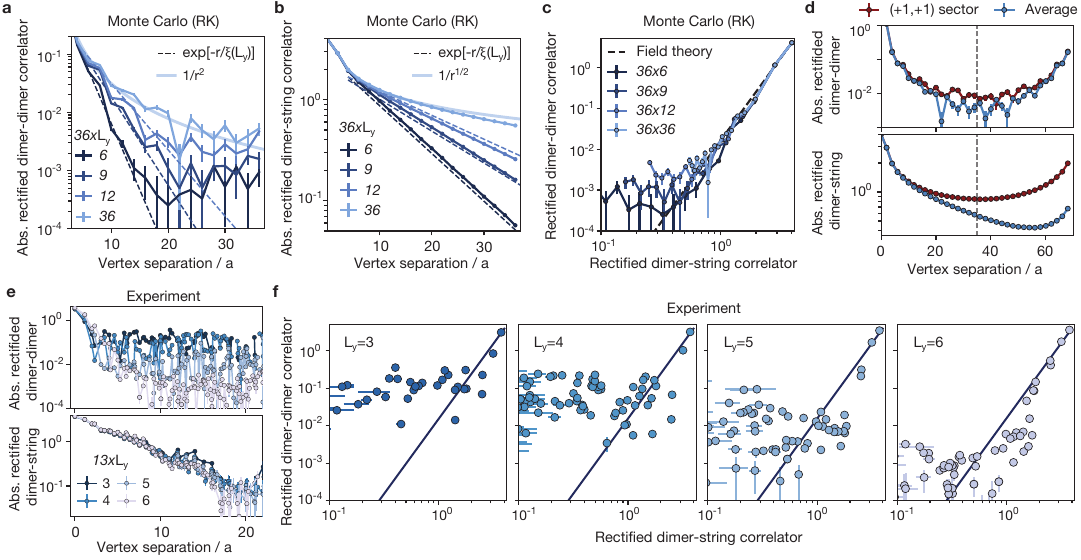}
\caption{\textbf{Finite-size scaling in numerics and experiment.}
\textbf{a,b} Finite-size scaling of rectified correlators in Monte Carlo simulations of the RK state. The correlators converge towards the predicted $1/r^2$ and $1/r^{1/2}$ decays at large system sizes. Dashed lines are guides to the eye showing exponential decays with correlation length $\xi(L_y)\propto L_y$. Experiment is well-captured by the $36\times 6$ data (ED Fig.~9).
\textbf{c,} Relation between the magnitudes of the rectified dimer-dimer and dimer-string correlators in Monte Carlo simulations. 
The ratio of 4 from $U(1)$ gauge theory holds for all system sizes until the dimer-dimer correlator plateaus.
\textbf{d,} Effect of topological sector in Monte Carlo with periodic boundary conditions.
Within the $(+1,+1)$ topological sector, defined as $+1$ eigenvalues of the two non-contractible $Z$-strings, both correlators are symmetric about half-length (dashed line), as enforced by translation and reflection symmetries. 
Averaging over all four topological sectors suppresses the correlators overall, and leads to a distance-dependent decay of the dimer–string correlator beyond half-length. 
In all other plots, correlators are averaged over sectors.
Data are for a $36\times 36 $ lattice.
\textbf{e,} Finite-size scaling in experiment.
Upon varying the lattice height from $L_y=3$ to $L_y=6$, at fixed width of $L_x=13$, we find the rectified dimer-string correlator is largely independent of system size while the rectified dimer-dimer decays until the correlations plateau.
This plateau arises from explicit breaking of the lattice's $\mathbb{Z}_3$ symmetry and decreases in magnitude as the lattice height increases.
For all other data in this work, $L_y=6$.
\textbf{f,} 
The field theory prediction (solid line) is in good agreement with the data at short distances for all system sizes until the plateau in the dimer-dimer correlations is reached.
}
\refstepcounter{EDfig}
\end{figure*}

\clearpage

\end{document}